\documentclass[superscriptaddress, showpacs, nofootinbib, amsmath, amssymb, aps, prd, notitlepage]{revtex4-1}
\usepackage{amsmath,amssymb,amsfonts,mathrsfs,bbold}
\usepackage{yfonts}

\usepackage[dvipsnames]{xcolor}

\usepackage{float}
\usepackage{graphics,graphicx}
\usepackage{dcolumn}
\usepackage{bm}
\usepackage{hyperref}
\usepackage{comment}
\usepackage{xcolor}
\usepackage{slashed}
\usepackage{pstricks}
\usepackage{color}
\usepackage{simplewick}
\usepackage{caption}
\usepackage{subcaption}

\usepackage{ragged2e} 
\DeclareCaptionJustification{justified}{\justifying}

\usepackage{multirow}
\usepackage{mathbbol}
\usepackage{mathtools}

\usepackage{slashed}
\usepackage{simpler-wick}

\usepackage[normalem]{ulem}

\usepackage{tikz}
\usetikzlibrary{decorations.pathmorphing, decorations.markings, shapes.geometric, shapes.misc, calc, math}
\tikzstyle{gluon}=[decorate, decoration={coil,aspect=0.8, amplitude=1.5pt,  segment length=3pt}]

\usepackage{stackrel}
\usepackage[most]{tcolorbox}

\allowdisplaybreaks[1]

\def\eq#1{{Eq.~(\ref{#1})}}
\def\fig#1{{Fig.~\ref{#1}}}
\newcommand{\ben}{\begin{eqnarray*}}
\newcommand{\een}{\end{eqnarray*}}
\newcommand{\un}[1]{\underline{#1}}

\newcommand{\pd}{\partial}
\newcommand{\xx}{\underline{x}}
\newcommand{\zz}{\underline{z}}

\newcommand{\bb}{\underline{b}}

\newcommand{\gt}{{\widetilde G}}
\newcommand{\gmt}{{\widetilde \Gamma}}
\newcommand{\ord}[2]{#1^{(#2)}}

\newcommand{\pol}{\text{pol}}

\newcommand{\llangle}{\Big\langle \!\! \Big\langle}
\newcommand{\rrangle}{\Big\rangle \!\! \Big\rangle}

\newcommand{\as}{\alpha_s}

\newcommand{\bas}{{\bar\alpha}_s}

\makeatletter
\DeclareRobustCommand{\cev}[1]{%
  {\mathpalette\do@cev{#1}}%
}
\newcommand{\do@cev}[2]{%
  \vbox{\offinterlineskip
    \sbox\z@{$\m@th#1 x$}%
    \ialign{##\cr
      \hidewidth\reflectbox{$\m@th#1\vec{}\mkern4mu$}\hidewidth\cr
      \noalign{\kern-\ht\z@}
      $\m@th#1#2$\cr
    }%
  }%
}
\makeatother

\begin{document}
\title{Helicity Evolution at Small $x$: Revised Asymptotic Results at Large $N_c\& N_f$}

\author{Daniel Adamiak} 
         \email[Email: ]{adamiak.5@osu.edu}
         \affiliation{Department of Physics, The Ohio State University, Columbus, OH 43210, USA}
         \affiliation{Jefferson Lab, Newport News, Virginia 23606, USA}

\author{Yuri V. Kovchegov} 
         \email[Email: ]{kovchegov.1@osu.edu}
         \affiliation{Department of Physics, The Ohio State University, Columbus, OH 43210, USA}

\author{Yossathorn Tawabutr}
         \email[Email: ]{yossathorn.j.tawabutr@jyu.fi}
         \affiliation{Department of Physics, The Ohio State University, Columbus, OH 43210, USA}
         \affiliation{Department of Physics, University of Jyv\"askyl\"a,  P.O. Box 35, 40014 University of Jyv\"askyl\"a, Finland}
         \affiliation{Helsinki Institute of Physics, P.O. Box 64, 00014 University of Helsinki, Finland}

\begin{abstract}
We present a numerical solution of the revised version \cite{Cougoulic:2022gbk} of the small-$x$ helicity evolution equations from \cite{Kovchegov:2015pbl, Kovchegov:2018znm} at large $N_c$ and $N_f$. (Here $N_c$ and $N_f$ are the numbers of quark colors and flavors, respectively.) The evolution equations are double-logarithmic in the Bjorken $x$ variable, resumming powers of $\as \, \ln^2 (1/x)$ with $\as$ the strong coupling constant. The large-$N_c \& N_f$ evolution we consider includes contributions of small-$x$ quark emissions and is thus more realistic than the large-$N_c$ one, which only involves gluon emissons. The evolution equations  \cite{Kovchegov:2015pbl, Kovchegov:2018znm,Cougoulic:2022gbk} are written for the so-called ``polarized dipole amplitudes", which are related to the helicity distribution functions and the $g_1$ structure function. Unlike the previously reported solution \cite{Kovchegov:2020hgb} of the earlier version of helicity evolution equations at large $N_c \& N_f$ \cite{Kovchegov:2015pbl, Kovchegov:2018znm}, our solution does not exhibit periodic oscillations in $\ln (1/x)$ for $N_f <  2 N_c$, while only showing occasional sign reversals. For $N_f = 2 N_c$, we report oscillations with $\ln (1/x)$, similar to those found earlier in \cite{Kovchegov:2020hgb}. We determine the intercept of our evolution for $N_f <  2 N_c$ as well as the parameters of the oscillatory behavior for $N_f = 2 N_c$. We compare our results to the existing resummation \cite{Bartels:1996wc} and finite-order calculations for helicity-dependent quantities in the literature.  
\end{abstract}

\maketitle
\tableofcontents


\section{Introduction}
\label{sec:intro}

Small Bjorken $x$ behavior of helicity-dependent observables has received a considerable amount of attention in recent years \cite{Kovchegov:2015pbl, Hatta:2016aoc, Kovchegov:2016zex, Kovchegov:2016weo, Kovchegov:2017jxc, Kovchegov:2017lsr, Kovchegov:2018znm, Kovchegov:2019rrz, Boussarie:2019icw, Cougoulic:2019aja, Kovchegov:2020hgb, Cougoulic:2020tbc, Chirilli:2021lif, Adamiak:2021ppq, Kovchegov:2021lvz, Cougoulic:2022gbk,Borden:2023ugd}, with the aim of constraining the amount of the proton spin carried by the partons in that region of phase space. Apart from being a question of general interest in our understanding of the proton structure, the knowledge of helicity distributions at low $x$ may help us solve the proton spin puzzle \cite{Aidala:2012mv,Accardi:2012qut,Leader:2013jra,Aschenauer:2013woa,Aschenauer:2015eha,Boer:2011fh,Proceedings:2020eah,Ji:2020ena,AbdulKhalek:2021gbh}. Indeed, the helicity parton distribution functions (hPDFs) for quarks ($\Delta\Sigma(x,Q^2)$, which is flavor-singlet) and gluons ($\Delta G(x,Q^2)$), when integrated over all $x$, give us the contributions of the quark ($S_q$) and gluon ($S_G$) helicities to the proton spin, 
\begin{align}
S_q(Q^2) = \frac{1}{2} \int\limits_0^1 dx \; \Delta\Sigma(x,Q^2), \ \ \  \ \
S_G(Q^2) = \int\limits_0^1 dx \; \Delta G(x,Q^2).
\label{eqn:SqSG}
\end{align}
These contributions, in turn, enter the spin sum rules \cite{Jaffe:1989jz} (see also \cite{Ji:1996ek})
\begin{equation}
S_q+L_q+S_G+L_G=\frac{1}{2},
\label{eqn:JM}
\end{equation}
along with the orbital angular momenta (OAM) carried by the quarks ($L_q$) and gluons ($L_G$). The proton spin puzzle is a question of determining the values of all the ingredients ($S_q$, $S_G$, $L_q$, and $L_G$) of the sum rule \eqref{eqn:JM} at some sufficiently high value of the resolution scale $Q^2$. In addition, it would be very interesting to determine the $x$-distributions of $\Delta\Sigma(x,Q^2)$ and $\Delta G(x,Q^2)$ (along with the similar $x$-dependent quantities for $L_q$ and $L_G$ \cite{Bashinsky:1998if,Hagler:1998kg,Harindranath:1998ve,Hatta:2012cs,Ji:2012ba}), to learn which regions in $x$ carry most of the proton spin. 

The efforts to develop a theoretical formalism describing and predicting helicity distributions at low $x$ began with the pioneering work by Bartels, Ermolaev and Ryskin (BER) \cite{Bartels:1995iu,Bartels:1996wc} employing the infrared evolution equations (IREE) approach  \cite{Gorshkov:1966ht,Kirschner:1983di,Kirschner:1994rq,Kirschner:1994vc,Griffiths:1999dj}. The calculation \cite{Bartels:1995iu,Bartels:1996wc} was done in the double-logarithmic approximation (DLA), resumming powers of $\as \ln^2 (1/x)$. It led to helicity phenomenology developed in \cite{Blumlein:1995jp,Blumlein:1996hb,Ermolaev:1999jx,Ermolaev:2000sg,Ermolaev:2003zx,Ermolaev:2009cq}. 

In the more recent works \cite{Kovchegov:2015pbl, Hatta:2016aoc, Kovchegov:2016zex, Kovchegov:2016weo, Kovchegov:2017jxc, Kovchegov:2017lsr, Kovchegov:2018znm, Kovchegov:2019rrz, Cougoulic:2019aja, Kovchegov:2020hgb, Cougoulic:2020tbc, Chirilli:2021lif, Kovchegov:2021lvz, Cougoulic:2022gbk}, the question of the small-$x$ helicity asymptotics was addressed using a different approach based on the shock wave/$s$-channel evolution formalism originally developed in \cite{Mueller:1994rr,Mueller:1994jq,Mueller:1995gb,Balitsky:1995ub,Balitsky:1998ya,Kovchegov:1999yj,Kovchegov:1999ua,Jalilian-Marian:1997dw,Jalilian-Marian:1997gr,Weigert:2000gi,Iancu:2001ad,Iancu:2000hn,Ferreiro:2001qy} to describe the unpolarized eikonal scattering (see \cite{Gribov:1984tu,Iancu:2003xm,Weigert:2005us,JalilianMarian:2005jf,Gelis:2010nm,Albacete:2014fwa,Kovchegov:2012mbw,Morreale:2021pnn} for reviews and \cite{Altinoluk:2014oxa,Balitsky:2015qba,Balitsky:2016dgz, Kovchegov:2017lsr, Kovchegov:2018znm, Chirilli:2018kkw, Jalilian-Marian:2018iui, Jalilian-Marian:2019kaf, Altinoluk:2020oyd, Kovchegov:2021iyc, Altinoluk:2021lvu, Kovchegov:2022kyy, Altinoluk:2022jkk, Altinoluk:2023qfr,Altinoluk:2023dww}  for generalizations of the formalism to the sub-eikonal and sub-sub-eikonal observables). Helicity evolution equations in DLA for the sub-eikonal ``polarized dipole amplitudes" were constructed in  \cite{Kovchegov:2015pbl, Kovchegov:2016zex, Kovchegov:2017lsr, Kovchegov:2018znm} (KPS) (see also \cite{Chirilli:2021lif}), with important corrections and modifications found more recently in \cite{Cougoulic:2022gbk} (KPS-CTT). The equations \cite{Kovchegov:2015pbl, Kovchegov:2016zex, Kovchegov:2017lsr, Kovchegov:2018znm,Cougoulic:2022gbk}  close in the large-$N_c$ (cf. \cite{Balitsky:1995ub,Balitsky:1998ya,Kovchegov:1999yj,Kovchegov:1999ua}) and the large-$N_c \& N_f$ limits. Beyond those limits, a helicity version of the Jalilian-Marian--Iancu--McLerran--Weigert--Leonidov--Kovner (JIMWLK) \cite{Jalilian-Marian:1997dw,Jalilian-Marian:1997gr,Weigert:2000gi,Iancu:2001ad,Iancu:2000hn,Ferreiro:2001qy} functional evolution equation was constructed in \cite{Cougoulic:2019aja}, though corrections from \cite{Cougoulic:2022gbk} have not yet been implemented for it. 

Numerical and analytic solutions of the large-$N_c$ version of KPS-CTT evolution were constructed in \cite{Cougoulic:2022gbk} and \cite{Borden:2023ugd}, respectively. The resulting small-$x$ behavior of hPDFs was found to be of the power-law type,
\begin{align}\label{asymptotics_general}
    \Delta \Sigma (x, Q^2) \sim \Delta G (x, Q^2) \sim \left( \frac{1}{x} \right)^{\alpha_h},
\end{align}
with the power $\alpha_h$, also known as the intercept, found explicitly, $\alpha_h \approx 3.66 \sqrt{\tfrac{\as \, N_c}{2 \pi}}$. While the numerical solution from \cite{Cougoulic:2022gbk} appeared to be in good quantitative agreement with the pure-glue limit of the earlier work by BER \cite{Bartels:1996wc}, the analytical solution found in \cite{Borden:2023ugd} revealed differences between KPS-CTT evolution and BER. The differences appeared in the third decimal point in the intercept (that is, beyond the numerical precision achieved in \cite{Cougoulic:2022gbk}) and starting from four loops in the resummed polarized anomalous dimension $\Delta \gamma_{GG} (\omega)$ \cite{Borden:2023ugd}.

The KPS version of the large-$N_c \& N_f$ helicity evolution was solved numerically in \cite{Kovchegov:2020hgb}. In that reference, the resulting hPDF were found to exhibit an oscillatory behavior as functions of $\ln (1/x)$, in addition to the power-of-$1/x$ growth, similar to that in \eq{asymptotics_general}, of the amplitude of those oscillations. The goal of this paper is to revise \cite{Kovchegov:2020hgb} in light of the corrected KPS-CTT evolution constructed in \cite{Cougoulic:2022gbk}. Below we solve the revised large-$N_c \& N_f$ helicity evolution equations numerically and obtain the resulting small-$x$ asymptotics for hPDFs and for the $g_1$ structure function. We find that the oscillations in $\ln (1/x)$, found in \cite{Kovchegov:2020hgb}, are absent in the solution of the modified evolution \cite{Cougoulic:2022gbk} for $N_f <  2 N_c$, but reappear for $N_f = 2 N_c$. This appears to be slightly different from BER, whose evolution does not lead to any oscillations in hPDFs for $N_f \le 2 N_c$. In addition, for all values of $N_f \le 2 N_c$, the polarized dipole amplitudes resulting from the revised evolution \cite{Cougoulic:2022gbk}, along with the corresponding hPDFs, do exhibit an occasional sign reversal, similar to that observed by BER in \cite{Bartels:1996wc} for the non-zero $N_c \& N_f$ version of their IREE. We also compare the intercepts we find in the large-$N_c \& N_f$ limit to the same limit of BER intercepts for different values of $N_f/N_c$: as shown in Table~\ref{tab:BER_comparison}, we find the two sets of intercepts very close, yet slightly different from each other, continuing the trend found in \cite{Borden:2023ugd} at large $N_c$. 

The paper is structured as follows. We present the equations we want to solve in Sec.~\ref{sec:evoleqn}, along with the expression relating the polarized dipole amplitudes to hPDFs and the $g_1$ structure function. The details of the numerical solution are given in Sec.~\ref{sec:numecalc}, where we show the results of our calculations and obtain the intercepts for different values of $N_f/N_c$. We also demonstrate that for $N_f = 2 N_c$ oscillations tend to re-appear in the revised evolution. We run several cross-checks in Sec.~\ref{sec:discuss}, by modifying the initial conditions and by verifying the target--projectile symmetry of the polarized dipole amplitudes. (The latter was never shown to be valid for the original KPS evolution.) We further compare our results to the finite-order exact perturbative calculations \cite{Altarelli:1977zs,Dokshitzer:1977sg,Zijlstra:1993sh,Mertig:1995ny,Moch:1999eb,vanNeerven:2000uj,Vermaseren:2005qc,Moch:2014sna,Blumlein:2021ryt,Blumlein:2021lmf,Davies:2022ofz,Blumlein:2022gpp} in Sec.~\ref{sec:DGLAP}. We conclude in Sec.~\ref{sec:conclusion}.


\section{Small-$x$ Helicity Evolution Equations}
\label{sec:evoleqn}

As we mentioned above, a set of small-$x$ evolution equations has been derived in \cite{Kovchegov:2015pbl,Kovchegov:2018znm,Cougoulic:2022gbk} by applying the saturation/~color glass condensate (CGC) framework \cite{Gribov:1984tu,Iancu:2003xm,Weigert:2005us,JalilianMarian:2005jf,Gelis:2010nm,Albacete:2014fwa,Kovchegov:2012mbw,Morreale:2021pnn} to the helicity-dependent sub-eikonal contributions to the amplitudes for a color-dipole scattering on a longitudinally polarized target proton. In particular, one of the (anti)quarks in the dipole, whose helicity is tracked through its interaction with the polarized target, interacts with the target with the $S$-matrix corresponding to the so-called sub-eikonal \emph{polarized} Wilson line, $V_{\xx',\xx;\,\sigma',\sigma}^{\pol}$. Here, the polarized (anti)quark is incident with helicity $\sigma$ and transverse position $\xx$, interacts with the target at the helicity-dependent, sub-eikonal level and leaves the interaction with helicity $\sigma'$ and transverse position $\xx'$. More details about the polarized Wilson line will be outlined below. On the other hand, the other (anti)quark in the dipole  interacts with the target with the $S$-matrix simply corresponding to the infinite eikonal light-cone Wilson line, $V_{\xx} = V_{\xx}[\infty,-\infty]$, with
\begin{align}\label{Vxunpol}
V_{\xx}[x_f^-,x_i^-] &= \mathcal{P}\exp \left[ig\int\limits_{x_i^-}^{x_f^-}dx^-A^+(0^+,x^-,\xx) \right],
\end{align}
where $\mathcal{P}$ is the path-ordering operator and $A^{\mu} = \sum_aA^{a\mu}t^a$ is the background gluon field with $t^a$'s the SU($N_c$) generators in the fundamental representation. 
We also define similarly the adjoint-representation counterpart of the finite-length light-cone Wilson line as
\begin{align}\label{Uxunpol}
U_{\xx}[x_f^-,x_i^-] &= \mathcal{P}\exp \left[ig\int\limits_{x_i^-}^{x_f^-}dx^-\mathcal{A}^+(0^+,x^-,\xx) \right],
\end{align}
where $\mathcal{A}^{\mu} = \sum_aA^{a\mu}T^a$ with $T^a$'s the SU($N_c$) generators in the adjoint representation. Out light-cone variables are defined as $x^\pm = (t \pm z)/\sqrt{2}$, while the transverse vectors are denoted by $\xx = (x^1, x^2)$.

The polarized Wilson line can be further decomposed into the \emph{type-1} and \emph{type-2} terms, with different helicity structure, such that \cite{Kovchegov:2018znm,Cougoulic:2022gbk}
\begin{align}\label{Vxpol}
V_{\xx',\xx;\,\sigma',\sigma}^{\pol} &= \sigma\,\delta_{\sigma\sigma'}V_{\xx}^{\pol[1]}\,\delta^2(\xx-\xx') + \delta_{\sigma\sigma'}V_{\xx',\xx}^{\text{G}[2]}.
\end{align}
The type-1 polarized Wilson line can be further decomposed into the quark- and gluon-exchange terms,
\begin{align}\label{Vxpol1}
V_{\xx}^{\pol[1]} &= V_{\xx}^{\text{q}[1]}+V_{\xx}^{\text{G}[1]}.
\end{align}
As for the type-2 counterpart, only the gluon-exchange term contributes to helicity. The three sub-eikonal terms in the polarized Wilson line can be written in terms of sub-eikonal operators and the finite-length light-cone Wilson lines as \cite{Kovchegov:2018znm,Cougoulic:2022gbk}
\begin{subequations}\label{VqG}
    \begin{align}
      &  V_{\xx}^{\text{q}[1]} = \frac{g^2P^+}{2s}\int\limits_{-\infty}^{\infty}dx_1^-\int\limits_{x_1^-}^{\infty}dx_2^-\,V_{\xx}[\infty,x^-_2]\,t^b\psi_{\beta}(x_2^-,\xx)\,U^{ba}_{\xx}[x_2^-,x_1^-]\left(\gamma^+\gamma^5\right)_{\alpha\beta}\Bar{\psi}_{\alpha}(x_1^-,\xx)\,t^aV_{\xx}[x_1^-,-\infty]\,,   \label{Vq1} \\
     &   V_{\xx}^{\text{G}[1]} = \frac{igP^+}{s}\int\limits_{-\infty}^{\infty}dx^-\,V_{\xx}[\infty,x^-]\,F^{12}(x^-,\xx)\,V_{\xx}[x^-,-\infty]\,,  \label{VG1} \\
     &   V_{\xx',\xx}^{\text{G}[2]} = -\frac{iP^+}{s}\int\limits_{-\infty}^{\infty}dx^- d^2z\,V_{\xx'}[\infty,x^-]\,\delta^2(\xx'-\zz)\,\cev{D}^i(x^-,\zz)\,\Vec{D}^i(x^-,\zz)\,V_{\xx}[x^-,-\infty]\,\delta^2(\xx-\zz)\,.  \label{VG2} 
    \end{align}
\end{subequations}
Here $F^{\mu\nu}$ is the gluon field strength tensor, $\vec{D}^i = \pd^i - i g A^i$ and $\cev{D}^i = \cev{\pd}^i + i g A^i$ are the right- and left-acting covariant derivatives ($i=1,2$), $P^+$ is the large momentum of the (proton) target, and $s$ is the center-of-mass energy squared for the projectile--target scattering. Furthermore, the type-2 polarized Wilson line relates to the following operator \cite{Cougoulic:2022gbk},
\begin{align}\label{ViG2}
    V_{\xx}^{i\,\text{G}[2]} &= \frac{P^+}{2s}\int_{-\infty}^{\infty}dx^-\,V_{\xx}[\infty,x^-]\left[\Vec{D}^i(x^-,\xx) - \cev{D}^i(x^-,\xx)\right] V_{\xx}[x^-,-\infty]\,.
\end{align}

The helicity evolution equations are derived in terms of these polarized Wilson lines and their adjoint counterparts. Similarly to the unpolarized Balitsky--Kovchegov (BK) \cite{Balitsky:1995ub,Balitsky:1998ya,Kovchegov:1999yj,Kovchegov:1999ua}/JIMWLK \cite{Jalilian-Marian:1997dw,Jalilian-Marian:1997gr,Weigert:2000gi,Iancu:2001ad,Iancu:2000hn,Ferreiro:2001qy} evolution equations, the helicity evolution equations do not close in general. Particularly, each iteration in the general evolution equations leads to more complicated forms of multipole operators, which differ from the original dipole operator. However, the evolution becomes closed and yields predictive power once the large-$N_c$ \cite{tHooft:1973alw} or the large-$N_c\& N_f$ limit \cite{Veneziano:1976wm} is taken. (Here, $N_f$ is the number of quark flavor.) While the former is studied in \cite{Cougoulic:2022gbk, Borden:2023ugd}, in this paper we focus on the large-$N_c\& N_f$ limit of the evolution equations.


\subsection{Large-$N_c\& N_f$ Limit}

In the large-$N_c\& N_f$ limit, we take $N_c$ and $N_f$ to be large, while their ratio, $N_f/N_c$ is fixed to a constant \cite{Veneziano:1976wm}.  
In the context of small-$x$ evolution, application of the large-$N_c\& N_f$ limit is outlined in \cite{Kovchegov:2015pbl,Kovchegov:2018znm,Cougoulic:2022gbk}. First, we define the fundamental dipole amplitudes of type 1 and type 2, which correspond to the respective types of polarized Wilson lines, as
\begin{subequations}\label{QG2}
    \begin{align}
        Q(x^2_{10},zs) &= \int d^2\left(\frac{x_0+x_1}{2}\right)Q_{10}(zs)  \label{Q} \\
        &= \int d^2\left(\frac{x_0+x_1}{2}\right) \frac{1}{2N_c}\,\text{Re}\,\llangle\text{T\,tr}\left[V_{\underline{0}}V_{\underline{1}}^{\text{pol}[1]\dagger}\right] + \text{T\,tr}\left[V_{\underline{1}}^{\text{pol}[1]}V_{\underline{0}}^{\dagger}\right]\rrangle(zs)\,, \notag \\
        G_2(x^2_{10},zs) &= \frac{\epsilon^{ik} x_{10}^k}{x^2_{10}}\int d^2\left(\frac{x_0+x_1}{2}\right) G_{10}^{i}(zs) \label{G2} \\
        &= \frac{\epsilon^{ik} x_{10}^k}{x^2_{10}}\int d^2\left(\frac{x_0+ x_1}{2}\right) \frac{1}{2N_c}\,\text{Re}\,\llangle\text{T\,tr}\left[V_{\underline{0}}V_{\underline{1}}^{i\,\text{G}[2]\dagger}\right] + \text{T\,tr}\left[V_{\underline{1}}^{i\,\text{G}[2]}V_{\underline{0}}^{\dagger}\right]\rrangle(zs)\,, \notag
    \end{align}
\end{subequations}
where $V_{\underline{i}}=V_{\xx_i}$ for $i=0,1,\ldots$, $\xx_{ij} = \xx_i-\xx_j$, and $x_{ij} = |\xx_{ij}|$. Here, the unpolarized (anti)quark in the dipole is located at $\xx_0$, while the polarized (anti)quark is located at $\xx_1$. The variable $z$ (roughly) denotes the minus momentum fraction of the softest parton in a dipole. The double angle brackets denote the standard CGC averaging applied to the longitudinally polarized target and multiplied by $z s$, that is $\langle\langle \ldots \rangle\rangle = z s \, \langle \ldots \rangle$; in this re-scaling $z$ denotes the longitudinal momentum fraction of the polarized parton in the dipole. As shown in \cite{Cougoulic:2022gbk}, the pre-factor, $\epsilon^{ik} x_{10}^k/x^2_{10}$, in Eq.~\eqref{G2} is chosen so as to project out the components that contribute to helicity evolution ($i,k = 1,2$). Furthermore, we also need to consider the adjoint dipole amplitude of type 1, whose operator structure in the quark sector and evolution differ from that of its fundamental counterpart in a nontrivial manner. In the large-$N_c\& N_f$ limit where gluons are approximated by color-octet quark-antiquark pairs, the ``fundamental part" of the adjoint type-1 dipole is defined by \cite{Cougoulic:2022gbk}
\begin{align}\label{Gtilde}
    \gt(x^2_{10},zs) &= \int d^2\left(\frac{\xx_0+\xx_1}{2}\right) \gt_{10}(zs) \\
    &= \int d^2\left(\frac{\xx_0+\xx_1}{2}\right)\frac{1}{2N_c}\,\text{Re}\,\llangle\text{T\,tr}\left[V_{\underline{0}}W_{\underline{1}}^{\text{pol}[1]\dagger}\right] + \text{T\,tr}\left[W_{\underline{1}}^{\text{pol}[1]}V_{\underline{0}}^{\dagger}\right]\rrangle(zs) \,, \notag
\end{align}
where
\begin{align}\label{Wpolx}
    W_{\xx}^{\pol[1]} &= V_{\xx}^{\text{G}[1]} + \frac{g^2P^+}{8s}\int_{-\infty}^{\infty}dx_1^-\int_{x_1^-}^{\infty}dx_2^-\,V_{\xx}[\infty,x^-_2]\,\psi_{\beta}(x_2^-,\xx)\left(\gamma^+\gamma_5\right)_{\alpha\beta}\Bar{\psi}_{\alpha}(x_1^-,\xx)\,V_{\xx}[x_1^-,-\infty]\,.
\end{align}

In addition to the three dipole amplitudes, the evolution will also lead to dipole amplitudes with the same operator definitions as $Q$, $\gt$, and $G_2$, but whose physical transverse size does not match the transverse size that determines the cutoff on the longitudinal lifetime for further evolution \cite{Kovchegov:2015pbl,Kovchegov:2018znm,Cougoulic:2022gbk}. This leads to the definition of \emph{neighbor dipole amplitudes} $\Bar{\Gamma}(x^2_{10},x^2_{32},zs)$, $\gmt(x^2_{10},x^2_{32},zs)$ and $\Gamma_2(x^2_{10},x^2_{32},zs)$ as the dipole amplitudes, $Q$, $\gt$ and $G_2$,  respectively, with physical dipole transverse size $x_{10}$ and lifetime $\sim x^2_{32}z$ \cite{Kovchegov:2015pbl,Kovchegov:2016zex,Kovchegov:2017lsr,Kovchegov:2018znm,Cougoulic:2022gbk}. Note that these neighbor dipole amplitudes only appear (become needed) in the region where $x_{10}\gg x_{32}$.

Altogether, with six amplitudes in total, the large-$N_c\& N_f$ evolution equations become a system of six linear integral equations, namely \cite{Cougoulic:2022gbk}
\begin{subequations}\label{evoleq}
\begin{align}
&Q(x^2_{10},zs) = Q^{(0)}(x^2_{10},zs) + \frac{\alpha_sN_c}{2\pi}   \int^z_{\max\left\{\Lambda^2/s,\,1/x^2_{10}s\right\}}\frac{dz'}{z'}\int_{1/z's}^{x^2_{10}} \frac{dx_{21}^2}{x_{21}^2}  \left[2\,{\widetilde \Gamma}(x^2_{10},x^2_{21},z's) +2\,{\widetilde G}(x^2_{21},z's)  \right. \label{evoleqQ} \\
&\;\;\;\;\;\;\;\;\;\;\;\;+ \left.  Q(x^2_{21},z's) - \overline{\Gamma}(x^2_{10},x^2_{21},z's) + 2\,\Gamma_2(x^2_{10},x^2_{21},z's) +2\,G_2(x^2_{21},z's) \right] \notag   \\
&\;\;\;+ \frac{\alpha_sN_c}{4\pi}\int^z_{\Lambda^2/s}\frac{dz'}{z'}\int_{1/z's}^{x^2_{10}z/z'}  \frac{dx_{21}^2}{x_{21}^2} \left[Q(x^2_{21},z's) + 2\,G_2(x^2_{21},z's)\right] ,\notag \\
&\overline{\Gamma}(x^2_{10},x^2_{21},z's) = Q^{(0)}(x^2_{10},z's) + \frac{\alpha_sN_c}{2\pi}   \int^{z'}_{\max\left\{\Lambda^2/s,\,1/x^2_{10}s\right\}}\frac{dz''}{z''}\int_{1/z''s}^{\min\left\{x^2_{10},\,x^2_{21}z'/z''\right\}} \frac{dx_{32}^2}{x_{32}^2}  \left[2\,{\widetilde \Gamma}(x^2_{10},x^2_{32},z''s)  \right. \label{evoleqGmb} \\
&\;\;\;\;\;\;\;\;\;\;\;\;+ \left. 2\,{\widetilde G}(x^2_{32},z''s) + Q(x^2_{32},z''s) - \overline{\Gamma}(x^2_{10},x^2_{32},z''s) + 2\,\Gamma_2(x^2_{10},x^2_{32},z''s) +2\,G_2(x^2_{32},z''s) \right]  \notag    \\
&\;\;\;+ \frac{\alpha_sN_c}{4\pi}\int^z_{\Lambda^2/s}\frac{dz''}{z''}\int_{1/z''s}^{x^2_{21}z'/z''}  \frac{dx_{32}^2}{x_{32}^2} \left[Q(x^2_{32},z''s) + 2\,G_2(x^2_{32},z''s)\right] ,\notag \\
&{\widetilde G}(x^2_{10},zs) = {\widetilde G}^{(0)}(x^2_{10},zs) + \frac{\alpha_sN_c}{2\pi}\int^{z}_{\max\left\{\Lambda^2/s,\,1/x^2_{10}s\right\}}\frac{dz'}{z'}\int_{1/z's}^{x^2_{10}} \frac{dx_{21}^2}{x_{21}^2} \left[{\widetilde \Gamma}(x^2_{10},x^2_{21},z's) +3\,{\widetilde G}(x^2_{21},z's)   \right.  \label{evoleqGt}  \\
&\;\;\;\;\;\;\;\;\;\;\;\;+ \left. 2\,G_2(x^2_{21},z's) + 2\,\Gamma_2(x^2_{10},x^2_{21},z's)  \right] \notag \\
&\;\;\;- \frac{\alpha_sN_f}{8\pi} \int_{\Lambda^2/s}^{z}\frac{dz'}{z'}\int_{1/z's}^{x^2_{10}z/z'} \frac{dx^2_{21}}{x^2_{21}}\left[ \, \overline{\Gamma}^{\text{gen}}(x^2_{20},x^2_{21},z's) + 2\, \Gamma^{\text{gen}}_2(x^2_{20},x^2_{21},z's)\right] ,  \notag \\
&{\widetilde \Gamma}(x^2_{10},x^2_{21},z's) = {\widetilde G}^{(0)}(x^2_{10},zs) + \frac{\alpha_sN_c}{2\pi}\int^{z'}_{\max\left\{\Lambda^2/s,\,1/x^2_{10}s\right\}}\frac{dz''}{z''}\int_{1/z''s}^{\min\left\{x^2_{10},\,x^2_{21}z'/z''\right\}} \frac{dx_{32}^2}{x_{32}^2}   \left[{\widetilde \Gamma}(x^2_{10},x^2_{32},z''s) \right. \label{evoleqGmt}  \\
&\;\;\;\;\;\;\;\;\;\;\;\;+ \left. 3\,{\widetilde G}(x^2_{32},z''s) + 2\,G_2(x^2_{32},z''s) + 2\,\Gamma_2(x^2_{10},x^2_{32},z''s)  \right] \notag \\
&\;\;\;- \frac{\alpha_sN_f}{8\pi} \int_{\Lambda^2/s}^{z'}\frac{dz''}{z''}\int_{1/z''s}^{x^2_{21}z'/z''} \frac{dx^2_{32}}{x^2_{32}}\left[ \, \overline{\Gamma}^{\text{gen}}(x^2_{30},x^2_{32},z''s) + 2\, \Gamma^{\text{gen}}_2(x^2_{30},x^2_{32},z''s)\right] ,  \notag \\
&G_2(x^2_{10},zs) = G^{(0)}_2(x^2_{10},zs) + \frac{\alpha_sN_c}{\pi}\int_{\Lambda^2/s}^z\frac{dz'}{z'}\int_{\max\left\{x^2_{10},\,1/z's\right\}}^{x^2_{10}z/z'} \frac{dx^2_{21}}{x^2_{21}} \left[ {\widetilde G}(x^2_{21},z's) + 2\,G_2(x^2_{21},z's)  \right] , \label{evoleqG2} \\
&\Gamma_2(x^2_{10},x^2_{21},z's) = G^{(0)}_2(x^2_{10},z's) + \frac{\alpha_sN_c}{\pi}\int_{\Lambda^2/s}^{z'x^2_{21}/x^2_{10}}\frac{dz''}{z''}\int_{\max\left\{x^2_{10},\,1/z''s\right\}}^{x^2_{21}z'/z''} \frac{dx^2_{32}}{x^2_{32}}   \left[ {\widetilde G}(x^2_{32},z''s) + 2\,G_2(x^2_{32},z''s)  \right] , \label{evoleqGm2} 
\end{align}
\end{subequations}
where we defined the \emph{generalized dipole amplitudes} as \cite{Kovchegov:2017lsr}
\begin{subequations}\label{gmgen}
\begin{align}
\overline{\Gamma}^{gen}(x^2_{10},x^2_{32},z''s) &= \theta(x_{32}-x_{10})\,Q(x^2_{10},z''s) + \theta(x_{10}-x_{32})\,\overline{\Gamma}(x^2_{10},x^2_{32},z''s) \label{gmbgen} \\
\Gamma_2^{gen}(x^2_{10},x^2_{32},z''s) &= \theta(x_{32}-x_{10})\,G_2(x^2_{10},z''s) + \theta(x_{10}-x_{32})\,\Gamma_2(x^2_{10},x^2_{32},z''s) \, . \label{gm2gen}
\end{align}
\end{subequations}
In the above equations $\Lambda$ is the scale characterizing the target proton, $s$ is the center-of-mass energy squared for the projectile-target scattering, $z$ is the fraction of the projectile's momentum carried by the softest parton in the dipole $10$, while $z'$ and $z''$ are similar fractions for the dipoles $21$ and $32$, respectively.

In our formalism, the $g_1$ structure function relates to the fundamental polarized dipole amplitudes $Q$ and $G_2$  through the relation \cite{Cougoulic:2022gbk}
\begin{align}\label{g1}
    &g_1(x,Q^2) = -\frac{N_c}{4\pi^3}\sum_fZ_f^2 \int\limits_{\Lambda^2/s}^1 \frac{dz}{z} \int\limits_{1/z s}^{\min \left\{ \frac{1}{z Q^2}, \frac{1}{\Lambda^2} \right\} } \frac{dx^2_{10}}{x_{10}^2} \left[Q(x^2_{10},zs) + 2 \, G_2(x^2_{10},zs) \right], 
\end{align}
where $Z_f$ is the fractional charge of the quarks and the sum goes over flavors. Furthermore, the flavor-singlet quark and gluon helicity PDFs relate to the polarized dipole amplitudes by \cite{Cougoulic:2022gbk}
\begin{subequations}\label{hPDFs}
    \begin{align}
        & \Delta\Sigma(x,Q^2) =  - \frac{N_cN_f}{2\pi^3} \int\limits_{\Lambda^2/s}^1 \frac{dz}{z} \int\limits_{1/z s}^{\min \left\{ \frac{1}{z Q^2}, \frac{1}{\Lambda^2} \right\} }  \frac{dx^2_{10}}{x_{10}^2} \left[Q(x^2_{10},zs) + 2 \, G_2(x^2_{10},zs) \right] , \label{qkhPDF} \\
        & \Delta G(x, Q^2) =   \frac{2N_c}{\alpha_s\pi^2} \left[1+x^2_{10}\frac{\partial}{\partial x^2_{10}} \right] G_2\left(x^2_{10},zs=\frac{Q^2}{x}\right) \bigg|_{x^2_{10} = 1/Q^2}  \, \approx \frac{2N_c}{\alpha_s\pi^2} \, G_2\left(x^2_{10},zs=\frac{Q^2}{x}\right) \bigg|_{x^2_{10} = 1/Q^2}. \label{glhPDF}
    \end{align}
\end{subequations}
When writing Eqs.~\eqref{g1} and \eqref{hPDFs} we have been treating $\Lambda$ as the infrared (IR) cutoff, as was done in \cite{Cougoulic:2022gbk}, and not as a scale characterizing the target, as was assumed in Eqs.~\eqref{evoleq}. The last step in \eq{glhPDF} is valid with the DLA accuracy. In \eq{qkhPDF} we have assumed for simplicity that the dipole amplitudes are independent of quark flavors.  

Eqs.~\eqref{g1} and \eqref{hPDFs} allow us to perform all the small-$x$ calculation in terms of the polarized dipole amplitudes, from which the $g_1$ structure function and the helicity PDFs can be computed in the final step.


\subsection{Initial Conditions}
\label{ssec:icintro}

In performing realistic phenomenology, the initial conditions for the small-$x$ helicity evolution have to be deduced from the data at moderate values of Bjorken $x$ \cite{Adamiak:2021ppq}. However, this is a complicated problem in its own right and is beyond the scope of this paper. Instead, in order to obtain the asymptotic behaviors of helicity PDFs and $g_1$ structure function at small $x$, we simplify the issue by employing two different approximations to the initial conditions -- the Born-level approximation and the constant approximation.

The Born-level approximation has been utilized in previous works of similar nature \cite{Kovchegov:2016weo,Kovchegov:2020hgb,Cougoulic:2022gbk}, both at large $N_c$ and large $N_c\& N_f$. However, to properly apply it to the large $N_c\& N_f$ evolution equations \eqref{evoleq}, we need to make a slight change described below.

In Eqs.~\eqref{evoleq}, $\Lambda$ is taken to be an energy scale in the process characterizing the typical transverse momentum in the target. This is in contrast to the treatment in the large-$N_c$ limit, where the corresponding distance scale $1/\Lambda$ is used as the infrared cutoff for the dipole size. Particularly, in our treatment of Eqs.~\eqref{evoleq} the dipole can grow larger than $1/\Lambda$. This leads to a necessary modification to the Born-level initial conditions. In particular, for fixed impact parameter the dipole amplitude of each type can be written at the Born level as \cite{Cougoulic:2022gbk,Kovchegov:2018znm,Kovchegov:2016zex,Kovchegov:2017lsr}
\begin{subequations}\label{dip010}
    \begin{align}
        Q_{10}^{(0)}(zs) &= \gt_{10}^{(0)}(zs) = \frac{\as^2C_F}{2 N_c}\left[\frac{C_F}{|\xx_1 - {\un b}_1|^2} - 2\pi\delta^2(\xx_1 - {\un b}_1)\,\ln(zsx^2_{10})\right] , \label{QG010}\\
        G_{10}^{i(0)}(zs) &= -\frac{\as^2C_F}{N_c}\epsilon^{ij}\frac{(\xx_1^j-\bb^j)}{|\xx_1-\bb|^2}\ln\frac{|\xx_1-\bb|}{|\xx_0-\bb|}\, . \label{Gi010}
    \end{align}
\end{subequations} 
These results follow from the calculation of the helicity-dependent amplitude of the quark- and gluon-exchange diagrams for the dipole projectile interacting with a quark target  located at position $\un b$ in the transverse plane. In the previous works, Eqs.~\eqref{dip010} were integrated over $b_{\perp}$ up to $1/\Lambda$, which acted as the infrared cutoff. In this work, however, we must allow $b_{\perp}$ to exceed $1/\Lambda$. In order to still keep one of the integrals finite, we introduce a lower energy scale, $\Lambda_{\text{IR}}\ll\Lambda$, such that the integrals over $b_{\perp}$ will go up to $1/\Lambda_{\text{IR}}$. Doing so and modifying the polarized target from being a single polarized quark to a polarized dipole of the transverse size $\approx 1/\Lambda$ averaged over all angles leads to the following impact parameter-integrated Born-level initial conditions we will employ in this work,
\begin{subequations}\label{dip0}
    \begin{align}
        Q^{(0)}(x^2_{10},zs) &= \gt^{(0)}(x^2_{10},zs) = \frac{\as^2C_F}{2N_c}\pi\left[C_F\ln\frac{zs}{\Lambda_{\text{IR}}^2} - 2\,\ln\left(zs\min\left\{x^2_{10},\frac{1}{\Lambda^2}\right\}\right)\right] , \label{QG0}\\
        G_2^{(0)}(x^2_{10},zs) &= - \frac{\as^2C_F}{2N_c}\pi \left[ \theta \left( \frac{1}{\Lambda} - x_{10} \right) \ln\frac{1}{x^2_{10}\Lambda^2} +\theta \left( x_{10} - \frac{1}{\Lambda} \right) \frac{1}{x^2_{10}\Lambda^2} \right] \approx - \frac{\as^2 \pi C_F}{2N_c} \theta \left( \frac{1}{\Lambda} - x_{10} \right) \ln\frac{1}{x^2_{10}\Lambda^2} . \label{G20}
    \end{align}
\end{subequations}
This provides a relatively realistic approximation to the initial conditions. The second term in \eq{G20} is not going to source DLA evolution in the equations \eqref{evoleq}, under the assumptions used in deriving these equations \cite{Cougoulic:2022gbk}: while a more comprehensive treatment would have required modifying Eqs.~\eqref{evoleq} for the dipoles with $x_{10} > 1/\Lambda$, here we will neglect this contribution since it is suppressed by at least one logarithm compared to the contribution we have kept in \eq{G20}. 

In the previous numerical solutions for the small-$x$ helicity evolution of similar nature \cite{Kovchegov:2016weo,Kovchegov:2020hgb,Cougoulic:2022gbk} it was observed that the asymptotic solution in the small-$x$ region quickly becomes independent of the initial conditions, with only the overall normalization of the solution dependent on the initial conditions. This inspires an even more simplified approximation to the initial conditions, which we will call the \emph{constant approximation} or the all-one approximation, in which we simply take
\begin{align}\label{dip0all1}
    Q^{(0)}(x^2_{10},zs) &= \gt^{(0)}(x^2_{10},zs) = G_2^{(0)}(x^2_{10},zs) = 1\,.
\end{align}
Most of the numerical computation in this work will be done using the constant approximation to the initial conditions. In order to verify that it produces the same asymptotic results as the more physical Born-level approximation, we will perform several cross checks below, the results of which will be presented in Sec.~\ref{ssec:icresults}.


\section{Numerical Solution}
\label{sec:numecalc}

As shown in \cite{Cougoulic:2022gbk}, parton helicity TMDs and PDFs, together with the $g_1$ structure function, all depend only on $Q(x^2_{10},zs)$ and $G_2(x^2_{10},zs)$. Hence, the main goal of this Section is to study the asymptotic form of these dipole amplitudes as $zs$ grows large \cite{Kovchegov:2016weo, Kovchegov:2020hgb}. Owing to the complexity of the large-$N_c\& N_f$ evolution equations \eqref{evoleq}, only a numerical solution has been developed.


\subsection{Discretization and Recursive Form}
\label{ssec:discretize}

Since the small-$x$ helicity evolution is double-logarithmic in $1/x$, with logarithms of $x$ coming from both the transverse and longitudinal integrals, we begin our numerical computation by defining the following variables,
\begin{align}\label{eta_s}
    \eta &= \sqrt{\frac{\alpha_sN_c}{2\pi}} \, \ln\frac{zs}{\Lambda^2} \;\;\text{,}\;\;\;\;\eta' = \sqrt{\frac{\alpha_sN_c}{2\pi}} \,\ln\frac{z's}{\Lambda^2}\;\;\;\;\text{and}\;\;\;\;\eta'' = \sqrt{\frac{\alpha_sN_c}{2\pi}} \,\ln\frac{z''s}{\Lambda^2} \; , \\
    s_{10} &= \sqrt{\frac{\alpha_sN_c}{2\pi}} \,\ln\frac{1}{x^2_{10}\Lambda^2} \;\;\text{,}\;\;\;\;s_{21} = \sqrt{\frac{\alpha_sN_c}{2\pi}} \,\ln\frac{1}{x^2_{21}\Lambda^2}\;\;\;\;\text{and}\;\;\;\;s_{32} = \sqrt{\frac{\alpha_sN_c}{2\pi}} \,\ln\frac{1}{x^2_{32}\Lambda^2}  \; . \notag
\end{align}
With these changes of variables applied to the longitudinal momentum fractions and the transverse dipole sizes, we rewrite Eqs.~\eqref{evoleq} as
\begin{subequations}\label{evol2}
\begin{align}
& Q(s_{10},\eta) = Q^{(0)}(s_{10},\eta) + \frac{1}{2} \, \int_{0}^{\eta} d\eta'   \int^{\eta'}_{s_{10}+\eta'-\eta}  ds_{21} \left[Q(s_{21},\eta') + 2 \, G_2(s_{21},\eta') \right]  \label{evol2Q}  \\
&\;\;\;+  \int_{\max\{0,\,s_{10}\}}^{\eta} d\eta' \int^{\eta'}_{s_{10}} ds_{21} \left[ 2 \, {\widetilde G}(s_{21},\eta') + 2 \, {\widetilde \Gamma}(s_{10},s_{21},\eta') + Q(s_{21},\eta') -  \overline{\Gamma}(s_{10},s_{21},\eta')  + 2 \, \Gamma_2(s_{10},s_{21},\eta')   \right. \notag   \\
&\;\;\;\;\;\;\;\;\;\;\;\;+ \left.   2 \, G_2(s_{21},\eta')   \right]   ,  \notag  \\
&\overline{\Gamma}(s_{10},s_{21},\eta') = Q^{(0)}(s_{10},\eta') + \frac{1}{2} \, \int_{0}^{\eta'} d\eta''   \int^{\eta''}_{s_{21}+\eta''-\eta'} ds_{32} \left[Q(s_{32},\eta'') + 2 \, G_2(s_{32},\eta'') \right]  \label{evol2Gmb} \\
&\;\;\;+   \int_{\max\{0,\,s_{10}\}}^{\eta'} d\eta''   \int^{\eta''}_{\max\{s_{10}, \, s_{21}+\eta''-\eta'\}}  ds_{32}  \left[ 2\, {\widetilde G} (s_{32},\eta'') + 2\, {\widetilde \Gamma} (s_{10},s_{32},\eta'') +  Q(s_{32},\eta'') -  \overline{\Gamma}(s_{10},s_{32},\eta'') \right.  \notag \\
&\;\;\;\;\;\;\;\;\;\;\;\;+ \left.   2 \, \Gamma_2(s_{10},s_{32},\eta'') + 2 \, G_2(s_{32},\eta'') \right]   , \notag \\
& {\widetilde G}(s_{10},\eta) = {\widetilde G}^{(0)}(s_{10},\eta) - \frac{N_f}{4N_c} \, \int_{0}^{\eta} d\eta' \int^{\min\{s_{10},\,\eta'\}}_{s_{10}+\eta'-\eta} ds_{21}   \left[  Q(s_{21},\eta') +     2 \, G_2(s_{21},\eta')  \right] \label{evol2Gt} \\
&\;\;\;+  \int_{\max\{0,\,s_{10}\}}^{\eta}d\eta' \int^{\eta'}_{s_{10}} ds_{21} \left[ 3 \, {\widetilde G}(s_{21},\eta') + {\widetilde \Gamma}(s_{10},s_{21},\eta') + 2\,G_2(s_{21},\eta') + 2\,\Gamma_2(s_{10},s_{21},\eta') \right.  \notag  \\
&\;\;\;\;\;\;\;\;\;\;\;\;- \left.  \frac{N_f}{4N_c} \, \overline{\Gamma}(s_{10},s_{21},\eta') - \frac{N_f}{2N_c} \,\Gamma_2(s_{10},s_{21},\eta')  \right]   , \notag \\
& {\widetilde \Gamma} (s_{10},s_{21},\eta') = {\widetilde G}^{(0)}(s_{10},\eta') - \frac{N_f}{4N_c} \,  \int_{0}^{\eta'+s_{10}-s_{21}} d\eta'' \int^{\min\{s_{10},\,\eta''\}}_{s_{21}+\eta''-\eta'} ds_{32}  \left[   Q(s_{32},\eta'') +  2  \,  G_2(s_{32},\eta'')  \right] \label{evol2Gmt} \\ 
&\;\;\;+  \int_{\max\{0,\,s_{10}\}}^{\eta'}d\eta'' \int^{\eta''}_{\max\{s_{10},\,s_{21}+\eta''-\eta'\}} ds_{32}  \left[ 3 \, {\widetilde G} (s_{32},\eta'') + {\widetilde \Gamma}(s_{10},s_{32},\eta'') + 2 \, G_2(s_{32},\eta'') +  2 \, \Gamma_2(s_{10},s_{32},\eta'') \right. \notag  \\
&\;\;\;\;\;\;\;\;\;\;\;\;- \left. \frac{N_f}{4N_c} \, \overline{\Gamma}(s_{10},s_{32},\eta'')  - \frac{N_f}{2N_c} \,\Gamma_2(s_{10},s_{32},\eta'') \right], \notag \\ 
& G_2(s_{10}, \eta)  =  G_2^{(0)} (s_{10}, \eta) + 2 \, \int_{0}^{\eta} d\eta'  \int^{\min\{s_{10} ,\, \eta'\}}_{s_{10}+\eta'-\eta} ds_{21} \left[ {\widetilde G} (s_{21} , \eta') + 2 \, G_2 (s_{21} , \eta')  \right] , \label{evol2G2} \\
& \Gamma_2 (s_{10},s_{21} , \eta')  =  G_2^{(0)} (s_{10},\eta') + 2 \, \int_{0}^{\eta'+s_{10}-s_{21}} d\eta''  \int^{\min\{s_{10} , \, \eta''\}}_{s_{21}+\eta''-\eta'} ds_{32} \left[ {\widetilde G} (s_{32} , \eta'') + 2 \, G_2(s_{32} , \eta'')  \right] . \label{evol2Gm2} 
\end{align}
\end{subequations}
In obtaining Eqs.~\eqref{evol2}, the ordering $s_{10}\leq s_{21}\leq\eta'$ was assumed in the arguments of $\overline{\Gamma}$, ${\widetilde \Gamma}$ and $\Gamma_2$. This is the only region where $\overline{\Gamma}$, ${\widetilde \Gamma}$ and $\Gamma_2$ appear in any large-$N_c\& N_f$ evolution kernel, where the subsequent evolution lifetime in the daughter dipole $20$ depends on the smallest transverse distance scale $x_{21}$ in the splitting. Along the way, we also separated the integrals from Eqs.~\eqref{evoleqGt} and \eqref{evoleqGmt} that involve the generalized dipole amplitudes, c.f. Eqs.~\eqref{gmgen}, into the regions where they reduce to the corresponding ordinary and neighbor dipole amplitudes. 

Now, we discretize the integrals in Eqs.~\eqref{evol2} with step size $\delta$ in both directions and define the discretized dipole amplitudes as  
\begin{align}\label{dipij}
        Q_{ij} &= Q\left(i\delta, j\delta\right) \;\;\;\,, \;\;\;\;\;\;\;\overline{\Gamma}_{ikj} = \overline{\Gamma}\left(i\delta, k\delta, j\delta\right) \; , \\
        {\widetilde G}_{ij} &= {\widetilde G}\left(i\delta, j\delta\right) \;\;\;\,, \;\;\;\;\;\;\;{\widetilde \Gamma}_{ikj} = {\widetilde \Gamma}\left(i\delta, k\delta, j\delta\right) \; , \notag \\
        G_{2,ij} &= G_2\left(i\delta, j\delta\right) \;\;,\;\;\;\;\;
        \Gamma_{2,ikj} = \Gamma_2\left(i\delta, k\delta, j\delta\right) . \notag
\end{align}
As a result, we obtain the following discretized evolution equations,
\begin{subequations}\label{evol3}
\begin{align}
& Q_{ij} = Q^{(0)}_{ij} + \frac{1}{2} \, \delta^2 \, \sum\limits_{j'=0}^{j-1} \sum\limits_{i'=i+j'-j}^{j'-1} \left[Q_{i'j'} + 2 \, G_{2,i'j'} \right] \label{evol3Q} \\
&\;\;\;+ \delta^2 \, \sum\limits_{j'=\max\{0,\,i\}}^{j-1} \sum\limits_{i'=i}^{j'-1}  \left[ 2 \, {\widetilde G}_{i'j'} + 2 \, {\widetilde \Gamma}_{ii'j'} + Q_{i'j'} -  \overline{\Gamma}_{ii'j'} + 2 \, \Gamma_{2,ii'j'} + 2 \, G_{2,i'j'}   \right]  ,  \notag  \\
&\overline{\Gamma}_{ikj} = Q^{(0)}_{ij} + \frac{1}{2} \, \delta^2 \, \sum\limits_{j'=0}^{j-1} \sum\limits_{i'=k+j'-j}^{j'-1}   \left[Q_{i'j'} + 2 \, G_{2,i'j'} \right] \label{evol3Gmb} \\
&\;\;\;+  \delta^2 \, \sum\limits_{j'=\max\{0,\,i\}}^{j-1} \sum\limits_{i'=\max\{i,\,k+j'-j\}}^{j'-1}  \left[ 2\, {\widetilde G}_{i'j'} + 2\, {\widetilde \Gamma}_{ii'j'}  +  Q_{i'j'} -  \overline{\Gamma}_{ii'j'}  + 2 \, \Gamma_{2,ii'j'}  + 2 \, G_{2,i'j'} \right]   , \notag \\
& {\widetilde G}_{ij} = {\widetilde G}^{(0)}_{ij} - \frac{N_f}{4N_c} \, \delta^2 \, \sum\limits_{j'=0}^{j-1} \sum\limits_{i'=i+j'-j}^{\min\{i,\,j'\}-1}  \left[  Q_{i'j'} +     2 \, G_{2,i'j'}  \right] \label{evol3Gt} \\ 
&\;\;\;+ \delta^2 \, \sum\limits_{j'=\max\{0,\,i\}}^{j-1} \sum\limits_{i'=i}^{j'-1}  \left[ 3 \, {\widetilde G}_{i'j'} + {\widetilde \Gamma}_{ii'j'} + 2\,G_{2,i'j'} + 2\,\Gamma_{2,ii'j'} - \frac{N_f}{4N_c} \, \overline{\Gamma}_{ii'j'} - \frac{N_f}{2N_c} \,\Gamma_{2,ii'j'}  \right]    , \notag \\
& {\widetilde \Gamma}_{ikj} = {\widetilde G}^{(0)}_{ij} - \frac{N_f}{4N_c} \,  \delta^2 \, \sum\limits_{j'=0}^{i+j-k-1}\sum\limits_{i'=k+j'-j}^{\min\{i,\,j'\}-1}  \left[   Q_{i'j'} +  2  \,  G_{2,i'j'}  \right]  \label{evol3Gmt}  \\
&\;\;\;+ \delta^2 \, \sum\limits_{j'=\max\{0,\,i\}}^{j-1} \sum\limits_{i'=\max\{i,\,k+j'-j\}}^{j'-1}  \left[ 3 \, {\widetilde G}_{i'j'} + {\widetilde \Gamma}_{ii'j'} + 2 \, G_{2,i'j'}+  2 \, \Gamma_{2,ii'j'} - \frac{N_f}{4N_c} \, \overline{\Gamma}_{ii'j'}  - \frac{N_f}{2N_c} \,\Gamma_{2,ii'j'} \right]   , \notag  \\
& G_{2,ij}  =  G_{2,ij}^{(0)} + 2 \, \delta^2 \, \sum\limits_{j'=0}^{j-1} \sum\limits_{i'=i+j'-j}^{\min\{i ,\, j'\}-1}  \left[ {\widetilde G}_{i'j'} + 2 \, G_{2,i'j'}  \right] , \label{evol3G2} \\
& \Gamma_{2,ikj}  =  G_{2,ij}^{(0)}  + 2 \, \delta^2 \, \sum\limits_{j'=0}^{i+j-k-1}\sum\limits_{i'=k+j'-j}^{\min\{i , \, j'\}-1}  \left[ {\widetilde G}_{i'j'} + 2 \, G_{2,i'j'}   \right] .  \label{evol3Gm2}
\end{align}
\end{subequations}
Through a careful consideration of Eqs.~\eqref{evol3}, we see that we need to know the values of the following dipole amplitudes in the following regions to determine the values of $Q_{ij}$, ${\widetilde G}_{ij}$ and $G_{2,ij}$ for $0\leq i \leq i_{\max}$ and $0\leq j\leq j_{\max}$:
\begin{itemize}
    \item $Q_{ij}$, ${\widetilde G}_{ij}$ and $G_{2,ij}$ such that $0\leq j \leq j_{\max}$ and $j-j_{\max} \leq i \leq j$, while also keeping $i\leq i_{\max}$.
    \item $\overline{\Gamma}_{ikj}$, ${\widetilde \Gamma}_{ikj}$ and $\Gamma_{2,ikj}$ such that $0\leq i\leq k\leq j$, with $0\leq j \leq j_{\max}$ and $j-j_{\max} \leq i \leq k \leq j$, while keeping $k\leq i_{\max}$. This is partly because the neighbor dipole amplitudes only appear in Eqs.~\eqref{evol3Q} - \eqref{evol3Gmt}.
\end{itemize}

Furthermore, the numerical computation becomes more efficient once we realize recursive relations coming from equations \eqref{evol3}. For $Q_{ij}$, ${\widetilde G}_{ij}$ and $G_{2,ij}$, we retrieve the initial conditions in the case where $i=j$.  For $i<j$, we can re-write Eqs.~\eqref{evol3Q}, \eqref{evol3Gt} and \eqref{evol3G2} recursively as
\begin{subequations}\label{evol4}
\begin{align}
& Q_{ij} = Q^{(0)}_{ij} - Q^{(0)}_{i(j-1)} + Q_{i(j-1)}  + \frac{1}{2} \, \delta^2 \,  \sum\limits_{i'=i-1}^{j-2} \left[Q_{i'(j-1)} + 2 \, G_{2,i'(j-1)} \right] +   \frac{1}{2} \, \delta^2 \, \sum\limits_{j'=0}^{j-2} \left[Q_{(i+j'-j)j'} + 2 \, G_{2,(i+j'-j)j'} \right]  \label{evol4Q} \\ 
&\;\;\;+ \delta^2 \,  \sum\limits_{i'=i}^{j-2}  \left[ 2 \, {\widetilde G}_{i'(j-1)} + 2 \, {\widetilde \Gamma}_{ii'(j-1)} + Q_{i'(j-1)} -  \overline{\Gamma}_{ii'(j-1)} + 2 \, \Gamma_{2,ii'(j-1)} + 2 \, G_{2,i'(j-1)}   \right]    ,  \notag  \\
& {\widetilde G}_{ij} = {\widetilde G}^{(0)}_{ij} - {\widetilde G}^{(0)}_{i(j-1)} +   {\widetilde G}_{i(j-1)} - \frac{N_f}{4N_c} \, \delta^2    \left[  Q_{(i-1)(j-1)} +     2 \, G_{2,(i-1)(j-1)}  \right] - \frac{N_f}{4N_c} \, \delta^2 \, \sum\limits_{j'=0}^{j-2}    \left[  Q_{(i+j'-j)j'} +     2 \, G_{2,(i+j'-j)j'}  \right]  \notag \\ 
&\;\;\;+ \delta^2 \,  \sum\limits_{i'=i}^{j-2} \left[ 3 \, {\widetilde G}_{i'(j-1)} + {\widetilde \Gamma}_{ii'(j-1)} + 2\,G_{2,i'(j-1)} + 2\,\Gamma_{2,ii'(j-1)} - \frac{N_f}{4N_c} \, \overline{\Gamma}_{ii'(j-1)} - \frac{N_f}{2N_c} \,\Gamma_{2,ii'(j-1)}  \right] , \label{evol4Gt}  \\
& G_{2,ij}  =  G_{2,ij}^{(0)} - G_{2,i(j-1)}^{(0)} + G_{2,i(j-1)} + 2 \, \delta^2  \left[ {\widetilde G}_{(i-1)(j-1)} + 2 \, G_{2,(i-1)(j-1)}  \right] \label{evol4G2}  \\
&\;\;\;+ 2 \, \delta^2 \, \sum\limits_{j'=0}^{j-2} \left[ {\widetilde G}_{(i+j'-j)j'} + 2 \, G_{2,(i+j'-j)j'}  \right] . \notag
\end{align}
\end{subequations}
Note that we need to have $j>0$ in order to have $i<j$. 

Turning to $\overline{\Gamma}_{ikj}$, ${\widetilde \Gamma}_{ikj}$, and $\Gamma_{2,ikj}$, we note that they similarly reduce to their ordinary dipole counterparts $Q_{ij}$, $ {\widetilde G}_{ij}$, and $G_{2,ij}$, respectively, when $i=k$. For $i<k$, Eqs.~\eqref{evol3Gmb}, \eqref{evol3Gmt} and \eqref{evol3Gm2} can be written in recursive forms as
\begin{subequations}\label{evol5}
\begin{align}
&\overline{\Gamma}_{ikj} = Q^{(0)}_{ij} - Q^{(0)}_{i(j-1)} + \overline{\Gamma}_{i(k-1)(j-1)}  + \frac{1}{2} \, \delta^2 \, \sum\limits_{i'=k-1}^{j-2}   \left[Q_{i'(j-1)} + 2 \, G_{2,i'(j-1)} \right]  \label{evol5Gmb}  \\
&\;\;\;+ \delta^2 \, \sum\limits_{i'=\max\{i,\,k-1\}}^{j-2} \left[ 2\, {\widetilde G}_{i'(j-1)} + 2\, {\widetilde \Gamma}_{ii'(j-1)}  +  Q_{i'(j-1)} -  \overline{\Gamma}_{ii'(j-1)}  + 2 \, \Gamma_{2,ii'(j-1)}  + 2 \, G_{2,i'(j-1)} \right]   ,  \notag  \\
& {\widetilde \Gamma}_{ikj} =   {\widetilde G}^{(0)}_{ij} - {\widetilde G}^{(0)}_{i(j-1)} + {\widetilde \Gamma}_{i(k-1)(j-1)}   \label{evol5Gmt}  \\
&\;\;\;+ \delta^2 \,   \sum\limits_{i'=\max\{i,\,k-1\}}^{j-2}  \left[ 3 \, {\widetilde G}_{i'(j-1)} + {\widetilde \Gamma}_{ii'(j-1)} + 2 \, G_{2,i'(j-1)}+  2 \, \Gamma_{2,ii'(j-1)} - \frac{N_f}{4N_c} \, \overline{\Gamma}_{ii'(j-1)}  - \frac{N_f}{2N_c} \,\Gamma_{2,ii'(j-1)} \right]    ,  \notag \\
& \Gamma_{2,ikj}  =      G_{2,ij}^{(0)} - G_{2,i(j-1)}^{(0)} + \Gamma_{2,i(k-1)(j-1)} \, . \label{evol5Gm2}
\end{align}
\end{subequations}
In the case where $j=0$, we have $0=i=k=j$. Consequently, the neighbor dipole amplitudes reduce to their respective initial conditions, as can also be seen directly from Eqs.~\eqref{evol3}.

Notice that the first sum in Eq.~\eqref{evol3Gmt}, containing the term $Q_{i'j'}+2\,G_{2,i'j'}$, does not survive in Eq.~\eqref{evol5Gmt}. There are two reasons for this, which depend on the values of $i$, $j$ and $k$. In particular, if $i+j-k-1\geq 0$, then the summation term remains the same once we simultaneously reduce $j$ and $k$ by $1$. On the other hand, in the case where $i+j-k-1 < 0$, which is possible for $i<0$, the specified term in Eq.~\eqref{evol3Gmt} simply vanishes because the upper limit of the sum over $j'$ is now below the corresponding lower limit. As a result, the recursive form in Eq.~\eqref{evol5Gmt} holds true in both regimes.

Similar to the computation at large $N_c$, for each step size, $\delta$, and maximum rapidity, $\eta_{\max}$, we start by computing each dipole amplitude at $j=0$ using the respective initial condition (the inhomogeneous term). Then, we compute their values at $j=1$ based on the values at $j=0$, with the help of Eqs.~\eqref{evol4} and \eqref{evol5}. Afterwards, the $j=2$ values can similarly be computed based on the $j=1$ values, {\sl etc.} The recursive relations \eqref{evol4} and \eqref{evol5} allow us to calculate the six polarized dipole amplitudes all the way up to $\eta = \eta_{\max}$. For each $\eta$, we need to compute the amplitudes for all $s_{10}$ (and $s_{21}$, if applicable) in the $\eta-\eta_{\max}\leq s_{10}\leq s_{21}\leq \eta$ region \cite{Kovchegov:2020hgb}. This is due to the fact that the soft-quark emission terms in Eqs.~\eqref{evol4Q}, \eqref{evol4Gt}, \eqref{evol5Gmb} and \eqref{evol5Gmt} have logarithmic divergence in different transverse position regions from their soft-gluon emission counterparts.

Our numerical solution indicates that the asymptotic behavior of the results differs qualitatively between the cases where $N_f\leq 5$ and $N_f= 6$ (both with $N_c = 3$). This warrants separating the discussion of the results into two respective sections, one for each case. In Sec.~\ref{ssec:lowNf} we discuss the numerical solution in the $N_f\leq 5$ case, while the $N_f= 6$ case is discussed in Sec.~\ref{ssec:highNf}. Note that we are solving the large-$N_c \& N_f$ evolution equations: in that limit both $N_c$ and $N_f$ are very large, and only their ratio $N_f/N_c$ is fixed.  Here and below, when we talk about particular values of $N_f$ and $N_c$, this should be understood only as fixing the ratio $N_f/N_c$ while assuming that both $N_c$ and $N_f$ are large. For instance, when we choose $N_f = 4$ and $N_c=3$, this only implies that $N_f/N_c = 4/3$, and should not be thought of as fixing the values of $N_c$ and $N_f$.


\subsection{Solution for $N_f\leq 5$}
\label{ssec:lowNf}

Performing the iterative computation outlined in Section~\ref{ssec:discretize}, we obtain the numerical values of $Q(s_{10},\eta)$, $G_2(s_{10},\eta)$ and ${\widetilde G}(s_{10},\eta)$ at various $s_{10}$ and $\eta$ on the grid. In order to extract their large-$\eta$ asymptotic form, which will later be useful to determine small-$x$ asymptotics of helicity PDFs, additional steps must be performed to the raw numerical results. To describe the subsequent process more clearly, we consider a sample run with step size $\delta = 0.1$ and maximum rapidity $\eta_{\max}=70$. As for the number of flavors, we first consider $N_f=4$, which is qualitatively similar to any $N_f\leq 5$. The procedure will be similar for other $\delta$ and $\eta_{\max}$ \cite{Cougoulic:2022gbk, Kovchegov:2020hgb, Kovchegov:2016weo}. Note that we fix the number of quark colors at $N_c=3$ throughout our large-$N_c\& N_f$ calculation, in the sense of the comment made above that only the ratio $N_f/N_c$ is truly fixed.

Furthermore, throughout this Section, we employ the constant ``all-one" approximation of initial condition, which upon discretization takes the form,
\begin{align}\label{asym1}
Q^{(0)}_{ij} &= {\widetilde G}^{(0)}_{ij} = G^{(0)}_{2,ij} = 1\,.
\end{align}
A more detailed discussion about the choices of initial conditions, including a justification for our simplified choice \eqref{asym1}, will be given in the next Section. 

The main goal of this Section is two-fold. First, we determine the asymptotic forms of $Q(0,\eta)$, $G_2(0,\eta)$ and ${\widetilde G}(0,\eta)$, that is, we only fit the polarized dipole amplitudes along the $s_{10}=0$ line. These results will provide sufficient ingredients for us to obtain the asymptotic form of helicity PDFs and the $g_1$ structure function at small $x$, which is the second part of our goal.

\begin{figure}[ht]
     \centering
     \begin{subfigure}[b]{0.32\textwidth}
         \centering
         \includegraphics[width=\textwidth]{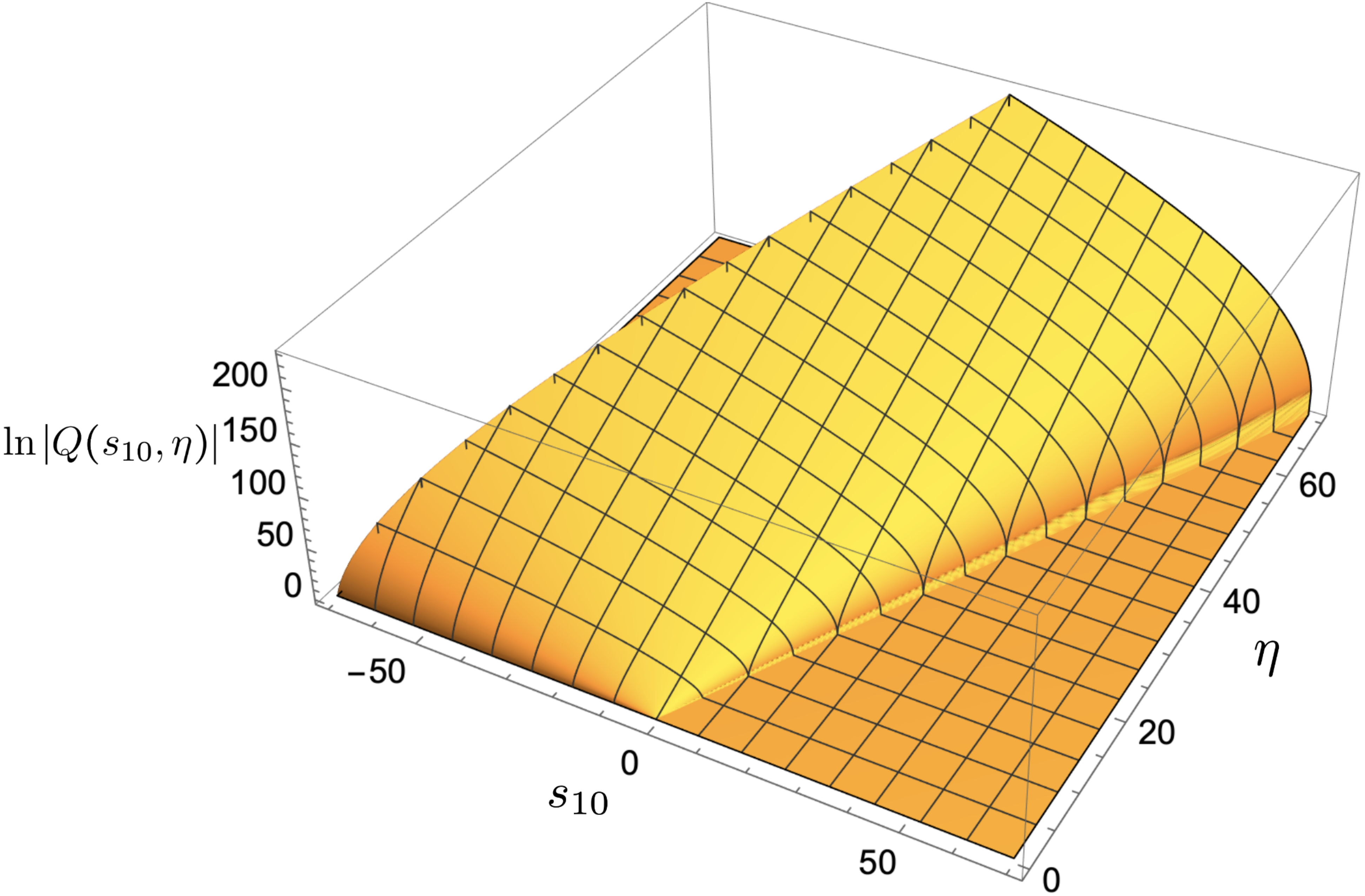}
         \caption{$\ln\left|Q(s_{10},\eta)\right|$}
         \label{fig:QGG3dNf4_Q}
     \end{subfigure} 
  \;
     \begin{subfigure}[b]{0.32\textwidth}
         \centering
         \includegraphics[width=\textwidth]{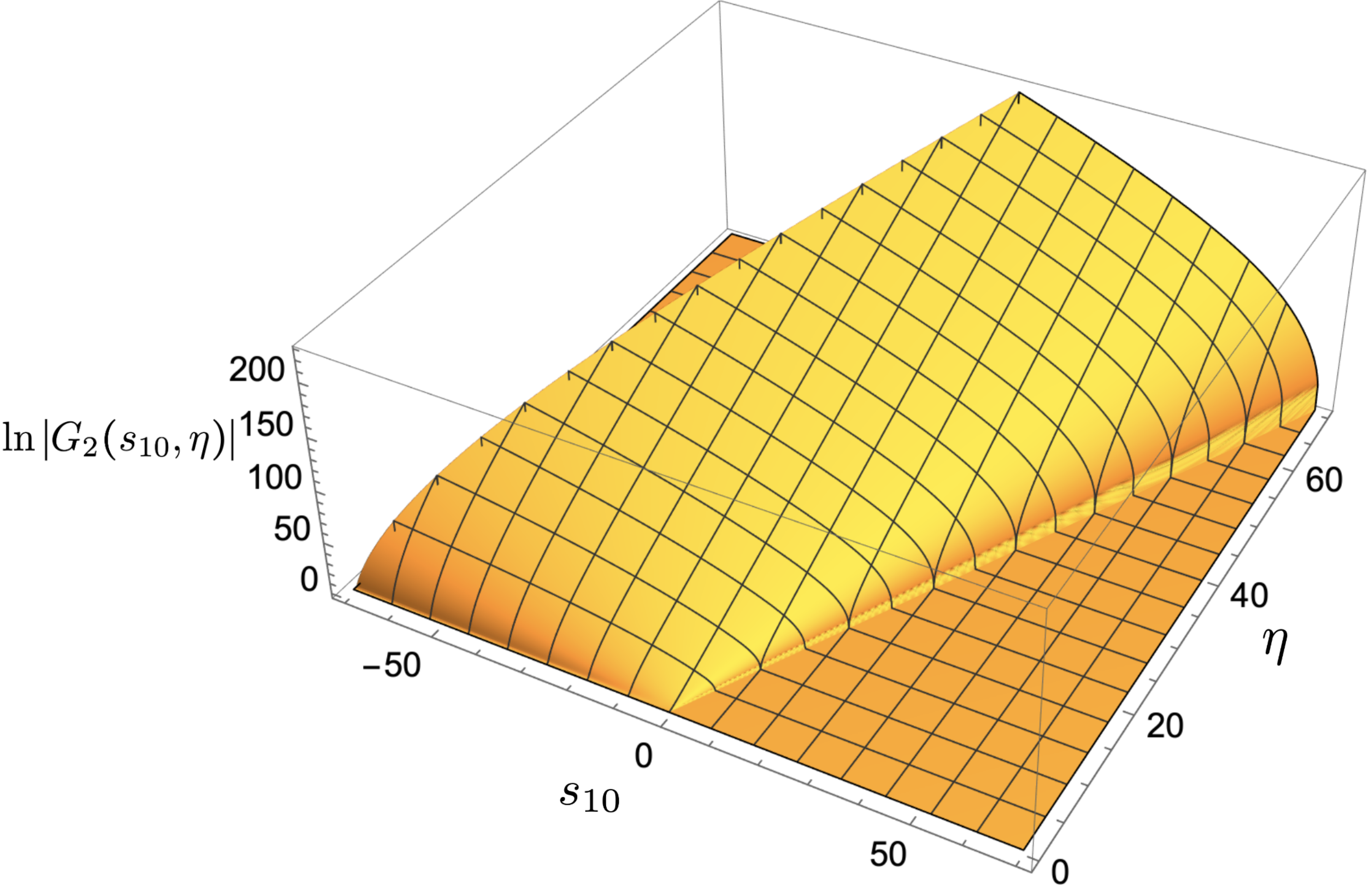}
         \caption{$\ln\left|G_2(s_{10},\eta)\right|$}
         \label{fig:QGG3dNf4_G2}
     \end{subfigure} 
 \;
     \begin{subfigure}[b]{0.32\textwidth}
         \centering
         \includegraphics[width=\textwidth]{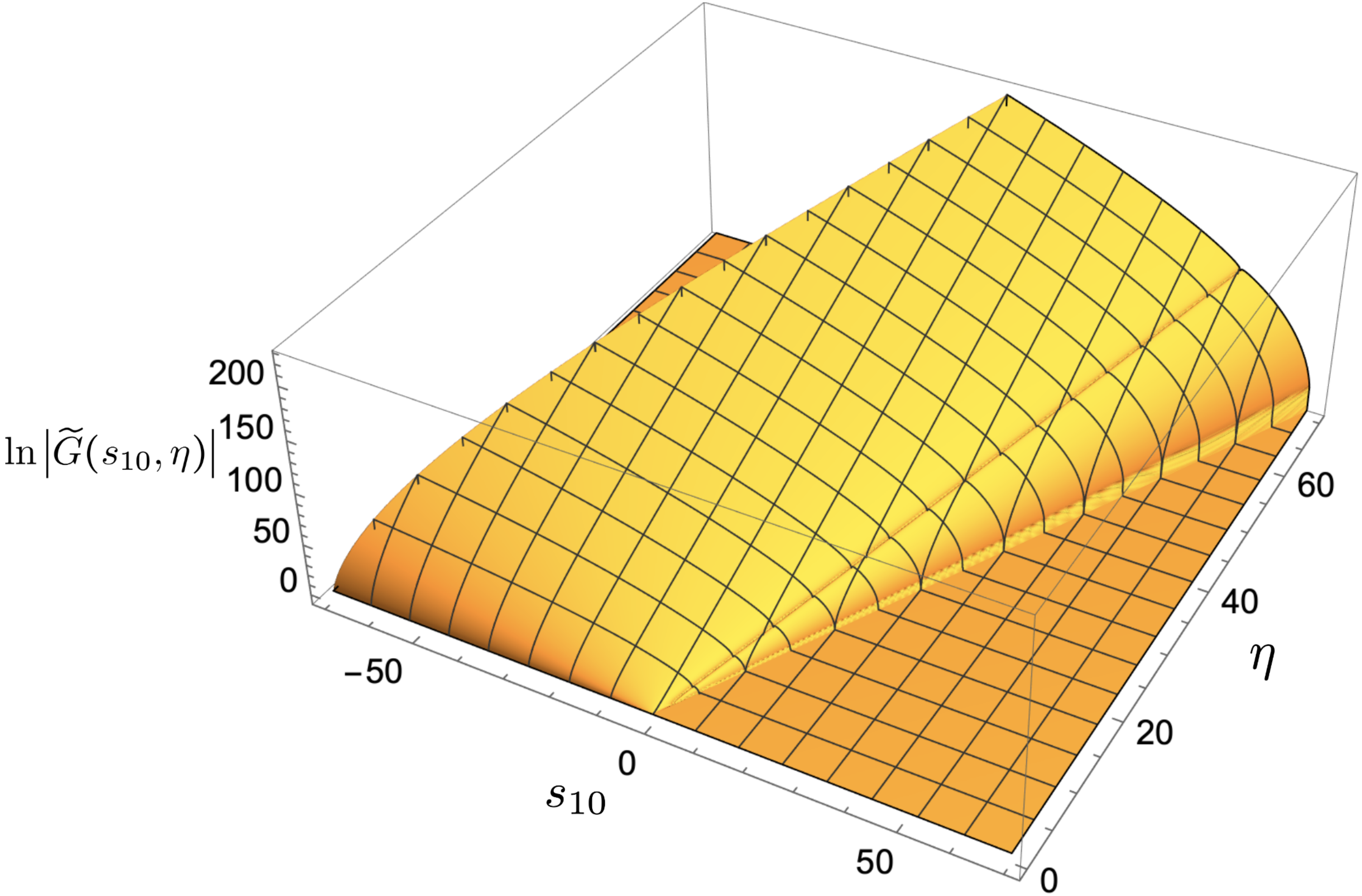}
         \caption{$\ln\left|{\widetilde G}(s_{10},\eta)\right|$}
         \label{fig:QGG3dNf4_G}
     \end{subfigure}
     \caption{The plots of logarithms of the absolute values of polarized dipole amplitudes $Q$, $G_2$ and ${\widetilde G}$ at $N_f=4$ and $N_c =3$ versus $s_{10}$ and $\eta$, for in the $-\eta_{\max}\leq s_{10}\leq \eta_{\max}$, $0\leq\eta\leq\eta_{\max}$ region with $\eta_{\max}=70$. The amplitudes are computed numerically using the step size $\delta = 0.1$.}
     \label{fig:QGG3dNf4}
\end{figure}

At $N_f=4$, we perform the numerical computation described above to obtain the dipole amplitudes $Q$, ${\widetilde G}$ and $G_{2}$. The logarithms of absolute values of these amplitudes are plotted in \fig{fig:QGG3dNf4} versus $\eta$ and $s_{10}$. As mentioned above, these amplitudes are calculated with step size $\delta = 0.1$ and maximum rapidity $\eta_{\max}=70$. Qualitatively, the plots in Figs.~\ref{fig:QGG3dNf4_Q} and \ref{fig:QGG3dNf4_G2} indicate the exponential growth with $\eta$ along the $s_{10}=0$ line. However, we see a line of cusp in the plot for ${\widetilde G}$, which is a new feature not seen in any of the amplitudes in previous similar works \cite{Cougoulic:2022gbk, Kovchegov:2020hgb, Kovchegov:2016weo}. This cusp only appears in the positive-$s_{10}$ region.

\begin{figure}[ht]
     \centering
     \begin{subfigure}[b]{0.3\textwidth}
         \centering
         \includegraphics[width=\textwidth]{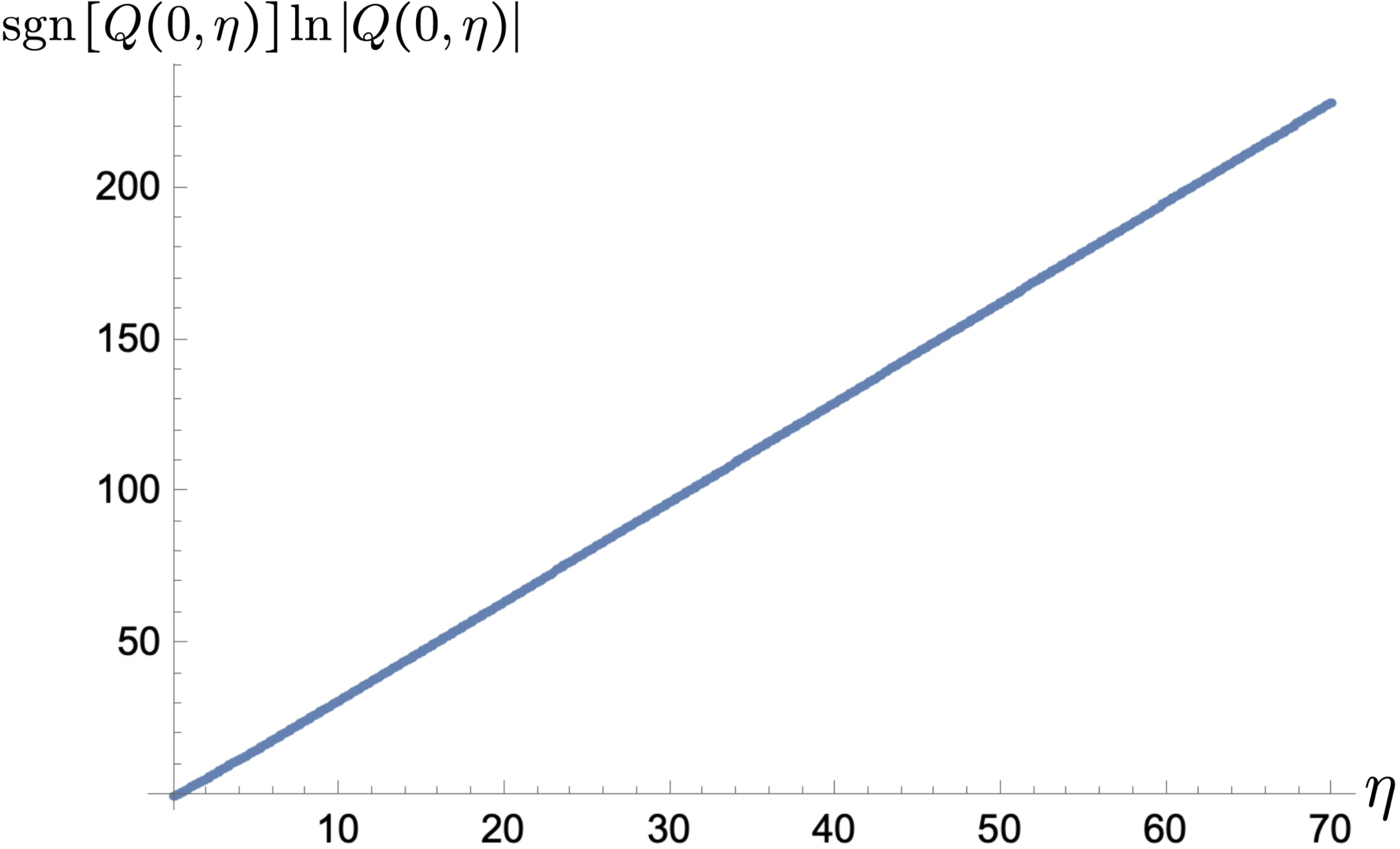}
         \caption{sgn$\left[Q(0,\eta)\right]\ln\left|Q(0,\eta)\right|$}
         \label{fig:signln2dNf4_Q}
     \end{subfigure} 
 \;
     \begin{subfigure}[b]{0.3\textwidth}
         \centering
         \includegraphics[width=\textwidth]{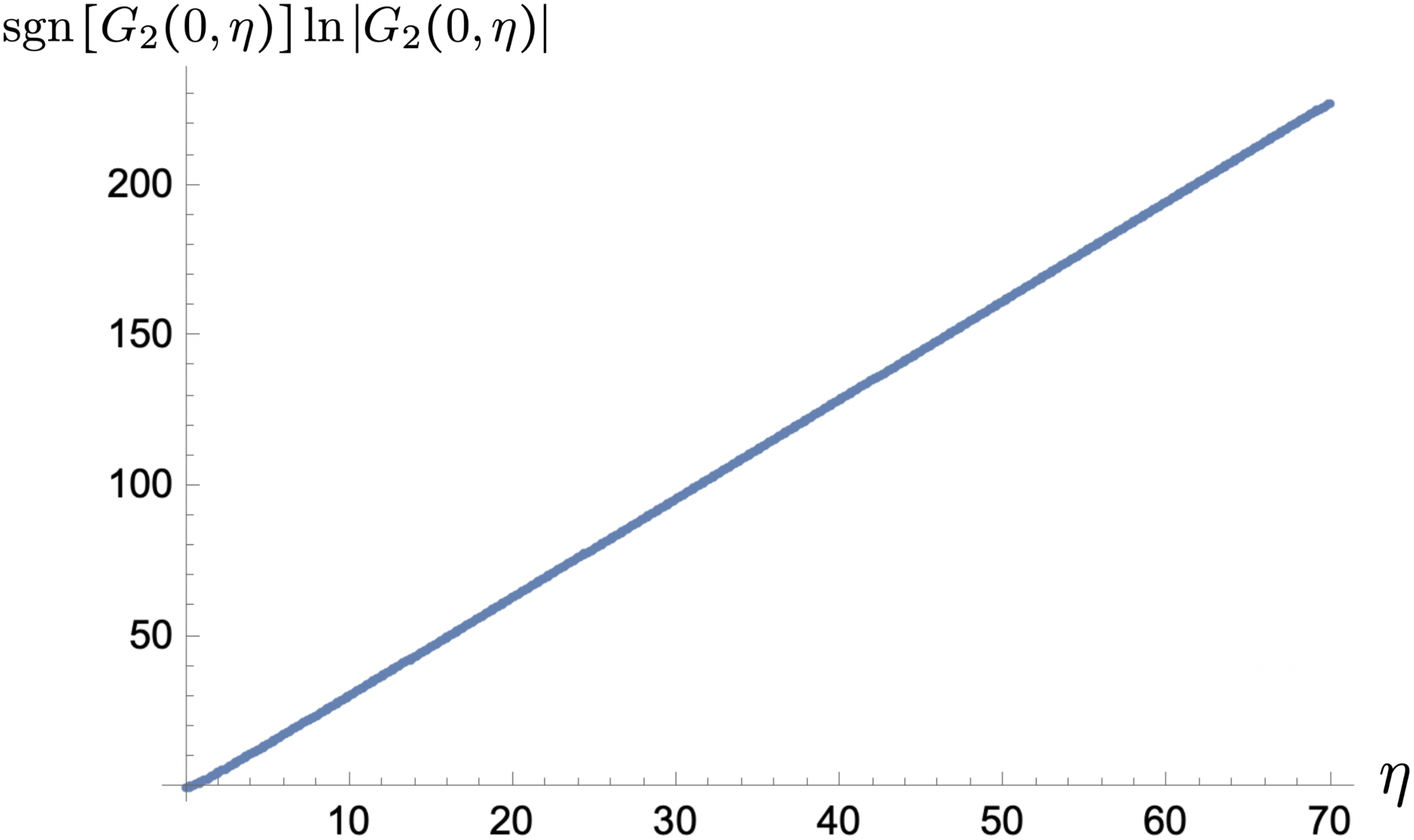}
         \caption{sgn$\left[G_2(0,\eta)\right]\ln\left|G_2(0,\eta)\right|$}
         \label{fig:signln2dNf4_G2}
     \end{subfigure} 
 \;
     \begin{subfigure}[b]{0.36\textwidth}
         \centering
         \includegraphics[width=\textwidth]{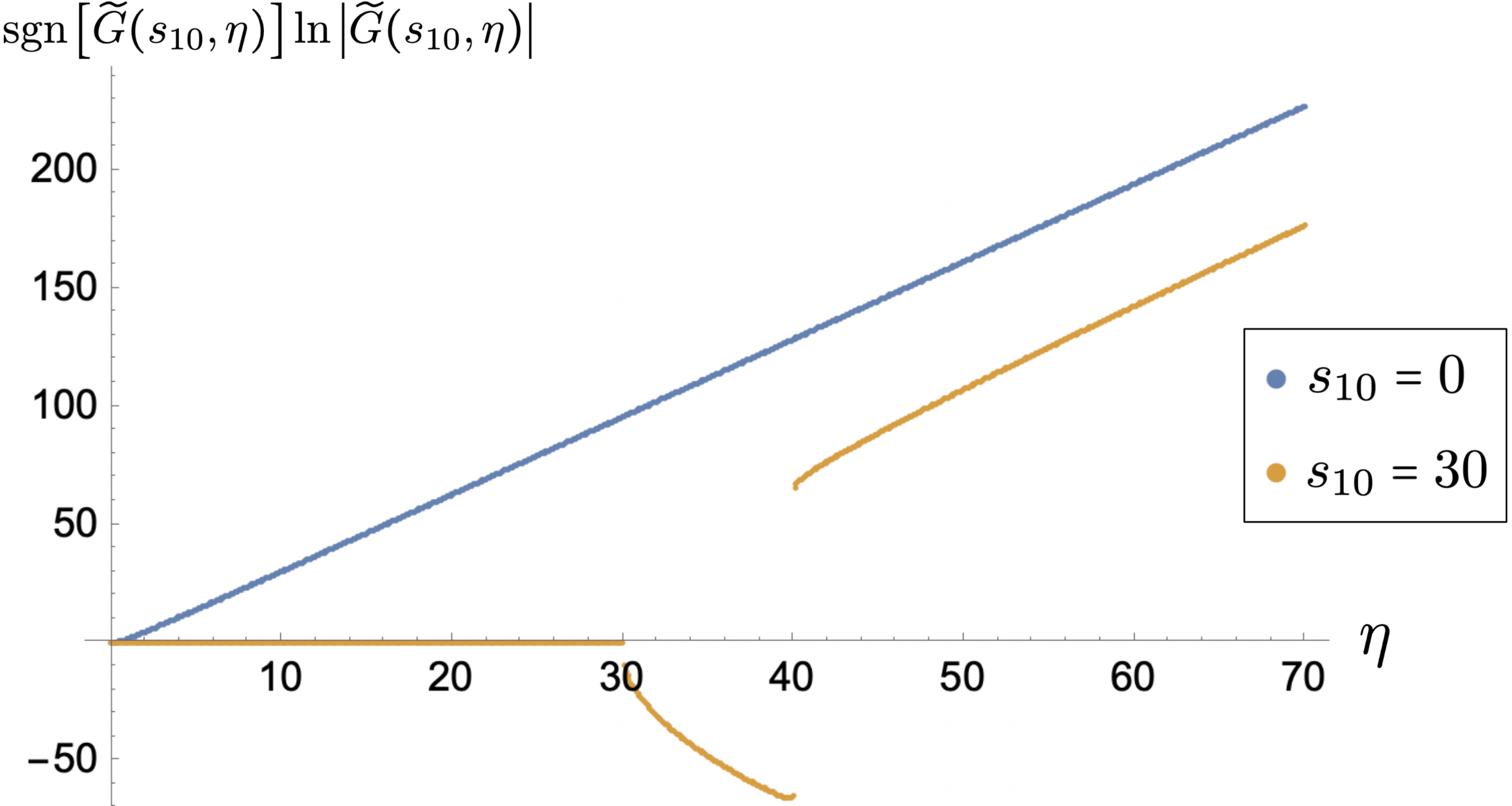}
         \caption{sgn$\left[{\widetilde G}(0,\eta)\right]\ln\left|{\widetilde G}(0,\eta)\right|$}
         \label{fig:signln2dNf4_G}
     \end{subfigure}
     \caption{The plots of logarithms of the absolute values of polarized dipole amplitudes $Q$, $G_2$ and ${\widetilde G}$, multiplied by the amplitudes' signs, along the $s_{10}=0$ line (for all amplitudes) and along the $s_{10}=30$ line for ${\widetilde G}$, versus the rapidity $\eta$. The amplitudes are computed numerically with $N_f=4, N_c = 3$ in the range $0\leq\eta\leq\eta_{\max} =70$ using step size $\delta = 0.1$.}
     \label{fig:signln2dNf4}
\end{figure}

\begin{figure}
    \centering
    \includegraphics[width=0.6\textwidth]{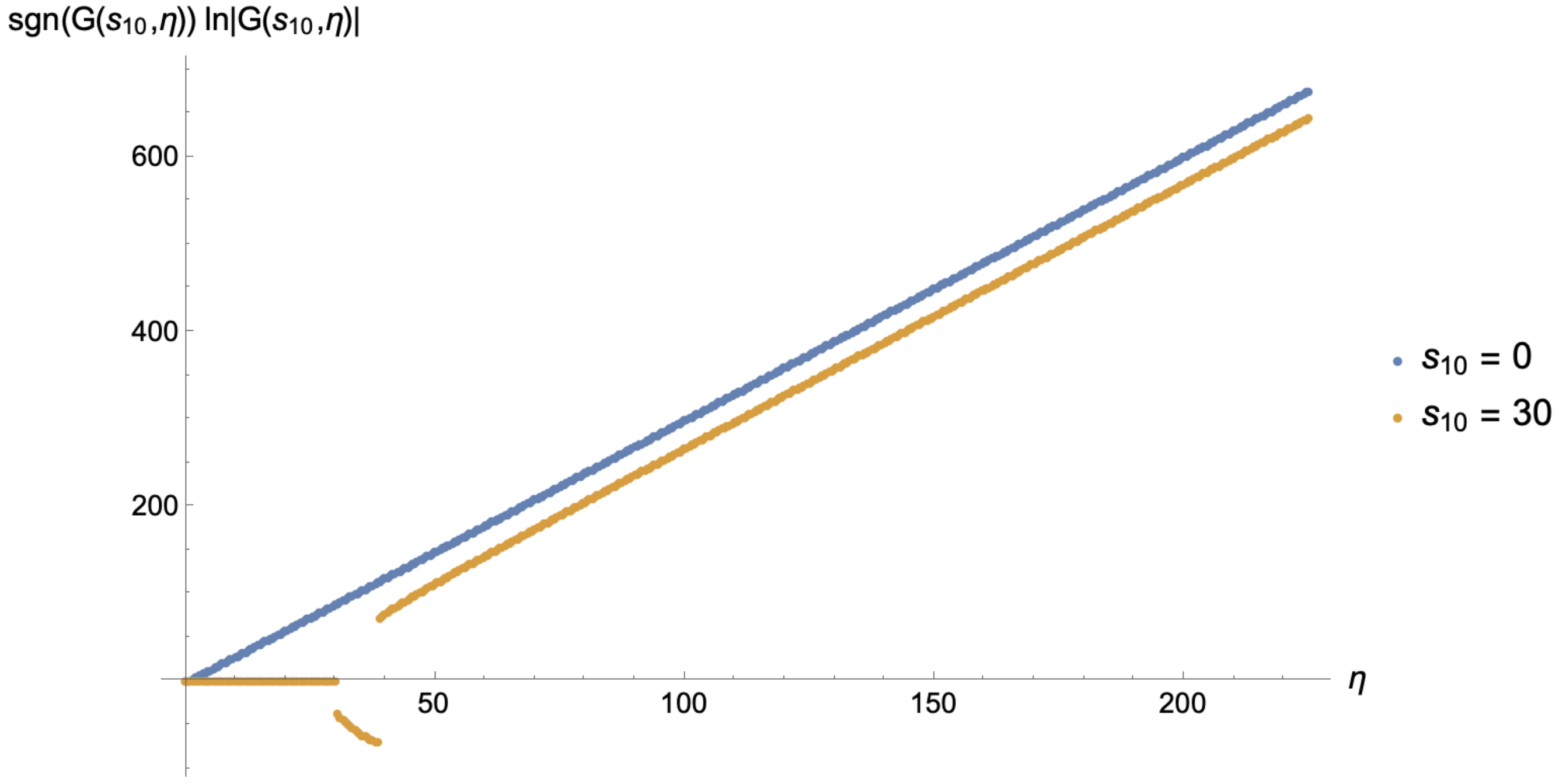}
    \caption{The plot of the logarithm of the absolute value of polarized dipole amplitude ${\widetilde G}$ multiplied by the amplitude's sign along the $s_{10}=0$ line (blue dots) and the $s_{10}=30$ line (orange dots), versus the rapidity, $\eta$. The amplitudes are computed numerically with $N_f=4, N_c = 3$ in the range $0\leq\eta\leq\eta_{\max} =225$ using the step size $\delta = 0.5$.}
    \label{fig:G05225Nf4}
\end{figure}

To further understand the results, we plot each amplitude at $s_{10}=0$ in \fig{fig:signln2dNf4}, with the exception of \fig{fig:signln2dNf4_G} for ${\widetilde G}$, which contains blue dots for $s_{10}=0$ and orange dots for $s_{10} = 30$. For each of the plots, the quantity in the vertical axis is the sign of the amplitude multiplied by the logarithm of the absolute value of the amplitude. In the plot for ${\widetilde G}$ at $s_{10}=30$, we see that ${\widetilde G}(30,\eta)$ grows from the initial condition toward negative values as $\eta > 30$. Now, as $\eta$ grows past a value near $40$, the sign of the amplitude flips and becomes positive once more, while its magnitude keeps growing exponentially. The sign flip appears only once at least up to the maximum rapidity of $\eta_{\max}=70$, as shown in \fig{fig:signln2dNf4_G}. Furthermore, we also confirm this fact of a single sign flip for up to $\eta = 225$ in a separate run with $\delta = 0.5$ and $\eta_{\max}=225$, whose result for ${\widetilde G}(30,\eta)$ is shown in \fig{fig:G05225Nf4}. A similar pattern is exhibited by the amplitude ${\widetilde G}$ at any $s_{10}>0$, with the location of the sign flip forming a straight line of the form $\eta = \kappa s_{10}$ for some positive constant $\kappa$. The source of this behavior likely comes from the $s_{10}$-dependence of the amplitude and does not appear to affect the asymptotic behavior at large $\eta$. Hence, understanding this sign reversal is beyond the scope and purpose of this work because, as we will see below, the large-$\eta$ asymptotics of $Q$ and $G_2$ at $s_{10}=0$ is sufficient for us to determine the small-$x$ asymptotics of the helicity PDFs and the $g_1$ structure function.

Now, along the $s_{10}=0$ line, \fig{fig:signln2dNf4} displays the linear increase in the logarithm of the absolute values of the amplitudes once we get sufficiently far away from $\eta=0$ for the initial conditions and the discretization errors not to be significant any longer. This justifies the following ans\"atze as large $\eta$,
\begin{subequations}\label{Nf101}
\begin{align}
Q(s_{10}=0,\eta) &\sim e^{\alpha_Q\eta}  ,   \\
G_2(s_{10}=0,\eta) &\sim e^{\alpha_{G_2}\eta}   ,  \\
{\widetilde G}(s_{10}=0,\eta) &\sim e^{\alpha_{{\widetilde G}}\eta}     .
\end{align}
\end{subequations}
Here, the parameters $\alpha_Q$, $\alpha_{G_2}$ and $\alpha_{{\widetilde G}}$ (the intercepts) correspond to the slopes of the respective plots in \fig{fig:signln2dNf4}. In particular, the slopes should be extracted from the log-amplitude data where the effects of discretization and initial condition are least significant. In this work, following \cite{Cougoulic:2022gbk, Kovchegov:2020hgb, Kovchegov:2016weo}, we choose to do so in the region where $\eta \in \left[0.75,1\right]\eta_{\max}$. 

We extract the intercepts following this recipe for $N_f=2,3,4$ with $\delta=0.1$ and $\eta_{\max}=70$. The results are listed in Table~\ref{tab:lowNfintercepts} together with their uncertainties, which are calculated from linear regression residual based on the 95\% confidence interval, for all the amplitudes and numbers of flavors. For each $N_f$, the intercepts appear to be the same within the uncertainty for all three polarized dipole amplitudes, $Q$, $G_2$ and ${\widetilde G}$. 

\begin{table}[h]
\begin{center}
\begin{tabular}{|c|c|c|c|}
\hline
\;Number of flavors\; 
& $\alpha_Q$
& $\alpha_{G_2}$
& $\alpha_{{\widetilde G}}$
\\ \hline 
$N_f=2$
& \;$3.48990 \pm 0.00004$\;
& \;$3.48989 \pm 0.00005$\;
& \;$3.48992 \pm 0.00004$\;
\\ \hline 
$N_f=3$
& $3.40163 \pm 0.00005$
& $3.40161 \pm 0.00005$
& $3.40166 \pm 0.00004$
\\ \hline 
$N_f=4$
& $3.29297 \pm 0.00005$
& $3.29296 \pm 0.00005$
& $3.29302 \pm 0.00004$
\\ \hline 
\end{tabular}
\caption{Summary of the intercept estimates and uncertainties for all types of polarized dipole amplitudes at $N_f=2,3,4$ along the $s_{10}=0$ line. Here, the number of quark colors is taken to be $N_c=3$. The computation is performed with step size, $\delta=0.1$, maximum rapidity, $\eta_{\max}=70$, and the all-one initial condition \eqref{asym1}.}
\label{tab:lowNfintercepts}
\end{center}
\end{table}

These intercept estimates calculated at $\delta=0.1$ and $\eta_{\max}=70$ only provide an initial estimate for the intercepts we would obtain by solving Eqs.~\eqref{evoleq} exactly. However, we will see later that they still differ significantly from the exact values, and the main source of error comes from the fact that we still work with a finite step size $\delta$ and a finite maximum rapidity $\eta_{\max}$. To resolve the mismatch, we repeat the computation for other choices of (finite) $\delta$ and $\eta_{\max}$. In particular, for each step size $\delta$ we numerically compute the intercepts for $\eta_{\max}\in\{10,20,\ldots,M(\delta)\}$, where $M(\delta)$ is given in Table~\ref{tab:M_delta_Nf234}. \footnote{We used $\eta_{\max}\in\{10,21\}$ for $\delta=0.0375$. Also, for $\delta = 0.05$, we set $M(0.05) = 40$ for $N_f=4$ and $M(0.05)=30$ for $N_f=2,3$.}

\begin{table}[h]
\begin{center}
\begin{tabular}{|c|c|c|c|c|c|c|c|c|c|c|}
\hline
$\delta$ 
& 0.016
& 0.025
& 0.0375
& \,0.05\,
& 0.0625
& \,0.08\,
& \,\,0.1\,\,
\\ \hline 
$M(\delta)$
& 10
& 20
& 21
& 30 or 40
& 40
& 50
& 70
\\ \hline
\end{tabular}
\caption{The maximum $M(\delta)$ of the $\eta_{\max}$-range computed for each step size $\delta$.}
\label{tab:M_delta_Nf234}
\end{center}
\end{table}

\begin{figure}
     \centering
     \begin{subfigure}[b]{0.32\textwidth}
         \centering
         \includegraphics[width=\textwidth]{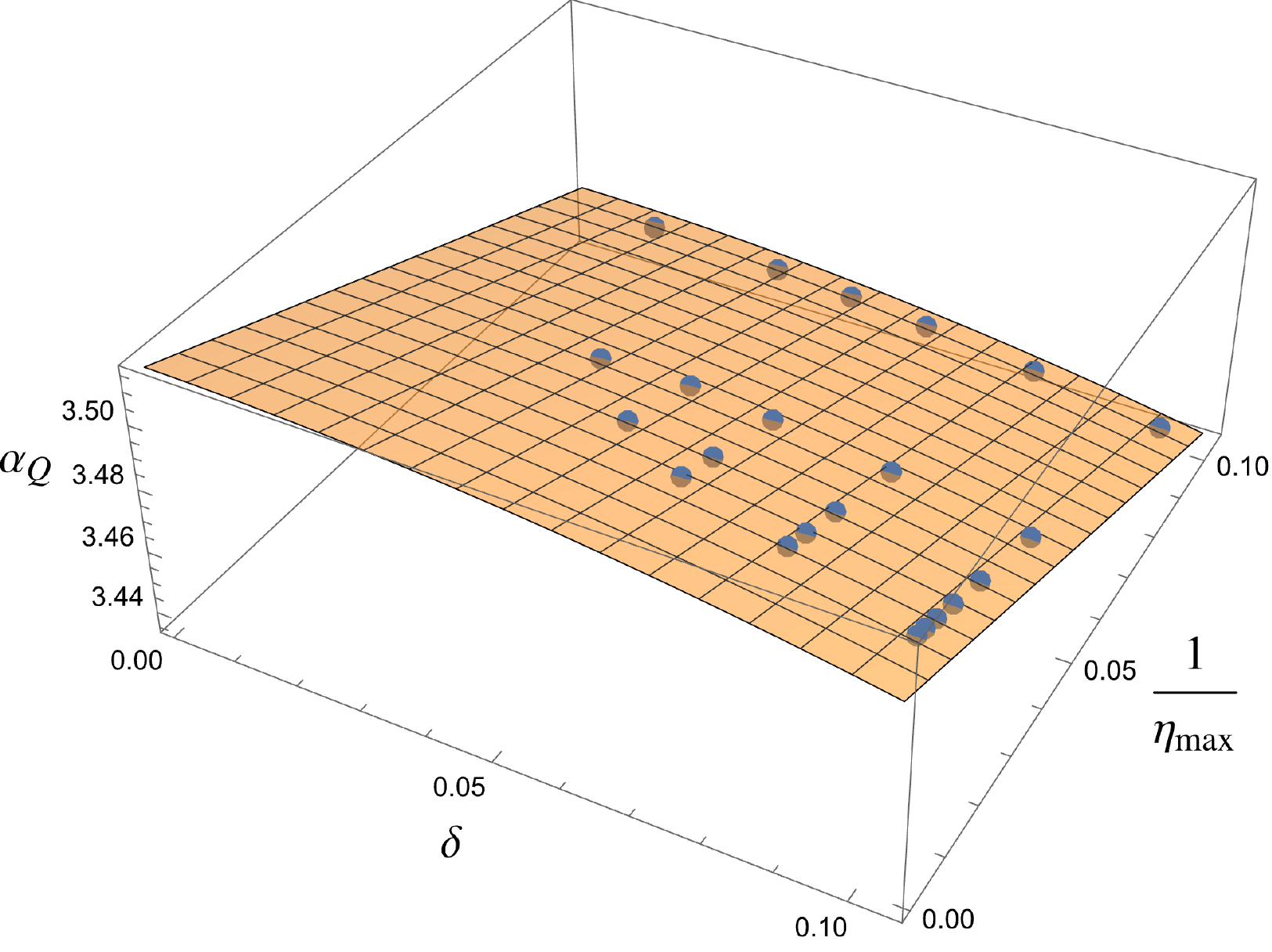}
         \caption{$N_f=2$}
         \label{fig:quad_surf_lowNf2}
     \end{subfigure}  \;
     \begin{subfigure}[b]{0.32\textwidth}
         \centering
         \includegraphics[width=\textwidth]{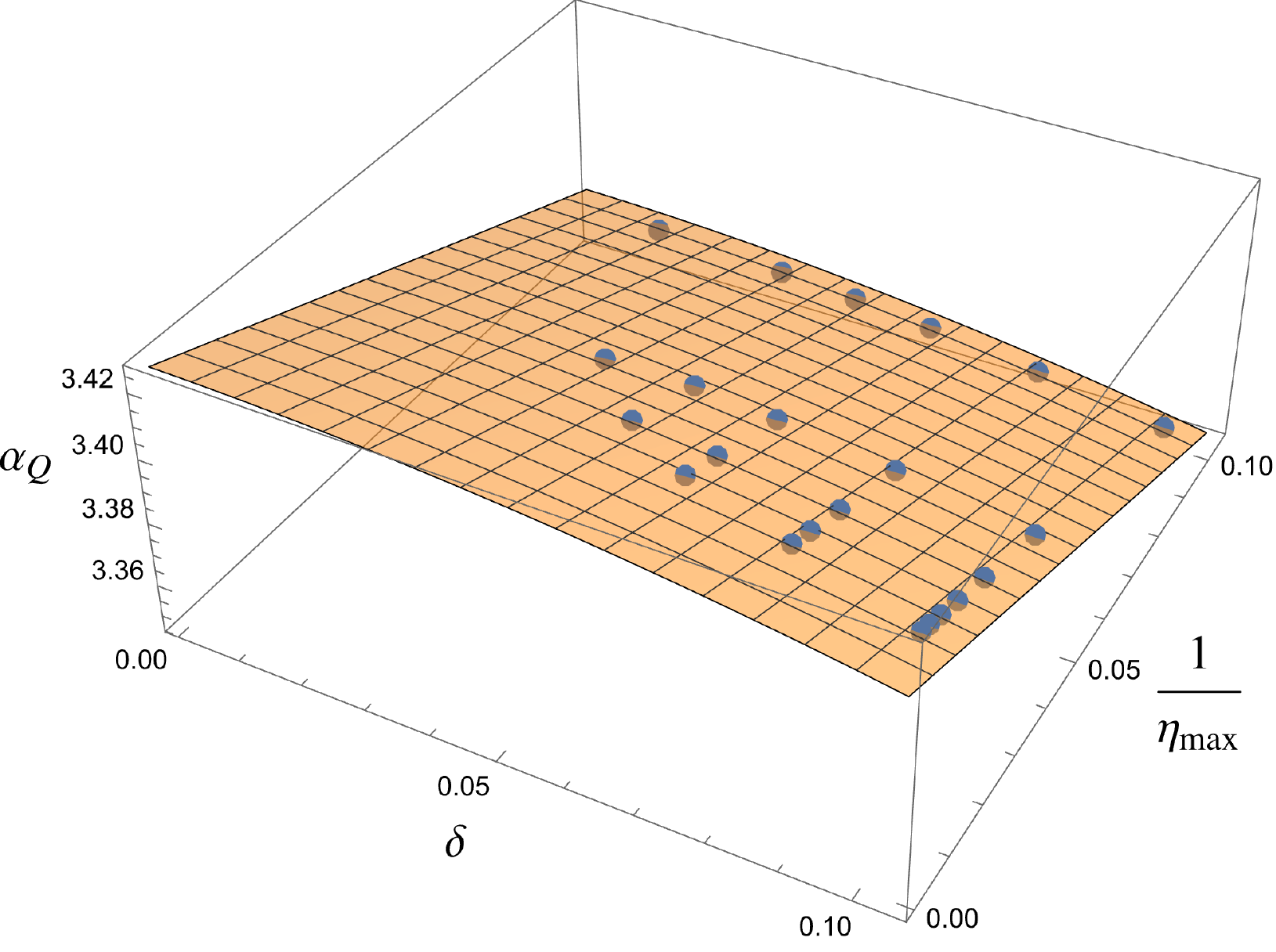}
         \caption{$N_f=3$}
         \label{fig:quad_surf_lowNf3}
     \end{subfigure}  \;
     \begin{subfigure}[b]{0.32\textwidth}
         \centering
         \includegraphics[width=\textwidth]{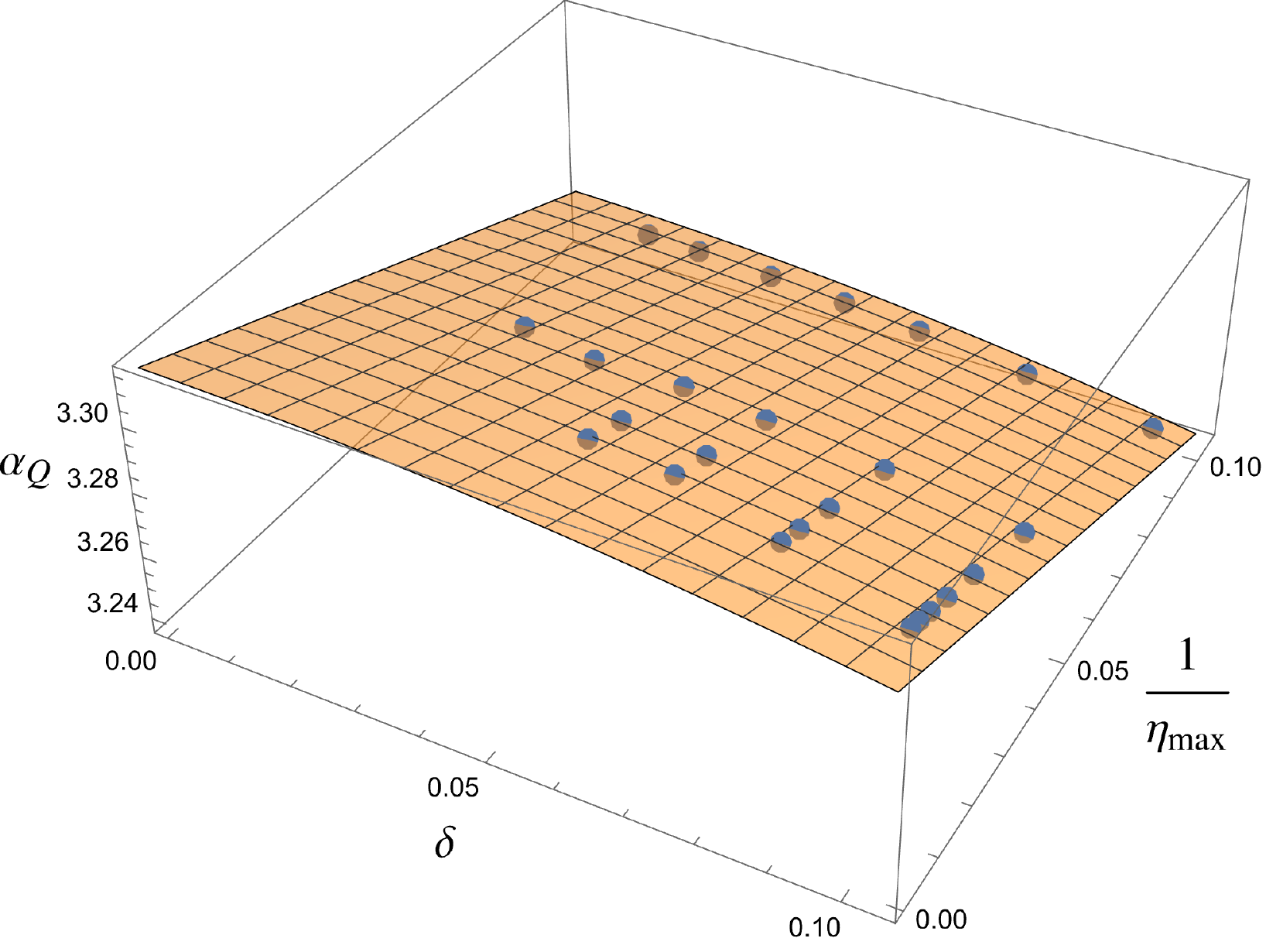}
         \caption{$N_f=4$}
         \label{fig:quad_surf_lowNf4}
     \end{subfigure}
     \caption{The plots of estimated intercepts, $\alpha_Q$, at each $N_f$, $\delta$ and $1/\eta_{\max}$ (blue dots), together with the corresponding best-fitted quadratic extrapolations (yellow surfaces). The continuum limit, $\delta=1/\eta_{\max}=0$, corresponds to the lower left corner of each plot.}
     \label{fig:quad_surf_lowNf}
\end{figure}

For each amplitude and at each $N_f$, we performed a weighted polynomial regression to fit the extracted intercepts as functions of $\delta$ and $1/\eta_{\max}$ using the linear, quadratic and cubic models \cite{Cougoulic:2022gbk}. As a result, the quadratic model performs best based on the Akaike information criterion (AIC) \cite{Akaike:1974} and the significance test on model parameters. Then, we take the model's prediction at $\delta=1/\eta_{\max}=0$ to be our estimate for the intercept in the continuum limit. The results are shown in Table~\ref{tab:lowNfinterceptsCont}. There, the uncertainty is deduced from the residue of the weighted quadratic regression model. In addition, we include in \fig{fig:quad_surf_lowNf} the plots showing the intercepts, $\alpha_Q$, versus $\delta$ and $1/\eta_{\max}$, together with the best-fit quadratic surface, for $N_f=2,3,4$ considered here and $N_c =3$. From Table~\ref{tab:lowNfinterceptsCont}, we see that the intercepts for $Q$, $G_2$ and ${\widetilde G}$ are the same within the uncertainty for each value of $N_f$. Furthermore, the intercepts decrease as $N_f$ increases. Recall that the large-$N_c$ intercept is 3.66 \cite{Cougoulic:2022gbk, Borden:2023ugd}, which is greater than the largest intercept in Table~\ref{tab:lowNfinterceptsCont}. This is in line with the fact that the large-$N_c$ limit from \cite{Cougoulic:2022gbk} neglects soft quark emissions and consequently puts $N_f\ll N_c$ or $N_f =0$ in our terminology. This further reinforces the observation that the intercept decreases with $N_f$.

\begin{table}[h]
\begin{center}
\begin{tabular}{|c|c|c|c|}
\hline
\;Number of flavors\; 
& $\alpha_Q$
& $\alpha_{G_2}$
& $\alpha_{{\widetilde G}}$
\\ \hline 
$N_f=2$
& \;$3.516 \pm 0.003$\;
& \;$3.516 \pm 0.003$\;
& \;$3.516 \pm 0.003$\;
\\ \hline 
$N_f=3$
& $3.427 \pm 0.003$
& $3.426 \pm 0.003$
& $3.427 \pm 0.003$
\\ \hline 
$N_f=4$
& $3.316 \pm 0.002$
& $3.316 \pm 0.002$
& $3.317 \pm 0.002$
\\ \hline 
\end{tabular}
\caption{Summary of estimates and uncertainties at the continuum limit ($\delta\to 0$ and $\eta_{\max}\to\infty$) for the intercepts of all types of polarized dipole amplitudes at $N_f=2,3,4$ along the $s_{10}=0$ line. Here, the number of quark colors is taken to be $N_c=3$. All the computations are performed with the all-one initial condition \eqref{asym1}.}
\label{tab:lowNfinterceptsCont}
\end{center}
\end{table}

Next we study what the large-$\eta$ asymptotics of the dipole amplitudes imply about the small-$x$ asymptotics of the helicity PDFs and the $g_1$ structure function. Starting with the gluon helicity PDF, we know that it is related to the type-2 polarized dipole amplitude as shown in Eq.~\eqref{glhPDF}, which in the double-logarithmic approximation (DLA) and at $Q^2 = \Lambda^2$ becomes \cite{Cougoulic:2022gbk}
\begin{align}\label{asym11}
\Delta G(x, Q^2 = \Lambda^2) &\approx  \frac{2N_c}{\alpha_s\pi^2} \ G_2 \left(s_{10}=0, \eta=\sqrt{\frac{\alpha_sN_c}{2\pi}}\,\ln\frac{1}{x}\right) .  
\end{align}
This is because the partial derivative with respect to $x^2_{10}$ in Eq.~\eqref{glhPDF} removes a single logarithm, and is, therefore, suppressed in the DLA. As a result, we see that the small-$x$ asymptotics of the gluon helicity PDF is exactly the same as the large-$\eta$ asymptotics of $G_2(0,\eta)$ with $\eta=\sqrt{\frac{\alpha_sN_c}{2\pi}}\,\ln\frac{1}{x}$, that is,
\begin{align}\label{glue_asympt}
    &\Delta G(x,Q^2)\Big|_{Q^2=\Lambda^2} \sim \left(\frac{1}{x}\right)^{\alpha_{G_2}\sqrt{\frac{\alpha_sN_c}{2\pi}}}.
\end{align}
Recall that $\alpha_{G_2}$ depends on $N_f$.

Next, we consider the flavor-singlet quark helicity PDF and the $g_1$ structure function. For simplicity, making the approximation that the dipole amplitudes are flavor-independent, we see that the sum over flavors in \eq{g1} becomes a multiplicative factor, leading to the two objects being simply proportional to each other, $g_1 \sim \Delta \Sigma$. Hence, from now on, we proceed with the calculation for the quark helicity PDF.

As a first step, since now $\Lambda$ is not the IR cutoff, but instead is the scale characterizing the target, we remove $1/\Lambda^2$ from the upper limit of the $x_{10}^2$-integral in \eq{qkhPDF}. (In principle, we should replace it by  $\Lambda_{\text{IR}}$: however, in this work, we simply take $\Lambda_{\text{IR}}$ to be sufficiently large so that the upper limit of $\frac{1}{zQ^2}$ suffices for the transverse integral.) We thus re-write \eq{qkhPDF} as
\begin{align}\label{asym12}
&\Delta\Sigma(x,Q^2) =  - \frac{N_cN_f}{2\pi^3} \int\limits_{\Lambda^2/s}^1 \frac{dz}{z} \int\limits_{1/zs}^{1/zQ^2} \frac{dx^2_{10}}{x_{10}^2} \left[Q(x^2_{10},zs) + 2 G_2(x^2_{10},zs) \right]. 
\end{align}
Here, the center-of-mass energy of the interaction is high, such that $x \simeq \frac{Q^2}{s}$ in the small-$x$ regime. Now, in terms of $s_{10}$ and $\eta$, the flavor-singlet quark helicity PDF from Eq.~\eqref{asym12} can be written as \cite{Kovchegov:2020hgb}
\begin{align}\label{asym13}
\Delta\Sigma(x,Q^2 = \Lambda^2) &=  - \frac{N_cN_f}{2\pi^3} \int\limits_{0}^{\sqrt{\frac{\alpha_sN_c}{2\pi}}\,\ln\frac{1}{x}} d\eta \int\limits_{\eta-\sqrt{\frac{\alpha_sN_c}{2\pi}}\,\ln\frac{1}{x} }^{\eta} ds_{10} \ \left[Q(s_{10},\eta) + 2 \, G_2(s_{10},\eta) \right].  
\end{align}
In Eq.~\eqref{asym13}, the quantity, $\sqrt{\frac{\alpha_sN_c}{2\pi}}\,\ln\frac{1}{x} \equiv \eta_b$, is large when $x$ is small. Here, $\eta_b$ can be viewed as the maximum rapidity limiting the integration region. 

The integral in \eq{asym13} can be evaluated as a Riemann sum using our numerical results for the fundamental dipole amplitudes,
\begin{align}\label{asym14}
\Delta\Sigma_j &\equiv \Delta\Sigma\left(x=\exp\left[-\sqrt{\frac{2\pi}{\alpha_sN_c}}\,j\,\delta\right],\,Q^2\sim\Lambda^2\right)  =  - \frac{N_cN_f}{2\pi^3}\,\delta^2 \sum_{j'=0}^{j-1}\sum_{i'=j'-j}^{j'-1} \left[Q_{i'j'} + 2 \, G_{2,i'j'} \right] ,  
\end{align}
which can be written recursively as
\begin{align}\label{asym15}
\Delta\Sigma_j &= \Delta\Sigma_{j-1} - \frac{N_cN_f}{2\pi^3}\,\delta^2 \sum_{j'=0}^{j-2} \left[Q_{(j'-j)j'} + 2 \, G_{2,(j'-j)j'} \right]  - \frac{N_cN_f}{2\pi^3}\,\delta^2 \sum_{i'=-1}^{j-2} \left[Q_{i'(j-1)} + 2 \, G_{2,i'(j-1)} \right] ,   
\end{align}
for $j\geq 1$. Note that $\Delta\Sigma_0 = 0$ in this notation. Physically, since $j=0$ corresponds to $x=1$, the value of $\Delta\Sigma_0$ simply implies that the quark helicity PDF at moderate $x$ is much smaller than its values at small $x$, as the latter is driven by the polarized dipole amplitudes that grow exponentially in magnitude with $\ln(1/x)$. In a phenomenological calculation, c.f. \cite{Adamiak:2021ppq, Kovchegov:2020hgb, Kovchegov:2016weo}, one may need to begin this iterative calculation at some $j=j_0>0$, corresponding to a value of Bjorken-$x$ that is small enough for our evolution to apply but large enough to have good experimental constraints. Then, the value of $\Delta\Sigma_{j_0}$ should also be deduced from experimental results. However, proper matching of small-$x$ evolution onto the large-$x$ physics is an open problem we are not going to address here.

\begin{figure}
\begin{center}
\includegraphics[width=0.5\textwidth]{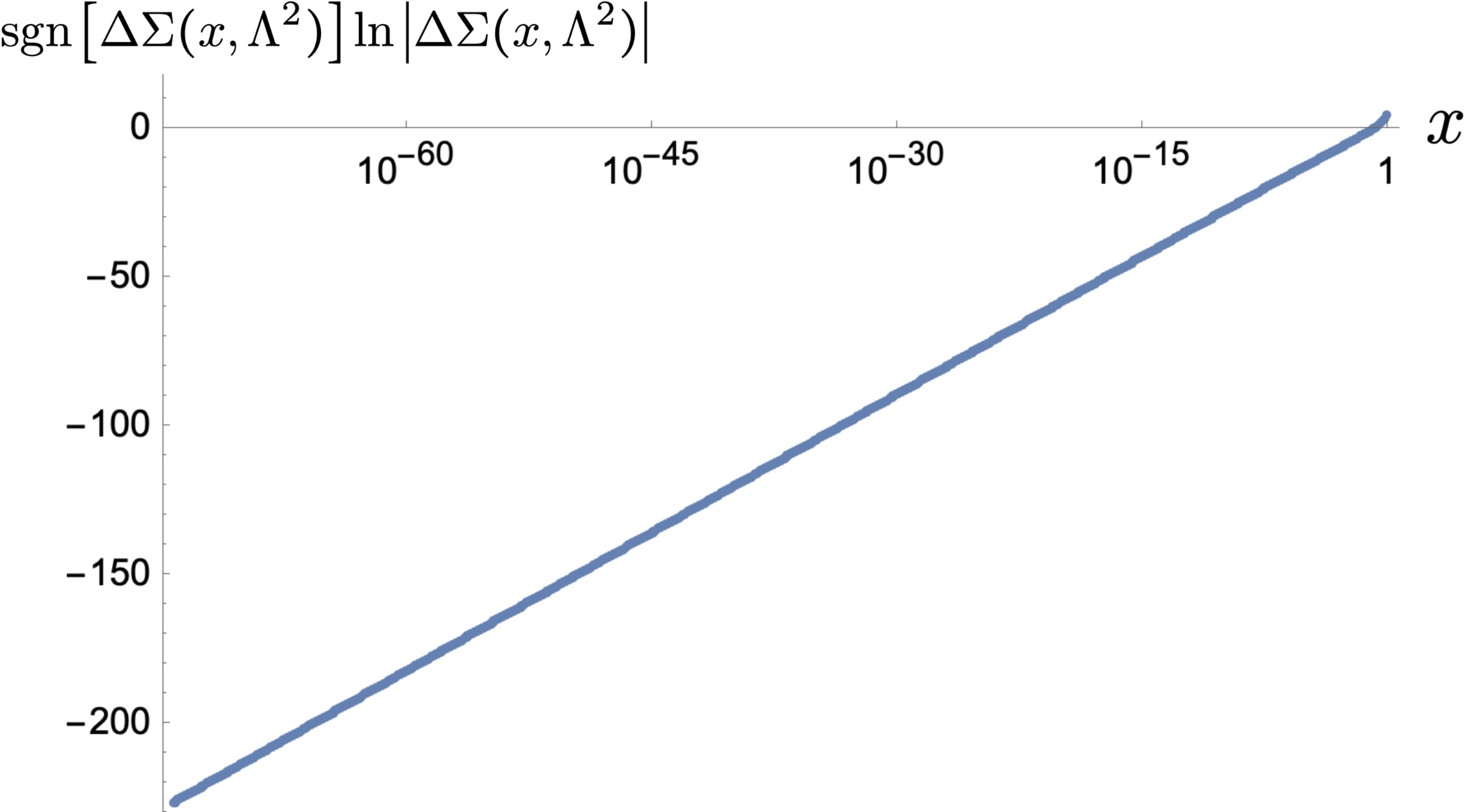}
\caption{The plot of sgn$\left[\Delta\Sigma(x,Q^2)\right]\ln\left|\Delta\Sigma(x,Q^2)\right|$, numerically computed at $Q^2=\Lambda^2$ using equation \eqref{asym15}, as a function of Bjorken $x$. In the calculation, we used the step size of $\delta=0.1$ and maximum rapidity of $\eta_{\max}=70$. Other parameters are $N_c=3$, $N_f=4$ and $\alpha_s=0.35$.}
\label{fig:qkhPDFNf4}
\end{center}
\end{figure}

For the purpose of this work, we perform the numerical integration starting from $j_0=0$ employing the dipole amplitudes we numerically obtained previously using $N_c=3$, $\delta=0.1$ and $\eta_{\max}=70$. For each of $N_f=2,3,4$, we obtain the values of quark helicity PDF at the values of Bjorken $x$ such that $\eta_b=\sqrt{\frac{\alpha_sN_c}{2\pi}}\,\ln\frac{1}{x} = 70,\,69.9,\,69.8, \, \ldots$. Here, we take the strong coupling constant to be $\alpha_s\simeq 0.35$ \cite{Kovchegov:2020hgb}. The result for $N_f=4$ is given by the plot in \fig{fig:qkhPDFNf4}. There, the vertical axis of the plot is the sign of the flavor-singlet quark helicity PDF, multiplied by the logarithm of its magnitude. The horizontal axis depicts $\ln x$. The clear linear trend implies that the magnitude of quark helicity PDF grows exponentially with $\ln(1/x)$, which is qualitatively the same as the large-$\eta$ asymptotics of polarized dipole amplitudes at $s_{10}=0$. Explicitly, we have 
\begin{align}\label{delSigmaLowNf}
&\Delta\Sigma(x,Q^2)\Big|_{Q^2=\Lambda^2} \sim g_1(x,Q^2)\Big|_{Q^2=\Lambda^2} \sim \left(\frac{1}{x}\right)^{\alpha_h^{(N_f)}} ,
\end{align}
where the intercept, $\alpha_h^{(N_f)}$, generally depends on the number of quark flavors, $N_f$. Next, we perform a linear regression on the data points for $\Delta\Sigma_j$ from \fig{fig:qkhPDFNf4} with $0.75 \, j_{\max}\leq j\leq j_{\max}$. As a result, the intercept for $N_f=4$, extracted at $\delta=0.1$ and $\eta_{\max}=70$, is equal to
\begin{align}\label{delSmLowNf2}
\alpha_h^{(4)}\Big|_{\delta=0.1,\eta_{\max}=70} &= \left(3.29294 \pm 0.00005 \right)\sqrt{\frac{\alpha_s N_c}{2\pi}} \, .
\end{align}
This value of intercept is within the uncertainties of the intercepts for the large-$\eta$ asymptotics of the polarized dipole amplitudes, $Q$, $G_2$ and ${\widetilde G}$, given in Table~\ref{tab:lowNfintercepts} for the same $N_f =4$. This result allows us to assume that a similar agreement would be obtained between the intercepts for $\Delta \Sigma$ and the polarized dipole amplitudes for other values of $\delta$ and $\eta_b$ (or $\eta_{\max}$). We, therefore, deduce the continuum-limit intercept ($\delta\to 0$ and $\eta_{\max}\to\infty$) for $\Delta \Sigma$ by reading off the corresponding continuum-limit results from the bottom row of Table~\ref{tab:lowNfinterceptsCont}. Approximately, this gives
\begin{align}\label{delSmLowNf21}
\alpha_h^{(4)} &= 3.32 \, \sqrt{\frac{\alpha_s N_c}{2\pi}} \, .
\end{align}

We repeat the above steps for $N_f=2$ and $N_f=3$, using again the amplitudes we computed at $\delta=0.1$ and $\eta_{\max}=70$. This results in the following intercepts for the discretized $\Delta \Sigma$, 
\begin{subequations}\label{delSmLowNf3}
\begin{align}
\alpha_h^{(2)}\Big|_{\delta=0.1,\eta_{\max}=70}  &= \left(3.48988 \pm 0.00005 \right)\sqrt{\frac{\alpha_s N_c}{2\pi}} \,,  \label{delSmLowNf3a} \\
\alpha_h^{(3)}\Big|_{\delta=0.1,\eta_{\max}=70}  &= \left(3.40160 \pm 0.00005 \right)\sqrt{\frac{\alpha_s N_c}{2\pi}} \, .  \label{delSmLowNf3b} 
\end{align}
\end{subequations}
For each $N_f$, the intercept of the small-$x$ asymptotics for the flavor-singlet quark helicity PDF is within the uncertainty from the intercepts of the respective large-$\eta$ asymptotics of the polarized dipole amplitudes we obtained in Table~\ref{tab:lowNfintercepts}.  This allows us to read off the corresponding continuum-limit results from Table~\ref{tab:lowNfinterceptsCont}, which gives
\begin{subequations}\label{delSmLowNf31}
\begin{align}
\alpha_h^{(2)} &= 3.52 \, \sqrt{\frac{\alpha_s N_c}{2\pi}} \,,  \label{delSmLowNf31a} \\
\alpha_h^{(3)} &= 3.43 \, \sqrt{\frac{\alpha_s N_c}{2\pi}} \, .  \label{delSmLowNf31b} 
\end{align}
\end{subequations}

Equations~\eqref{delSmLowNf21} and \eqref{delSmLowNf31} giving us the intercepts of the the $g_1$ structure function and the quark helicity PDFs are the main results of this Subsection. Note that the same intercepts drive the asymptotics of the gluon helicity PDFs, per \eq{glue_asympt}.

Let us cross check these intercepts against the results obtained using the IREE formalism of BER \cite{Bartels:1996wc}. We calculated the BER intercepts numerically using the formalism from \cite{Bartels:1996wc} while applying the large-$N_c \& N_f$ limit to them. The results are given in Table~\ref{tab:BER_comparison} for $N_f=2,3,4$ we considered in this Section. The KPS-CTT intercepts we found above are also listed for comparison. 

\begin{table}[h]
\begin{center}
\begin{tabular}{|c|c|c|c|}
\hline
\;\;$N_f$\;\; & BER intercept & KPS-CTT intercept \\ \hline
2     & 3.55                    & 3.52                            \\ \hline
3     & 3.48                    & 3.43                            \\ \hline
4     & 3.41                   & 3.32                            \\ \hline
\end{tabular}
\caption{The intercepts at large $N_c\& N_f$ for $N_f=2,3,4$ and $N_c =3$ (that is, for $N_f / N_c = 2/3, 3/3, 4/3$) according to the BER and KPS-CTT evolutions.}
\label{tab:BER_comparison}
\end{center}
\end{table}

Table \ref{tab:BER_comparison} shows that the intercepts of the KPS-CTT evolution calculated at large-$N_c \& N_f$ differ from the corresponding intercepts of the BER evolution \cite{Bartels:1996wc}. The discrepancy is at the $2-3\%$ level and increases with $N_f$. This is a significant improvement relative to the large mismatches found in \cite{Kovchegov:2020hgb} using the KPS evolution \cite{Kovchegov:2015pbl, Kovchegov:2018znm} that did not include the type-2 polarized dipole amplitude. Furthermore, as shown in \cite{Borden:2023ugd}, the analytic expression for the large-$N_c$ intercept of the KPS-CTT small-$x$ helicity evolution is also different from that of BER, although the difference in that case is only at the 0.1\% level. The root cause behind these remaining discrepancies deserves a further study, which is left for a future work.


\subsection{Solution for $N_f = 6$}

\label{ssec:highNf}

\begin{figure}
     \centering
     \begin{subfigure}[b]{0.32\textwidth}
         \centering
         \includegraphics[width=\textwidth]{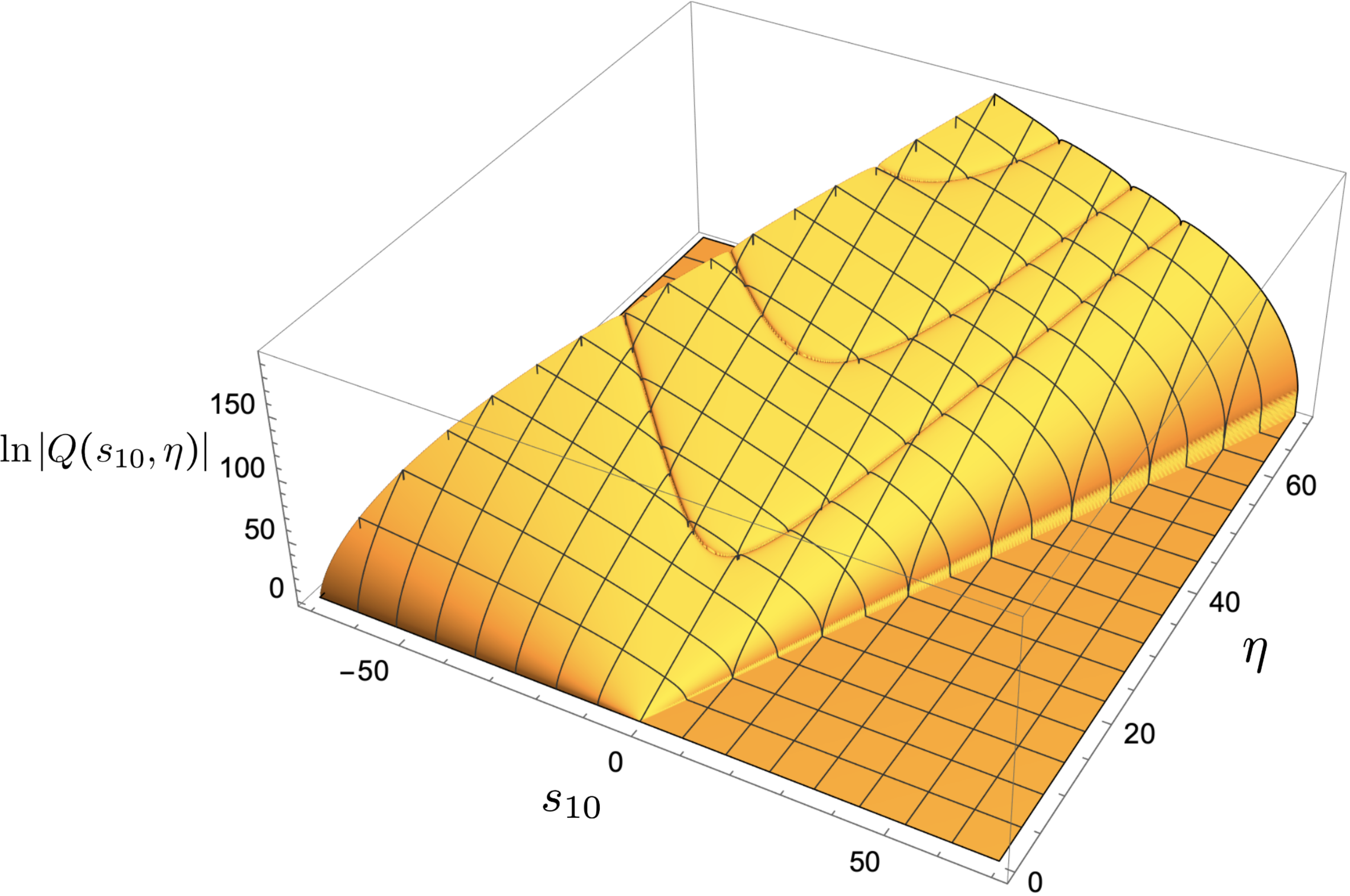}
         \caption{$\ln\left|Q(s_{10},\eta)\right|$}
         \label{fig:QGG3dNf_Q}
     \end{subfigure} \;
     \begin{subfigure}[b]{0.32\textwidth}
         \centering
         \includegraphics[width=\textwidth]{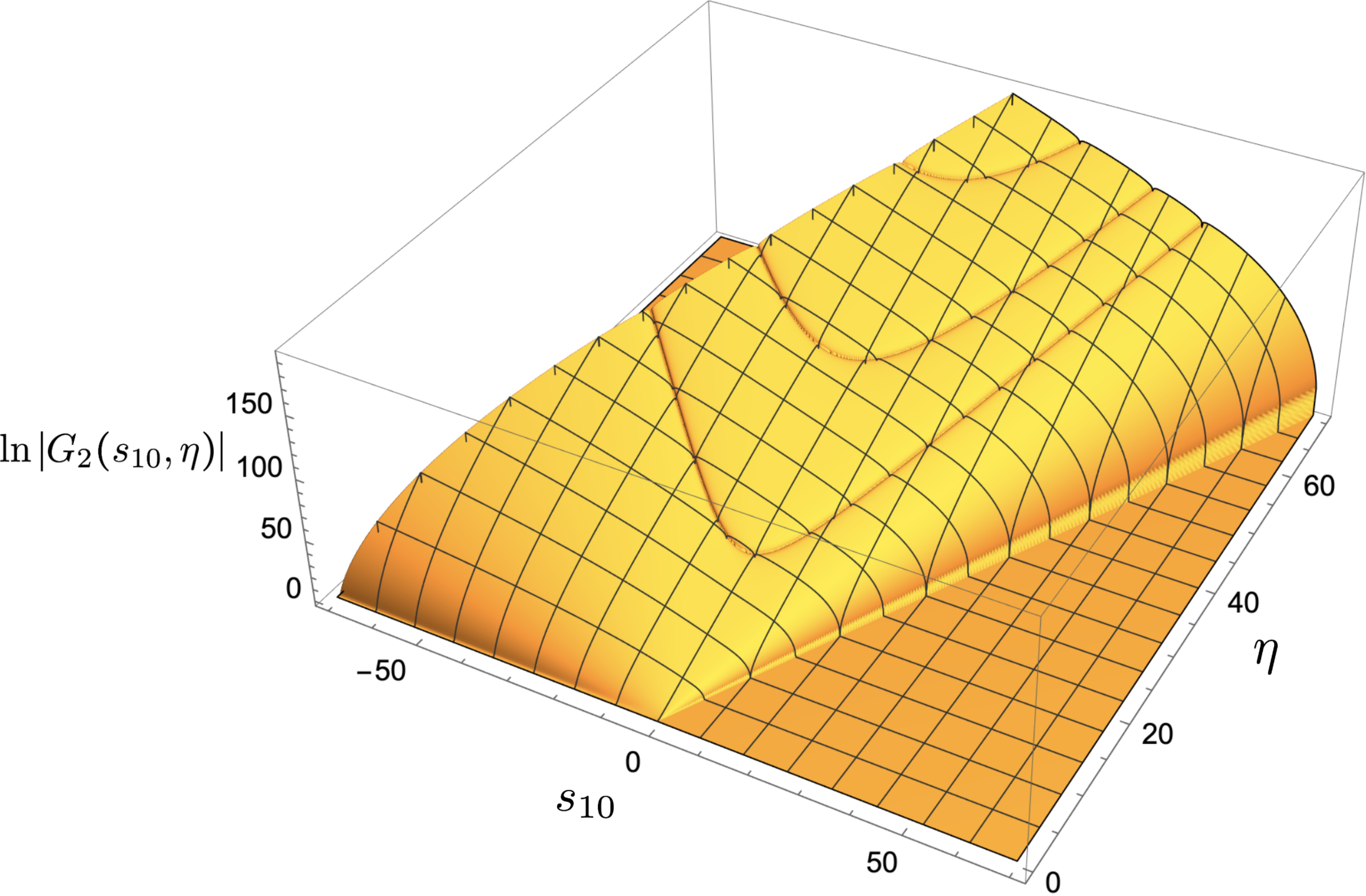}
         \caption{$\ln\left|G_2(s_{10},\eta)\right|$}
         \label{fig:QGG3dNf_G2}
     \end{subfigure}  \;
     \begin{subfigure}[b]{0.32\textwidth}
         \centering
         \includegraphics[width=\textwidth]{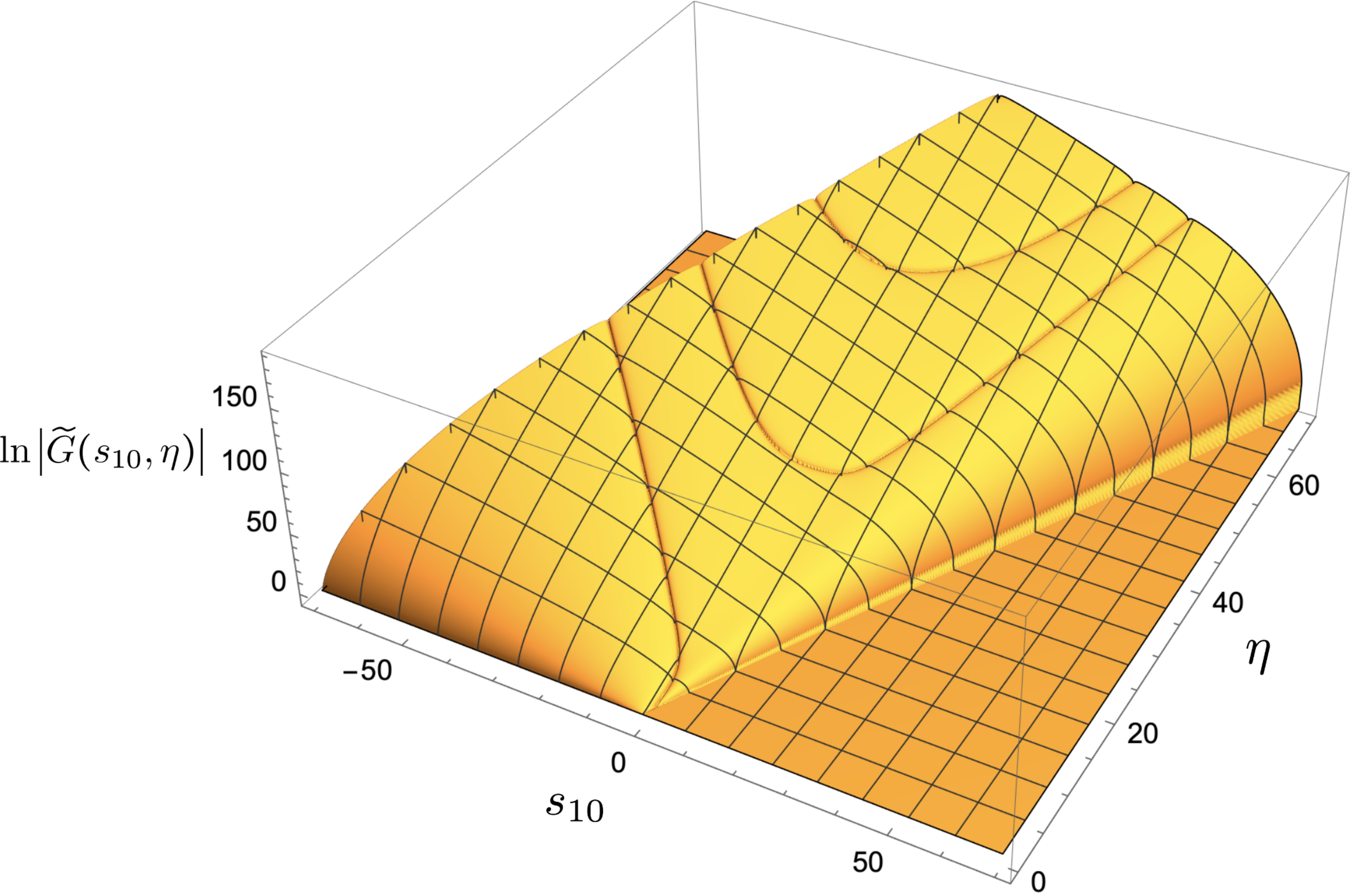}
         \caption{$\ln\left|{\widetilde G}(s_{10},\eta)\right|$}
         \label{fig:QGG3dNf_G}
     \end{subfigure}
     \caption{The plots of logarithms of the absolute values of polarized dipole amplitudes $Q$, $G_2$ and ${\widetilde G}$ at $N_f=6, N_c =3$ versus $s_{10}$ and $\eta$, in the $-\eta_{\max}\leq s_{10}\leq \eta_{\max}$, $0\leq\eta\leq\eta_{\max}$ region with $\eta_{\max}=70$. The amplitudes were computed numerically using the step size $\delta = 0.1$.}
     \label{fig:QGG3dNf}
\end{figure}

Now, we consider the $N_f=6$ case. Again, we begin with the step size $\delta = 0.1$ and maximum rapidity $\eta_{\max}=70$. This gives us the polarized dipole amplitudes plotted in \fig{fig:QGG3dNf}. There, each plot shows the logarithm of the absolute value of the labeled amplitude. The plots demonstrate an approximately linear rise of $\ln|Q(s_{10},\eta)|$, $\ln|G_2(s_{10},\eta)|$ and $\ln|{\widetilde G}(s_{10},\eta)|$ with $\eta$, similar to the lower-$N_f$ case presented in Sec.~\ref{ssec:lowNf}. The only difference is that the rise is no longer monotonic and appears to be periodically interrupted by lines of sharp local minima. 

\begin{figure}
     \centering
     \begin{subfigure}[b]{0.32\textwidth}
         \centering
         \includegraphics[width=\textwidth]{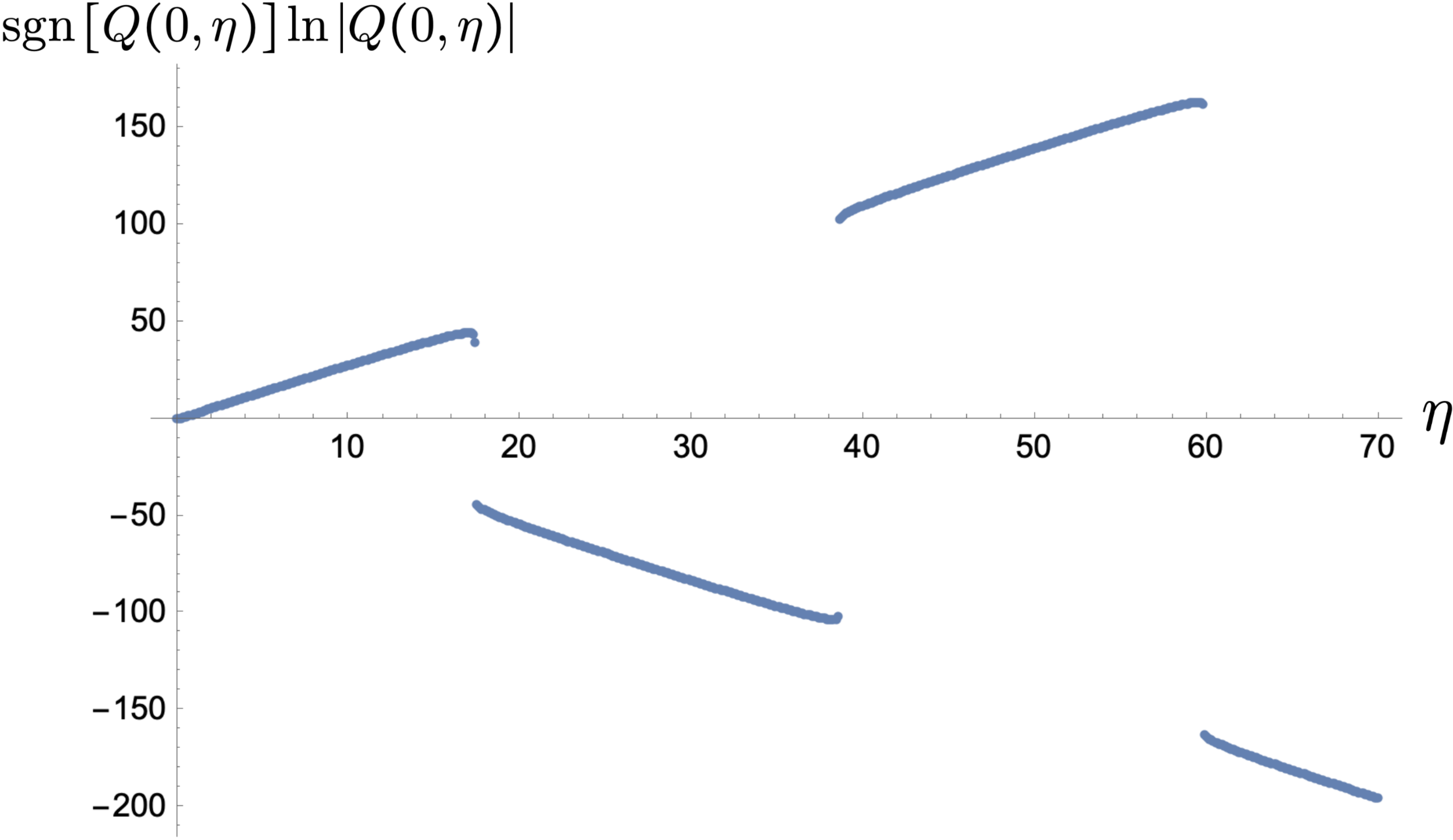}
         \caption{sgn$\left[Q(0,\eta)\right]\ln\left|Q(0,\eta)\right|$}
         \label{fig:signln2dNf_Q}
     \end{subfigure}  \;
     \begin{subfigure}[b]{0.32\textwidth}
         \centering
         \includegraphics[width=\textwidth]{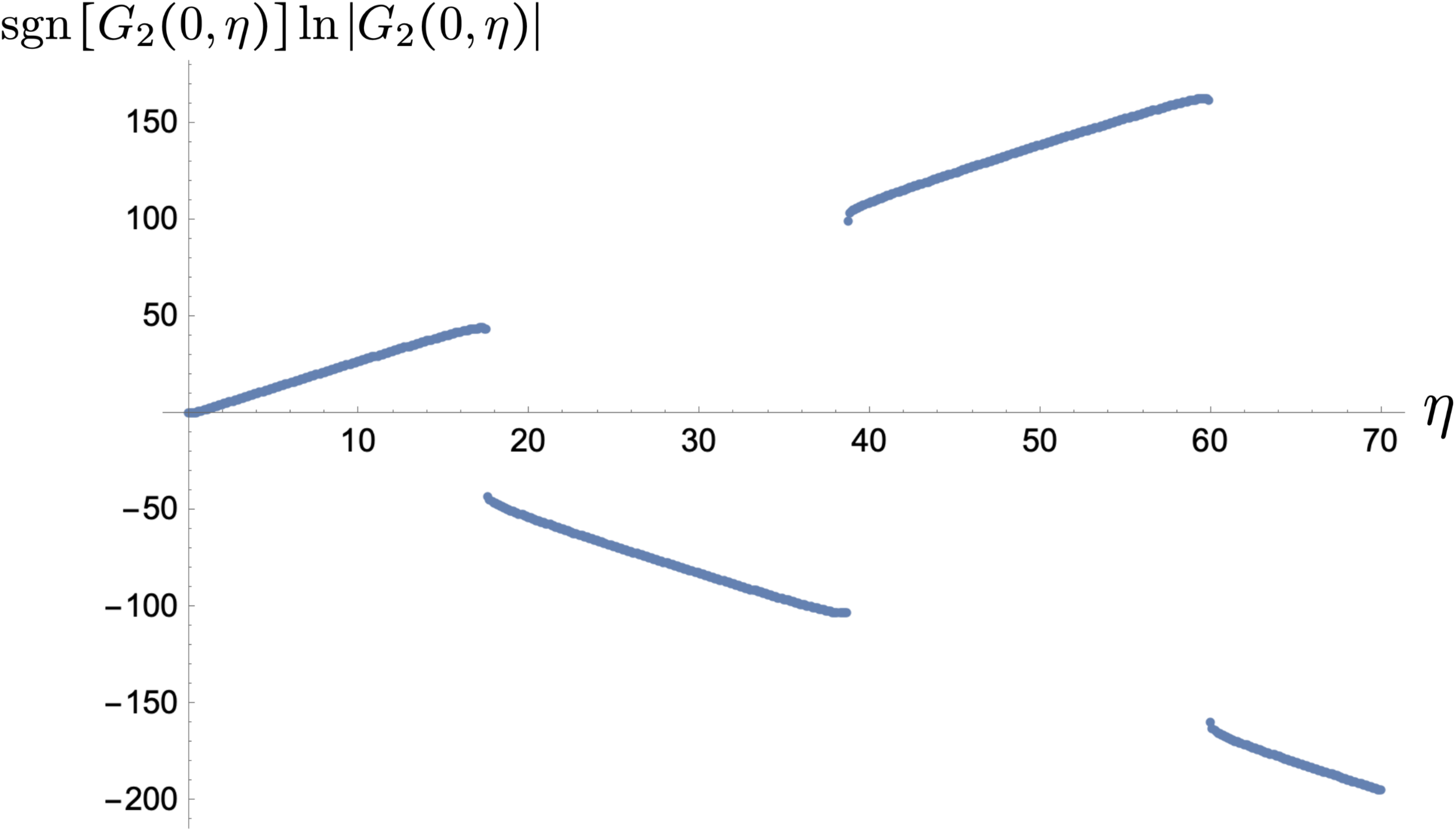}
         \caption{sgn$\left[G_2(0,\eta)\right]\ln\left|G_2(0,\eta)\right|$}
         \label{fig:signln2dNf_G2}
     \end{subfigure}  \;
     \begin{subfigure}[b]{0.32\textwidth}
         \centering
         \includegraphics[width=\textwidth]{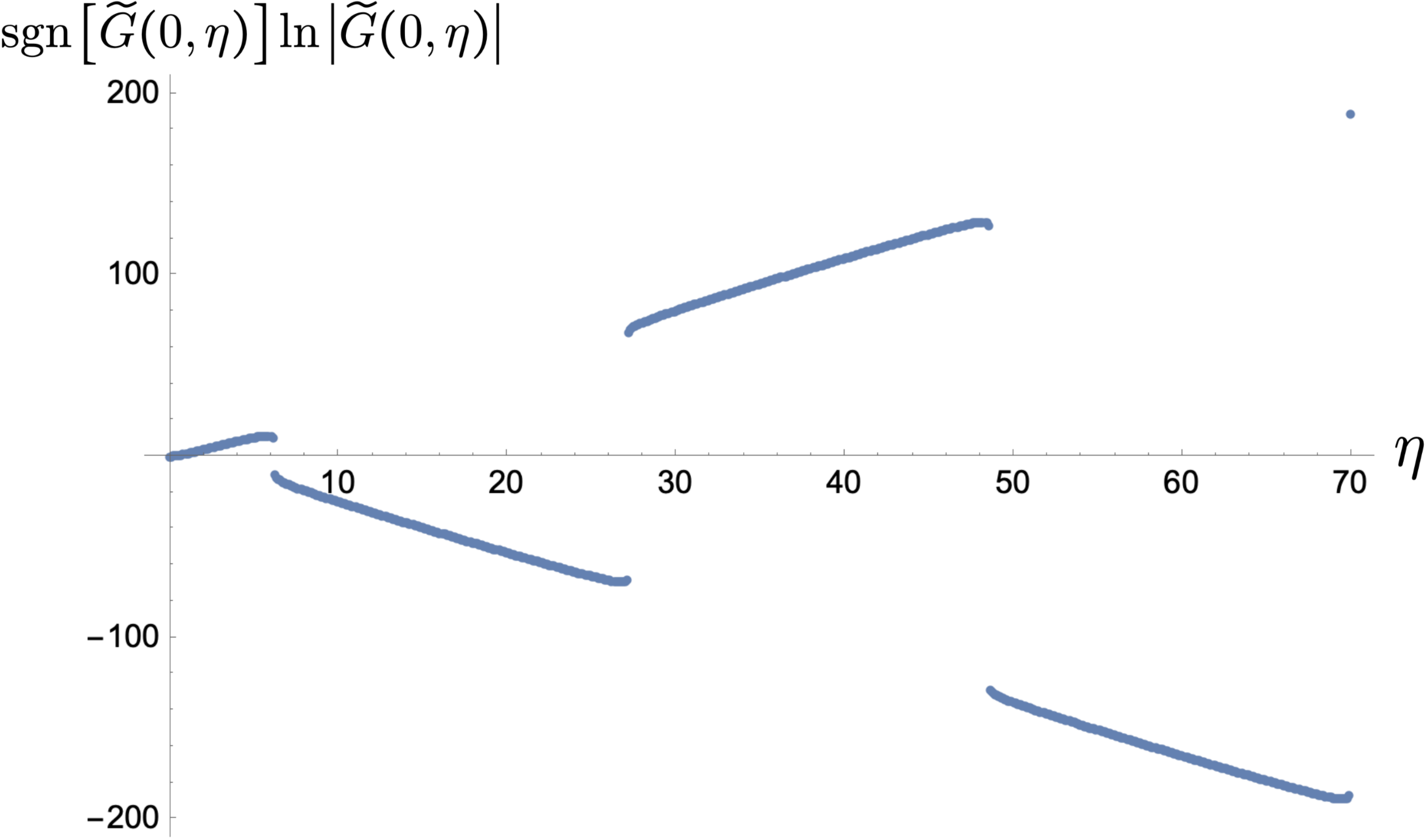}
         \caption{sgn$\left[{\widetilde G}(0,\eta)\right]\ln\left|{\widetilde G}(0,\eta)\right|$}
         \label{fig:signln2dNf_G}
     \end{subfigure}
     \caption{The plots of the logarithms of the absolute values of polarized dipole amplitudes $Q$, $G_2$ and ${\widetilde G}$, multiplied by the signs of the amplitudes, versus the rapidity $\eta$ along the $s_{10}=0$ line. The amplitudes are computed numerically in the range $0\leq\eta\leq\eta_{\max} =70$ using step size $\delta = 0.1$ at $N_f=6, N_c = 3$.}
     \label{fig:signln2dNf}
\end{figure}

To illustrate the origin of this non-monotonicity, we plot sgn$[Q(0,\eta)] \, \ln|Q(0,\eta)|$, sgn$[G_2(0,\eta)] \, \ln|G_2(0,\eta)|$ and sgn$[{\widetilde G}(0,\eta)] \, \ln|{\widetilde G}(0,\eta)|$ as functions of $\eta$ in \fig{fig:signln2dNf}. From these plots we see that $Q(0,\eta)$, $G_2(0,\eta)$ and ${\widetilde G}(0,\eta)$ oscillate with $\eta$: the oscillations explain the non-monotonic behavior we saw in \fig{fig:QGG3dNf}. This emergence of the oscillatory behavior in the amplitude marks the main qualitative difference between the case of $N_f=6$ and those with fewer quark flavors, for the calculation of small-$x$ asymptotics for the quark helicity distribution at large $N_c \& N_f$. (More precisely, we see that in the large-$N_c \& N_f$ limit, the oscillations are absent for $N_f/N_c < 2$ and set in at $N_f /N_c =2$.) In contrast, a similar study \cite{Kovchegov:2020hgb} performed at large $N_c\& N_f$ for the evolution equations without the type-2 polarized dipole amplitude saw the oscillatory behavior for any $N_f$ between 2 and 6. 

While an analytic solution of Eqs.~\eqref{evoleq} is beyond the scope of this work, we can attempt to find analytic formulas approximating our numerical results, at least in the large-$\eta$ asymptotics. Combining the oscillations with the exponential growth of the maxima of $|Q(0,\eta)|$, $|G_2(0,\eta)|$ and $|{\widetilde G}(0,\eta)|$ with $\eta$, we propose the following large-$\eta$ asymptotic forms for the polarized dipole amplitudes \cite{Kovchegov:2020hgb}:
\begin{subequations}\label{asym2}
\begin{align}
Q(0,\eta) &\sim e^{\alpha_Q\eta}\cos\left(\omega_Q\eta+\varphi_Q\right)    , \label{asym2Q} \\
G_2(0,\eta) &\sim e^{\alpha_{G_2}\eta}\cos\left(\omega_{G_2}\eta+\varphi_{G_2}\right) ,  \label{asym2G2}  \\
{\widetilde G}(0,\eta) &\sim e^{\alpha_{{\widetilde G}}\eta}\cos\left(\omega_{{\widetilde G}}\eta+\varphi_{{\widetilde G}}\right)    . \label{asym2G}
\end{align}
\end{subequations}
For the respective dipole amplitudes, the oscillation frequencies are denoted by $\omega_Q$, $\omega_{G_2}$ and $\omega_{{\widetilde G}}$, while the initial phases are denoted by $\varphi_Q$, $\varphi_{G_2}$ and $\varphi_{{\widetilde G}}$. Furthermore, the amplitudes of oscillations in $Q(0,\eta)$, $G_2(0,\eta)$ and ${\widetilde G}(0,\eta)$ grow exponentially with $\eta$, with the exponents $\alpha_Q$, $\alpha_{G_2}$ and $\alpha_{{\widetilde G}}$, respectively. 

To extract the parameters from Eqs.~\eqref{asym2}, consider a general function of the form
\begin{equation}
f(\eta) = Ke^{\alpha \eta}\cos\left(\omega \, \eta+ \varphi \right)
\label{eqn:Analysis1}
\end{equation}
with some parameters $\alpha$, $\omega$, $\varphi$, and $K$. This is the asymptotic form proposed at large $\eta$ in Eqs.~\eqref{asym2} for the polarized dipole amplitudes. We see that
\begin{align}
\frac{d^2}{d\eta^2}\ln\left|f(\eta)\right| &= \frac{d}{d\eta}\left[\alpha - \omega \, \tan\left(\omega \, \eta+ \varphi \right)\right] = -\frac{\omega^2}{\cos^2\left(\omega \, \eta+ \varphi \right)}\;.
\label{eqn:Analysis2}
\end{align}
Then, this second derivative contains local maxima where $\cos\left(\omega \, \eta+ \varphi \right) = \pm 1$. As a result, in the context of a numerical calculation, the frequency, $\omega$, can be found from the value of the numerically-obtained second derivative at the maximum,
\begin{align}\label{max_omega}
\max \left[ \frac{d^2}{d\eta^2}\ln\left|f(\eta)\right| \right] = - \omega^2 \, .
\end{align}
Here, we adopt a convention in which $\omega >0$ when deducing $\omega$ from a local maximum in Eq.~\eqref{max_omega}. In particular, we use the largest-$\eta$ maximum available in our numerical results to extract $\omega$, in order to get as close as possible to the large-$\eta$ regime. The extracted value of $\omega$ can be cross-checked by comparing $\pi/\omega$ to the spacing between the positions of the local maxima along the $\eta$-axis in the numerical solution. 

The phase, $\varphi$, can then be determined from the second derivative maximum condition $\omega \, \eta^* + \varphi = \pi \, n$, where $\eta^*$ is the numerically-extracted position of the largest-$\eta$ local maximum from Eq.~\eqref{max_omega}. Here, $n$ is an integer whose value is adjusted so that $\varphi \in (-\pi, \pi]$. In particular, the choice between $\varphi \in (0, \pi]$ and $\varphi \in (-\pi, 0]$ is made by making sure that the sign of $f(\eta^*)$ we calculated numerically matches that of $\cos(\omega\eta^*+\varphi)$.

Finally, we notice that 
\begin{equation}
\ln\left|\frac{f(\eta)}{\cos\left(\omega  \eta+ \varphi \right)}\right| = \alpha \, \eta + \ln K \,.
\label{eqn:Analysis4}
\end{equation}
This allows us to obtain an estimate for the parameter $\alpha$ by performing a linear regression on $\ln|f(\eta)/\cos(\omega\eta+\varphi)|$ as a function of $\eta$ to determine the slope. Here, it is appropriate to use the parameter estimates of $\omega$ and $\varphi$ that we obtained from the previous step. The uncertainties from the estimates of $\omega$ and $\varphi$ are also propagated to each data point, leading to the weights for the linear regression mentioned above. Given $\eta_{max}$, once we determine the numerical values of the amplitudes found in the range $\eta \in [0, \eta_{max}]$, there is always the largest local maximum, $\eta^*$, which leads to the definition, $d = \min\left\{\eta_{\max}-\eta^*, \frac{\pi}{10\omega}\right\}$. Then, we extract $\alpha$ from the range $\eta \in [\eta^*-d,\eta^*+d]$ and associate $\alpha$, together with $\omega$ and $\varphi$ we found earlier, with $\eta^*+d$ (instead of $\eta_{\max}$). This eliminates the bias in $\alpha$ that may arise from extracting the slope at different phases in the oscillation, while still allowing us to perform the extraction in the large-$\eta$ region where the asymptotic behavior dominates. Furthermore, one avoids extracting the slope near the zeros of the cosine function, where Eq.~\eqref{eqn:Analysis4} becomes less accurate due to the divergence.

To estimate the uncertainties for these parameters, we first notice that the intercept receives uncertainty from residuals of the linear regression fit to the function in Eq.~\eqref{eqn:Analysis4}. This provides the uncertainty estimate for $\alpha$. As for $\omega$ and $\varphi$, their uncertainty estimates require a more careful consideration.

The oscillation frequency $\omega$ receives an uncertainty from the fact that it is estimated by the quantity in Eq.~\eqref{max_omega}, whose values come in discrete steps. To estimate its uncertainty, consider the case where the true local maximum, $\eta_{\text{true}}$, is off from the estimated location, $\eta^*$, by a distance, $\Delta\eta$. We also approximate the function in Eq.~\eqref{eqn:Analysis2} to be quadratic around the local maximum, taking the form of $-a(\eta - \eta_{\text{true}})^2 - \omega_{\text{true}}^2$ for some constant $a>0$. In this notation, we would make the exactly correct frequency estimate, $\hat{\omega}=\omega_{\text{true}}$, when $\eta^*=\eta_{\text{true}}$. Otherwise, the estimated frequency would be 
\begin{align}\label{asym6}
\hat{\omega}&=\omega_{\text{true}}+\Delta\omega=\sqrt{a(\eta^* - \eta_{\text{true}})^2 + \omega_{\text{true}}^2}\; .
\end{align}
If we assume that $\Delta\omega$ is small relative to $\omega_{\text{true}}$, then Eq.~\eqref{asym6} yields 
\begin{align}\label{asym6a}
\Delta\omega&\approx\frac{a}{2 \, \omega_{\text{true}}}\left(\eta^* - \eta_{\text{true}}\right)^2 .
\end{align}
Eq.~\eqref{asym6a} would lead to the uncertainty, $\Delta\omega$, if we knew the values of $a$ and $\omega_{\text{true}}$. To determine these parameters, we first assume that $\eta^*<\eta_{\text{true}}$ without loss of generality. Then, consider the values, $-\omega_{1}^2$ and $-\omega_{2}^2$, of $\frac{d^2}{d\eta^2}\ln\left|f(\eta)\right|$ at $\eta^*+\delta$ and $\eta^*-\delta$, respectively. Both $\omega_1$ and $\omega_2$ are calculable from the numerical results. Then, through a calculation similar to Eq.~\eqref{asym6}, it follows that
\begin{align}\label{asym6b}
a &= \frac{1}{2\delta^2}\left[\left(\omega_1^2-\hat{\omega}^2\right) + \left(\omega_2^2-\hat{\omega}^2\right)\right] .
\end{align}
Plugging Eq.~\eqref{asym6b} into Eq.~\eqref{asym6a} and approximating $\omega_{\text{true}}$ by $\hat{\omega}$, we obtain
\begin{align}\label{asym6c}
\Delta\omega&=\frac{1}{4 \, \hat{\omega} \, \delta^2}\left(\eta^* - \eta_{\text{true}}\right)^2\left[\left(\omega_1^2-\hat{\omega}^2\right) + \left(\omega_2^2-\hat{\omega}^2\right)\right] .
\end{align}
Finally, we assume a uniform distribution from $-\frac{\delta}{2}$ to $\frac{\delta}{2}$ for $\eta^*-\eta_{\text{true}}$, which is reasonable given that the true local maximum is equally likely to fall anywhere within the grid. Then, with probability $P$, we have that $\eta^*-\eta_{\text{true}}$ falls within $\left(\Delta\eta\right)_P=\frac{P\delta}{2}$ from zero. As a result, with the same probability, $\Delta\omega$ falls within 
\begin{align}\label{asym6d}
\left(\Delta\omega\right)_P &= \frac{P^2}{16 \, \hat{\omega}} \left[\left(\omega_1^2-\hat{\omega}^2\right) + \left(\omega_2^2-\hat{\omega}^2\right)\right] .
\end{align}
Then, we take this range with $P=0.95$ to be the uncertainty for our frequency estimate, $\hat{\omega}$. This corresponds to the $95\%$ confidence interval.

Finally, for the initial phase, the method discussed above implies that its uncertainty is equal to that of the product, $\omega_{\text{true}}\,\eta_{\text{true}}$, which is approximated by $\hat{\omega}\,\eta^*$ in the notations we used previously. Then, the uncertainty is simply
\begin{align}\label{asym7a}
\left(\Delta\varphi\right)_P &= \hat{\omega}\left(\Delta\eta\right)_P + \eta^*\left(\Delta\omega\right)_P \,,
\end{align}
where $\left(\Delta\eta\right)_P$ and $\left(\Delta\omega\right)_P$ can be read off from above. Again, we use $P=0.95$ to estimate the uncertainty of our best-fitted initial phase, in order to be consistently using $95\%$ confidence interval. 

\begin{figure} 
	\centering
	\begin{subfigure}{.32\textwidth}
		\includegraphics[width=\textwidth]{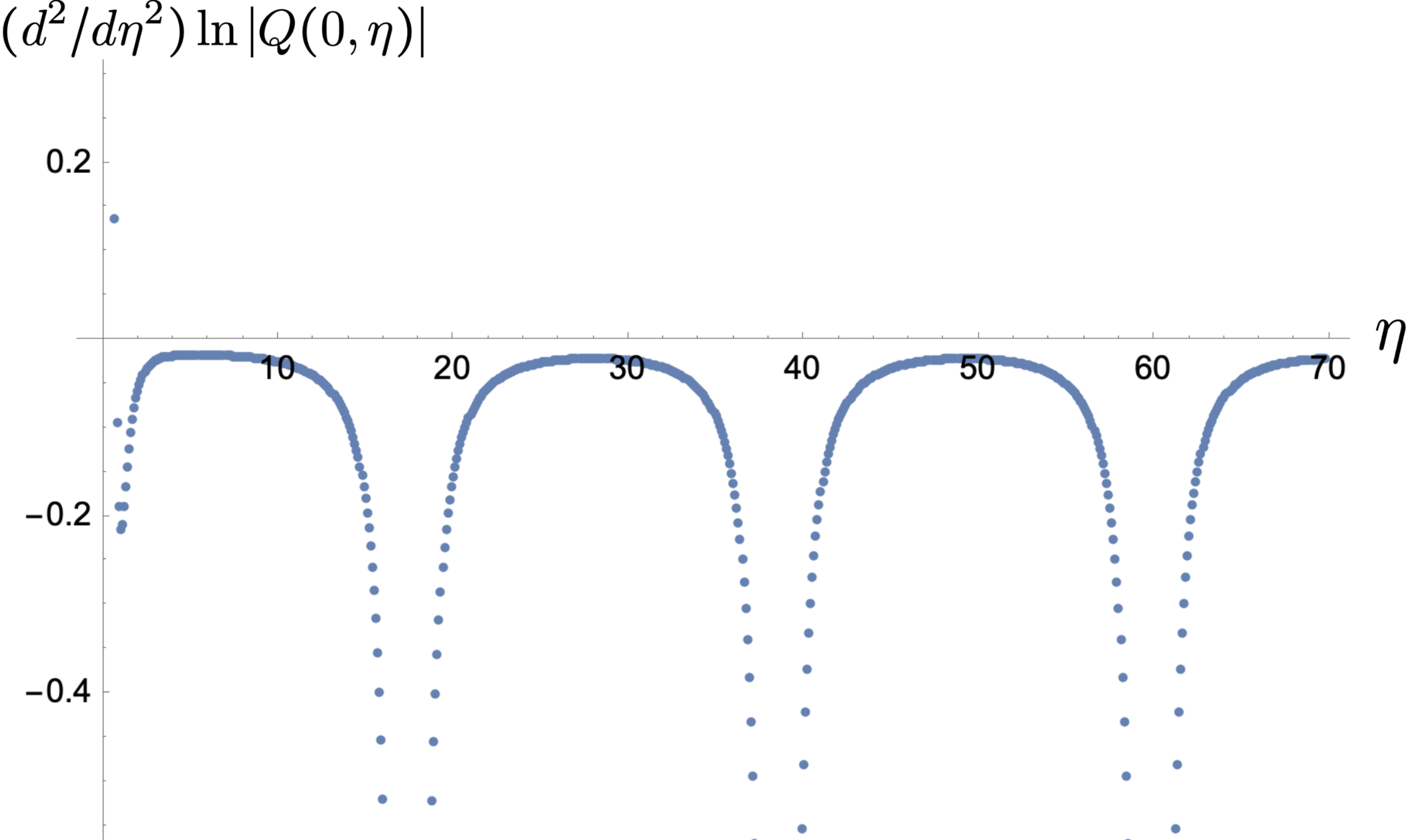}
		\caption{$\frac{d^2}{d\eta^2}\ln|Q(0,\eta)|$}
	\end{subfigure} \;
	\begin{subfigure}{.32\textwidth}
		\includegraphics[width=\textwidth]{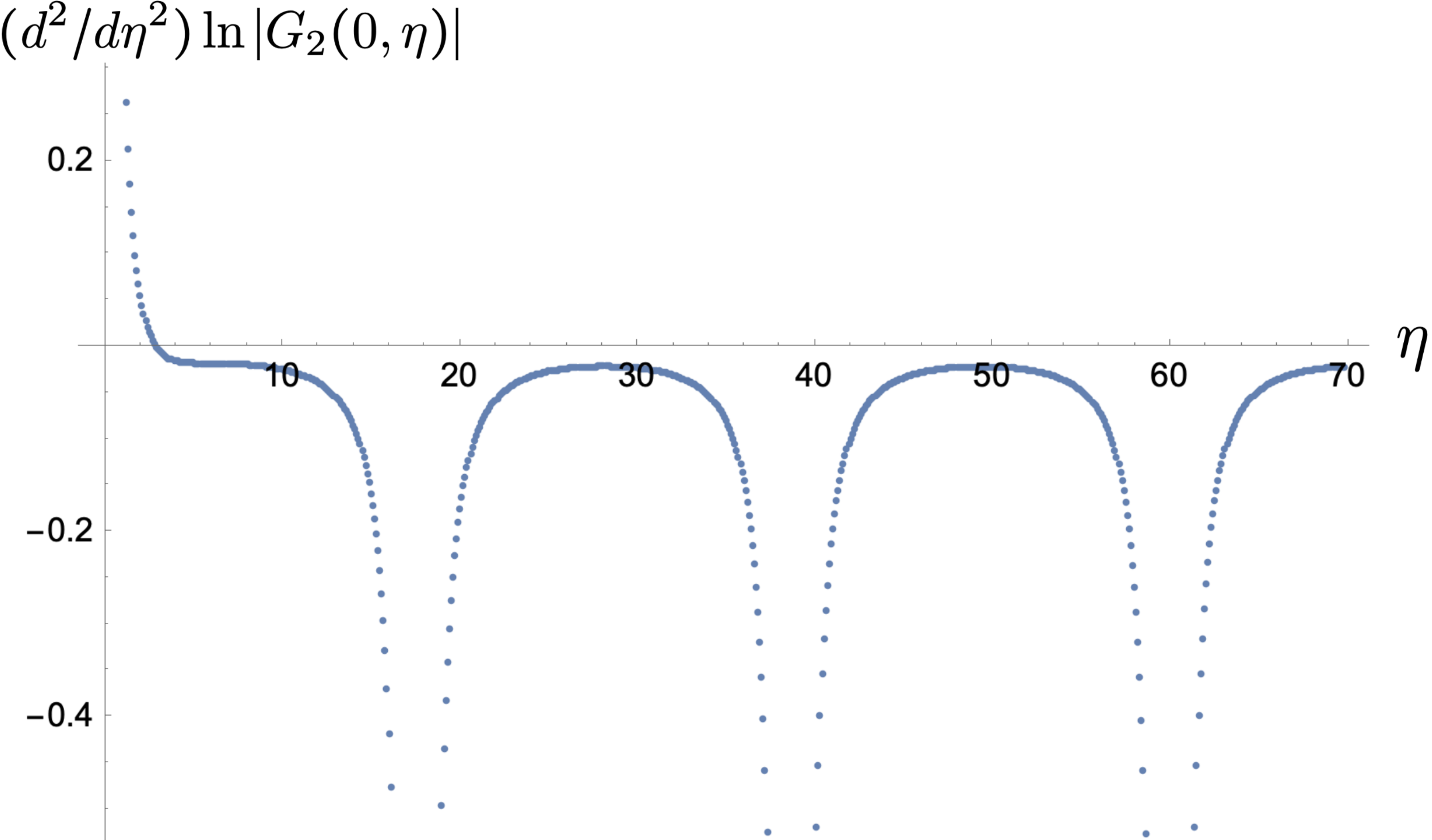}
		\caption{$\frac{d^2}{d\eta^2}\ln|G_2(0,\eta)|$}
	\end{subfigure} \;
	\begin{subfigure}{.32\textwidth}
		\includegraphics[width=\textwidth]{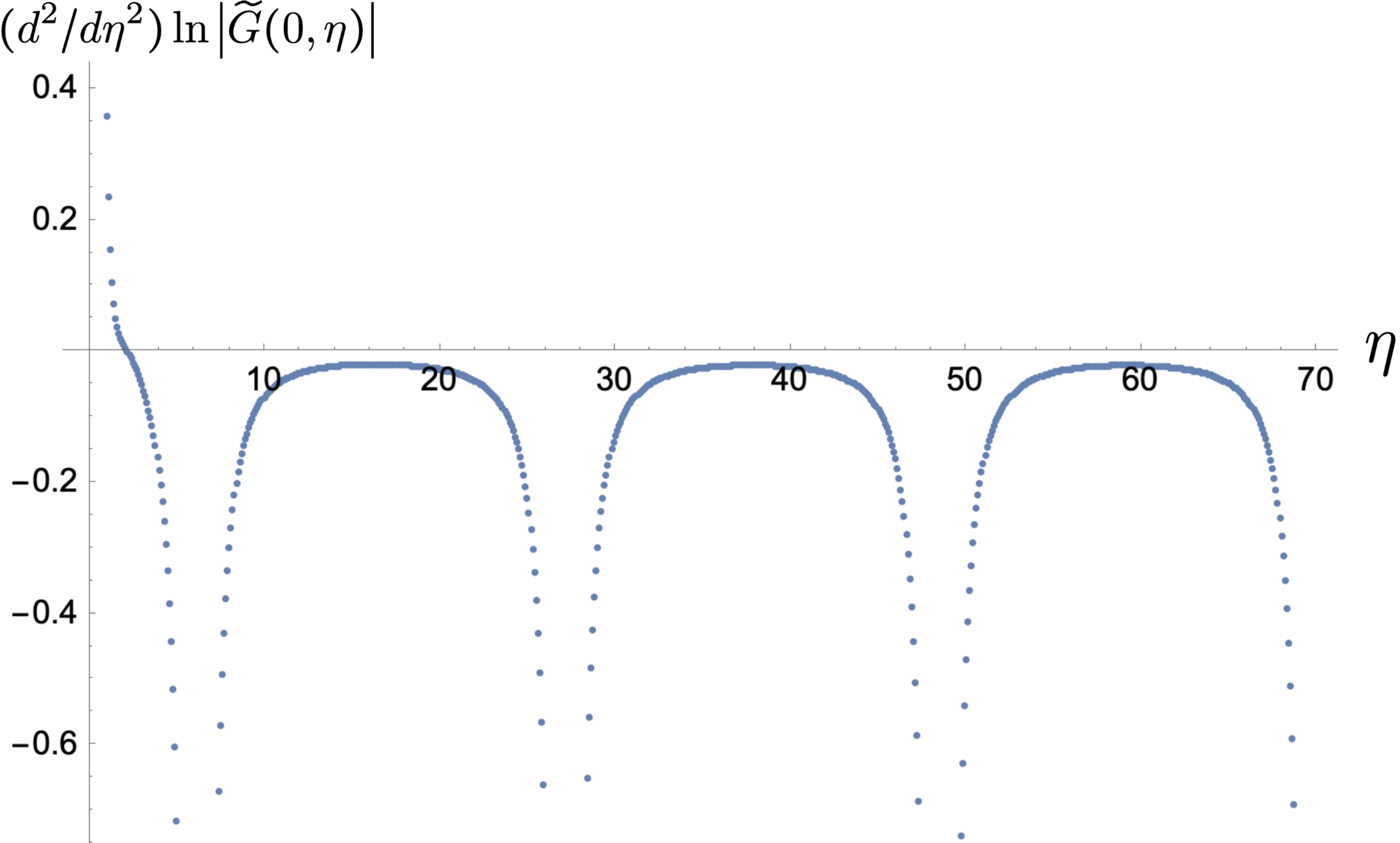}
		\caption{$\frac{d^2}{d\eta^2}\ln|{\widetilde G}(0,\eta)|$}
	\end{subfigure}
	\caption{Plots of $\frac{d^2}{d\eta^2}\ln\left|Q(0,\eta)\right|$, $\frac{d^2}{d\eta^2}\ln\left|G_2(0,\eta)\right|$ and $\frac{d^2}{d\eta^2}\ln\left|{\widetilde G}(0,\eta)\right|$ versus $\eta$ for $N_f=6$ and $N_c =3$. All graphs result from our numerical computation with step size $\delta = 0.1$ and $\eta_{\max} = 70$.}
\label{fig:ddlog}
\end{figure}

As a first cross check, we plot in \fig{fig:ddlog} the functions inspired by Eqs.~\eqref{asym2}, \eqref{eqn:Analysis1} and \eqref{eqn:Analysis2} applied to the polarized dipole amplitudes, namely $\frac{d^2}{d\eta^2}\ln\left|Q(0,\eta)\right|$, $\frac{d^2}{d\eta^2}\ln\left|G_2(0,\eta)\right|$ and $\frac{d^2}{d\eta^2}\ln\left|{\widetilde G}(0,\eta)\right|$ for $N_f=6$ as functions of $\eta$. For $\eta$ above 10, the shape of each graph qualitatively looks like that of the function $\frac{d^2}{d\eta^2}\ln\left|f(\eta)\right|$ in Eq.~\eqref{eqn:Analysis2}, displaying periodic local maxima below the $\eta$-axis. This provides another justification for the proposed asymptotic forms \eqref{asym2}, which resemble the definition of $f(\eta)$ given in Eq.~\eqref{eqn:Analysis1}.

\begin{table}[h]
\begin{center}
\begin{tabular}{|c|c|c|c|}
\hline
\;Dipole Amplitudes\;
& Intercept ($\alpha$)
& Frequency ($\omega$)
& \;Initial phase ($\varphi$)\;
\\ \hline 
$Q(0,\eta)$
& \;$2.801 \pm 0.007$\;
& \;$0.146803\pm 0.000004$\;
& $-0.940 \pm 0.007$
\\ \hline 
$G_2(0,\eta)$
& $2.802 \pm 0.007$
& $0.146821\pm 0.000004$
& $-0.955 \pm 0.007$
\\ \hline 
${\widetilde G}(0,\eta)$
& $2.802 \pm 0.006$
& $0.146294\pm 0.000004$
& $0.764 \pm 0.007$
\\ \hline 
\end{tabular}
\caption{Summary of the parameter estimates and uncertainties for all types of polarized dipole amplitudes along the $s_{10}=0$ line. Here, the number of quark flavors and colors are taken to be $N_f=6$ and $N_c=3$, respectively. The computation is performed with step size $\delta=0.1$, maximum rapidity $\eta_{\max}=70$, and the all-one initial condition~\eqref{asym1}.}
\label{tab:Nf6results}
\end{center}
\end{table}

Next, we fit our numerical results at $N_f=6$ with $\delta=0.1$ and $\eta_{\max}=70$, following the method outlined above. This leads to the parameter estimates and uncertainties listed in Table~\ref{tab:Nf6results}. There, the intercepts seem to be within the uncertainties from one another. In addition, the significant discrepancy in the frequency estimates could be a result of the discretization error that will be addressed shortly. However, the clear discrepancy is in the initial phase for $\gt$ compared to those of $Q$ and $G_2$, which is unlikely to be caused by the discretization error alone. This is entirely possible given that $\gt$ is defined based on an adjoint polarized dipole while $Q$ and $G_2$ are defined based on the fundamental counterparts. As far as the application of the results is concerned, the initial phase is a parameter that depends not only on the choice of initial condition but also on the value of Bjorken $x$ at which the small-$x$ evolution begins to dominate. In an actual phenomenological fit, one would be able to determine the proper initial phases by the moderate-$x$ data.

\begin{figure} 
	\centering
     \begin{subfigure}[b]{0.32\textwidth}
         \centering
         \includegraphics[width=\textwidth]{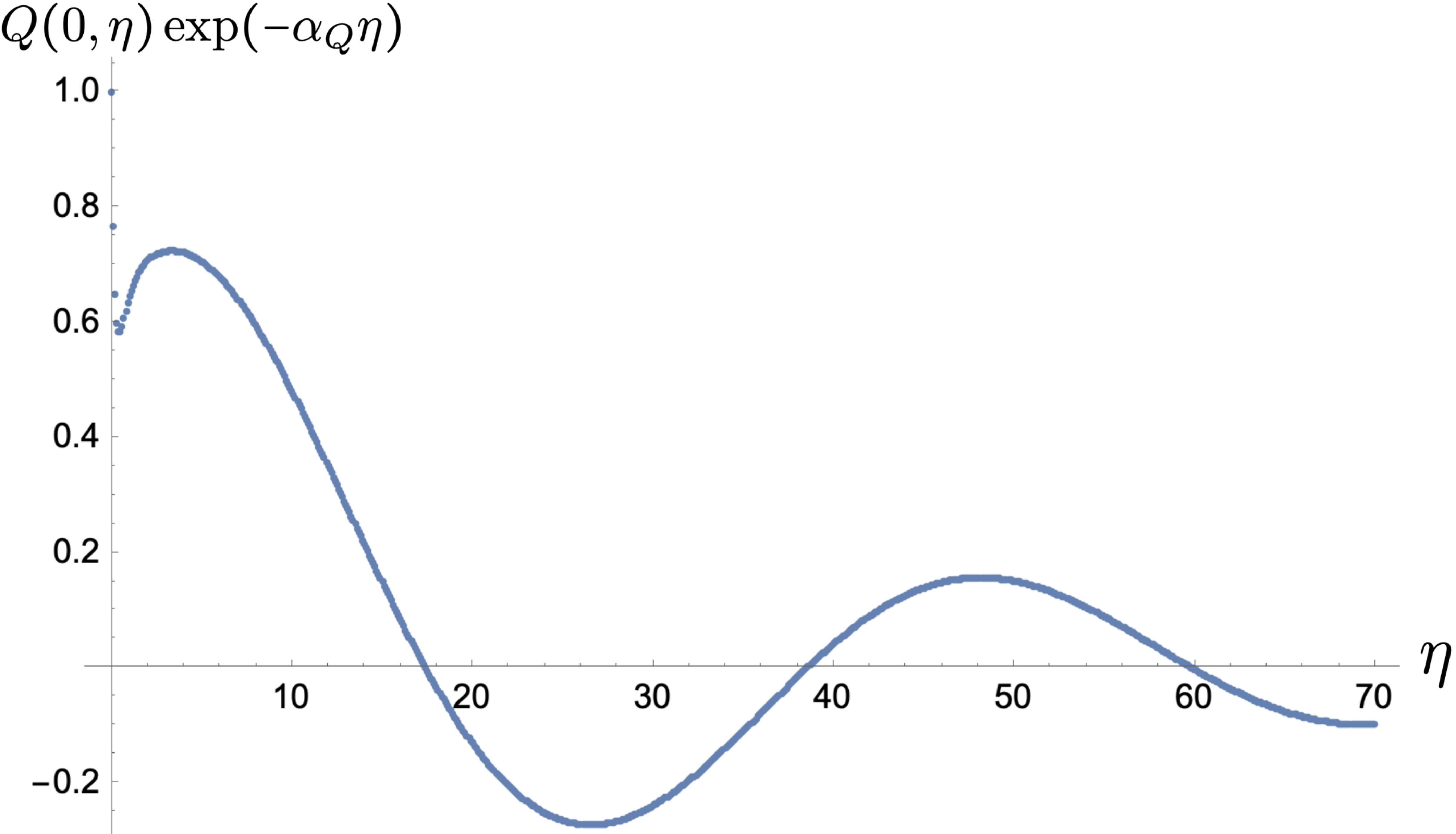}
         \caption{$e^{-\alpha_Q\eta}Q(0,\eta)$}
         \label{fig:osc_Q}
     \end{subfigure}  \;
     \begin{subfigure}[b]{0.32\textwidth}
         \centering
         \includegraphics[width=\textwidth]{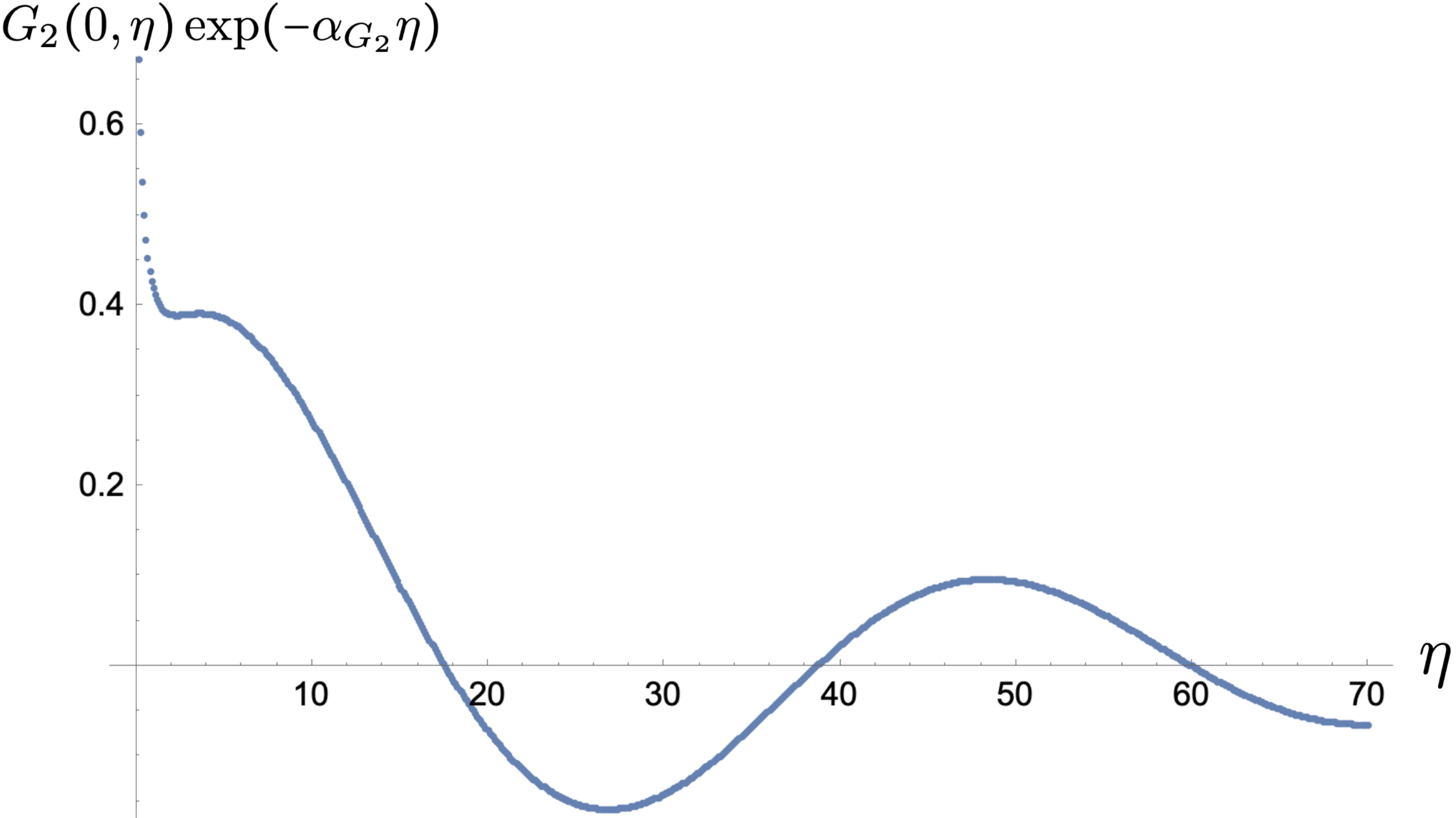}
         \caption{$e^{-\alpha_{G_2}\eta}G_2(0,\eta)$}
         \label{fig:osc_G2}
     \end{subfigure}  \;
     \begin{subfigure}[b]{0.32\textwidth}
         \centering
         \includegraphics[width=\textwidth]{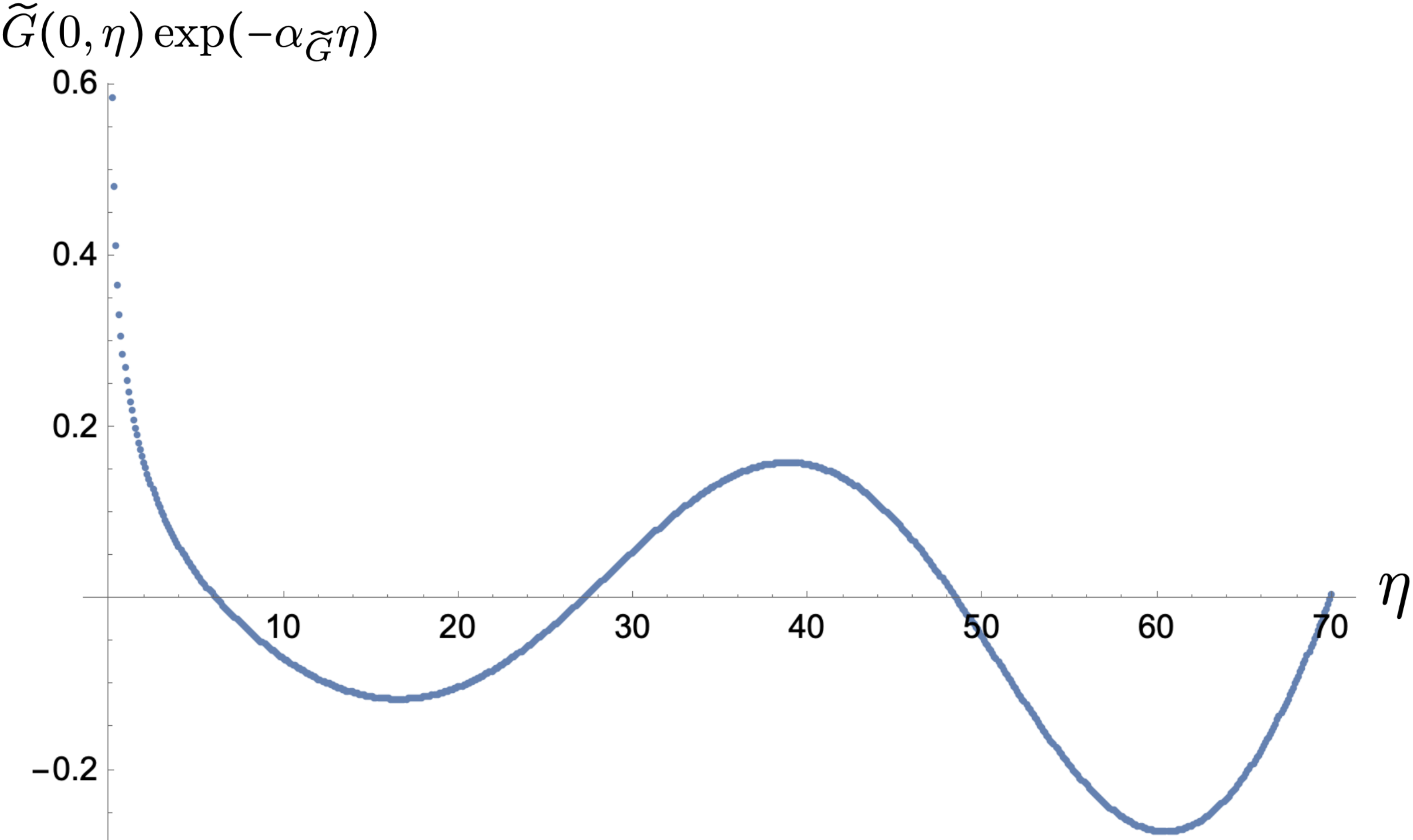}
         \caption{$e^{-\alpha_{\widetilde G}\eta}{\widetilde G}(0,\eta)$}
         \label{fig:osc_G}
     \end{subfigure}
	\caption{Plots of $e^{-\alpha_Q\eta}Q(0,\eta)$, $e^{-\alpha_{G_2}\eta}G_2(0,\eta)$ and $e^{-\alpha_{\widetilde G}\eta}{\widetilde G}(0,\eta)$ versus $\eta$ at $N_f=6$ and $N_c =3$. All the graphs are numerically computed with step size $\delta = 0.1$ and $\eta_{\max} = 70$.}
\label{fig:oscillating}
\end{figure}

\begin{figure} 
	\centering
     \begin{subfigure}[b]{0.32\textwidth}
         \centering
         \includegraphics[width=\textwidth]{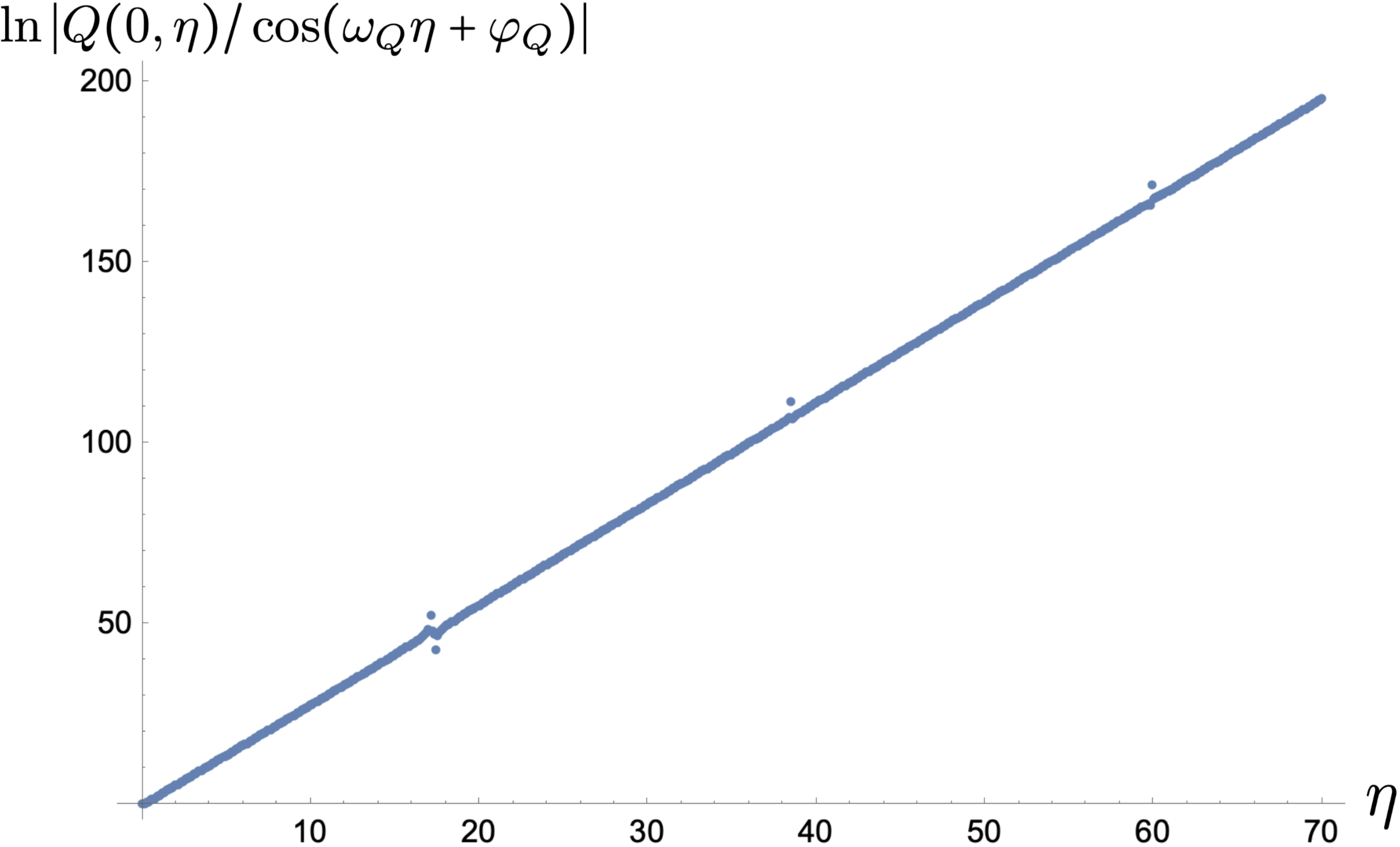}
         \caption{$\ln\left|\frac{Q(0,\eta)}{\cos(\omega_Q\eta+\varphi_Q)}\right|$}
         \label{exponential_Q}
     \end{subfigure}  \;
     \begin{subfigure}[b]{0.32\textwidth}
         \centering
         \includegraphics[width=\textwidth]{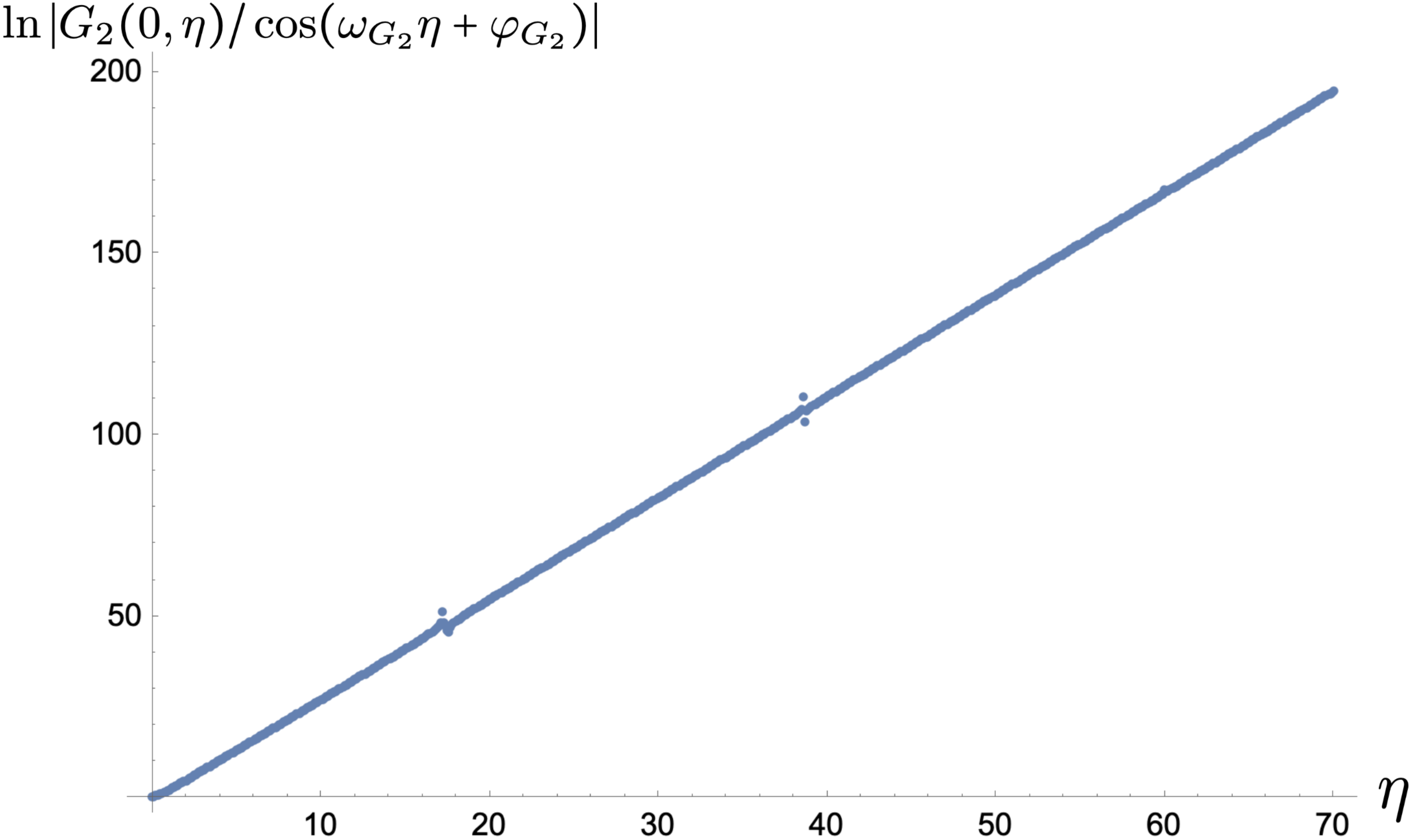}
         \caption{$\ln\left|\frac{G_2(0,\eta)}{\cos(\omega_{G_2}\eta+\varphi_{G_2})}\right|$}
         \label{exponential_G2}
     \end{subfigure}  \;
     \begin{subfigure}[b]{0.32\textwidth}
         \centering
         \includegraphics[width=\textwidth]{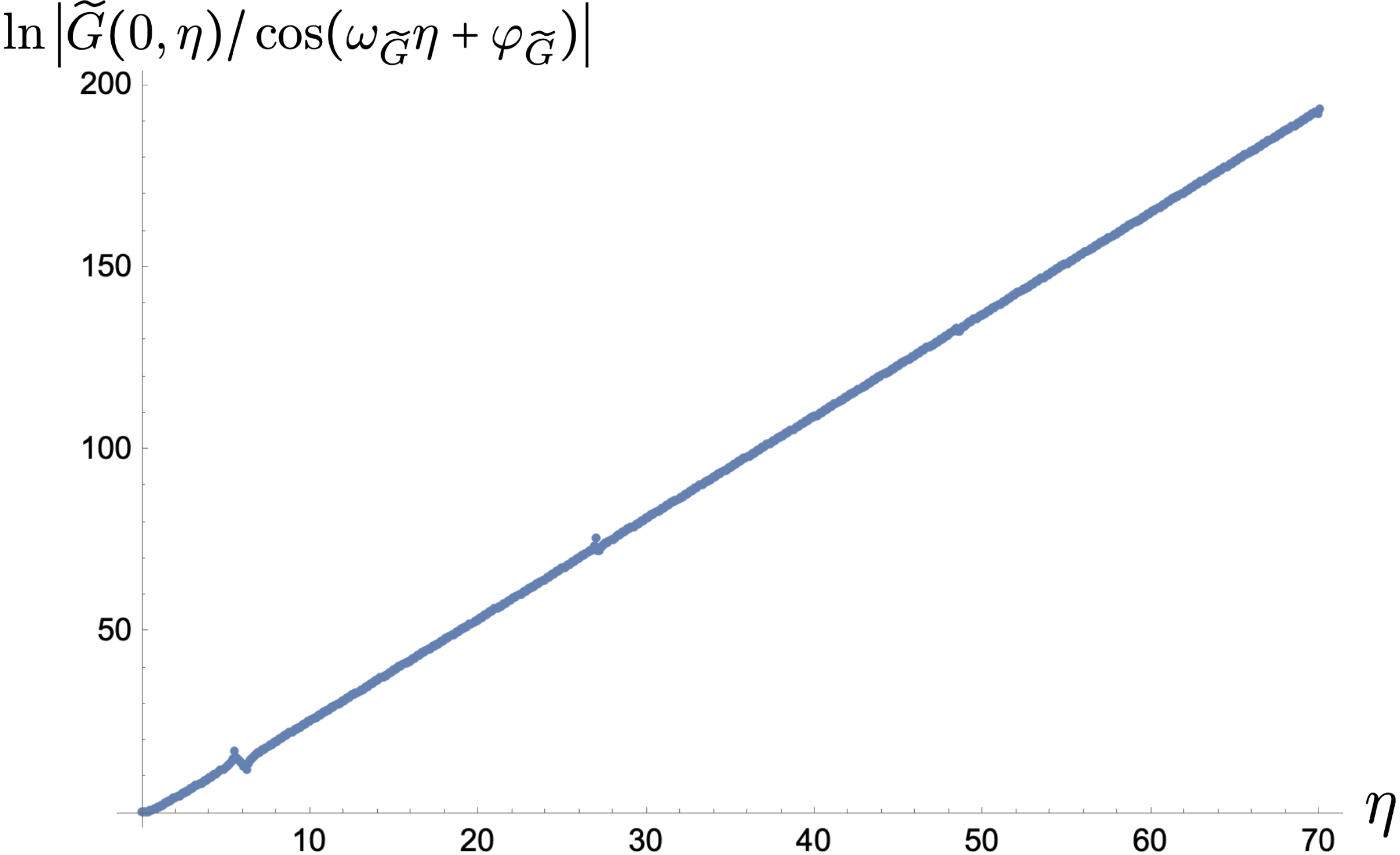}
         \caption{$\ln\left|\frac{{\widetilde G}(0,\eta)}{\cos(\omega_{\widetilde G}\eta+\varphi_{\widetilde G})}\right|$}
         \label{exponential_G}
     \end{subfigure}
	\caption{Plots of $\ln\left|\frac{Q(0,\eta)}{\cos(\omega_Q\eta+\varphi_Q)}\right|$, $\ln\left|\frac{G_2(0,\eta)}{\cos(\omega_{G_2}\eta+\varphi_{G_2})}\right|$ and $\ln\left|\frac{{\widetilde G}(0,\eta)}{\cos(\omega_{\widetilde G}\eta+\varphi_{\widetilde G})}\right|$ versus $\eta$ at $N_f=6$ and $N_c =3$. All the graphs are numerically computed with step size $\delta = 0.1$ and $\eta_{\max} = 70$.}
\label{exponential}
\end{figure}

Before we address the potential discretization error, we consider as final cross checks for the asymptotic forms \eqref{asym2} the functions, $e^{-\alpha_Q\eta}Q(0,\eta)$, $e^{-\alpha_{G_2}\eta}G_2(0,\eta)$ and $e^{-\alpha_{\widetilde G}\eta}{\widetilde G}(0,\eta)$. In \fig{fig:oscillating}, these functions are plotted against $\eta$. We see that the functions display a clear sinusoidal pattern for $\eta \gtrsim 30$, demonstrating oscillatory behavior in the large-$\eta$ asymptotics, as expected from our ans\"atze \eqref{asym2}. Furthermore, we plot $\ln\left|\frac{Q(0,\eta)}{\cos(\omega_Q\eta+\varphi_Q)}\right|$, $\ln\left|\frac{G_2(0,\eta)}{\cos(\omega_{G_2}\eta+\varphi_{G_2})}\right|$ and $\ln\left|\frac{{\widetilde G}(0,\eta)}{\cos(\omega_{\widetilde G}\eta+\varphi_{\widetilde G})}\right|$ versus $\eta$ in \fig{exponential}. From this second set of plots, we see that the logarithms grow roughly linearly with $\eta$, except for minor periodic bumps that occur near the sinusoidal nodes. This implies that the amplitudes divided by the corresponding cosine functions grow exponentially with $\eta$. Again, this is consistent with the asymptotic forms \eqref{asym2} proposed earlier.

To address potential biases coming from discretization, we repeat the calculation for different values of step size, $\delta$, and maximum rapidity, $\eta_{\max}$. In particular, we use $\delta = 0.0375, 0.05, 0.0625, 0.08, 0.1, 0.16, 0.2, 0.25, 0.5$. For each $\delta$, we obtain the parameter estimates using the method outlined above at each of the oscillation antinodes below $M(\delta)$, which is listed for each $\delta$ in Table~\ref{tab:M_delta_Nf6}. Note that the parameter estimation method also implies the corresponding $\eta^*+d$ to be associated with the estimated $\alpha$, $\omega$ and $\varphi$. We then perform a weighted polynomial regression against $\delta$ and $\eta^*+d$ for each fitted parameter, similar to what we did in Section~\ref{ssec:lowNf}. For all the parameters and the amplitudes, the quadratic model performs the best, resulting in the continuum-limit estimates, which is the model's prediction at $\delta = 1/\eta_{\max} = 0$, shown in Table~\ref{tab:Nf6resultsCont}.

\begin{table}[h]
\begin{center}
\begin{tabular}{|c|c|c|c|c|c|c|c|c|c|c|}
\hline
$\delta$ 
& 0.0375
& \,0.05\,
& 0.0625
& \,0.08\,
& \,\,0.1\,\,
& \,0.16\,
& \,\,0.2\,\,
& \,0.25\,
& \,\,0.5\,\,
\\ \hline 
$M(\delta)$
& 30
& 40
& 40
& 50
& 70
& 100
& 120
& 150
& 200
\\ \hline
\end{tabular}
\caption{The maximum, $M(\delta)$, of $\eta_{\max}$ computed for each step size $\delta$ in the case where $N_f=6, N_c =3$.}
\label{tab:M_delta_Nf6}
\end{center}
\end{table}

\begin{table}[h]
\begin{center}
\begin{tabular}{|c|c|c|c|}
\hline
\;Dipole Amplitudes\; 
& Intercept ($\alpha$)
& Frequency ($\omega$)
& \;Initial phase ($\varphi$)\;
\\ \hline 
$Q(0,\eta)$
& \;$2.82 \pm 0.04$\;
& \;$0.15074 \pm 0.00008$\;
& $-0.94 \pm 0.10$
\\ \hline 
$G_2(0,\eta)$
& $2.83 \pm 0.04$
& $0.15041 \pm 0.00008$
& $-0.90 \pm 0.10$
\\ \hline 
${\widetilde G}(0,\eta)$
& $2.82 \pm 0.04$
& $0.14807\pm 0.00005$
& $0.74 \pm 0.07$
\\ \hline 
\end{tabular}
\caption{Summary of the parameter estimates and uncertainties at the continuum limit ($\delta\to 0$ and $\eta_{\max}\to \infty$) for all types of polarized dipole amplitudes along the $s_{10}=0$ line. Here, the number of quark flavors and colors are taken to be $N_f=6$ and $N_c=3$, respectively. The computation is performed with the all-one initial condition \eqref{asym1}.}
\label{tab:Nf6resultsCont}
\end{center}
\end{table}

From Table~\ref{tab:Nf6resultsCont}, we see that the intercepts, $\alpha$, are the same within the uncertainty for all the amplitudes, while the frequencies, $\omega$, exhibit statistically significant but very small differences. As for the initial phase, $\varphi$, it is the same within the uncertainty for $Q$ and $G_2$, but it is significantly different for ${\widetilde G}$. Furthermore, the intercepts are below those for $N_f=4$, continuing the trend we observed earlier that the intercept decreases as we add more quark flavors.

\begin{figure}[ht] 
	\centering
     \begin{subfigure}[b]{0.32\textwidth}
         \centering
         \includegraphics[width=\textwidth]{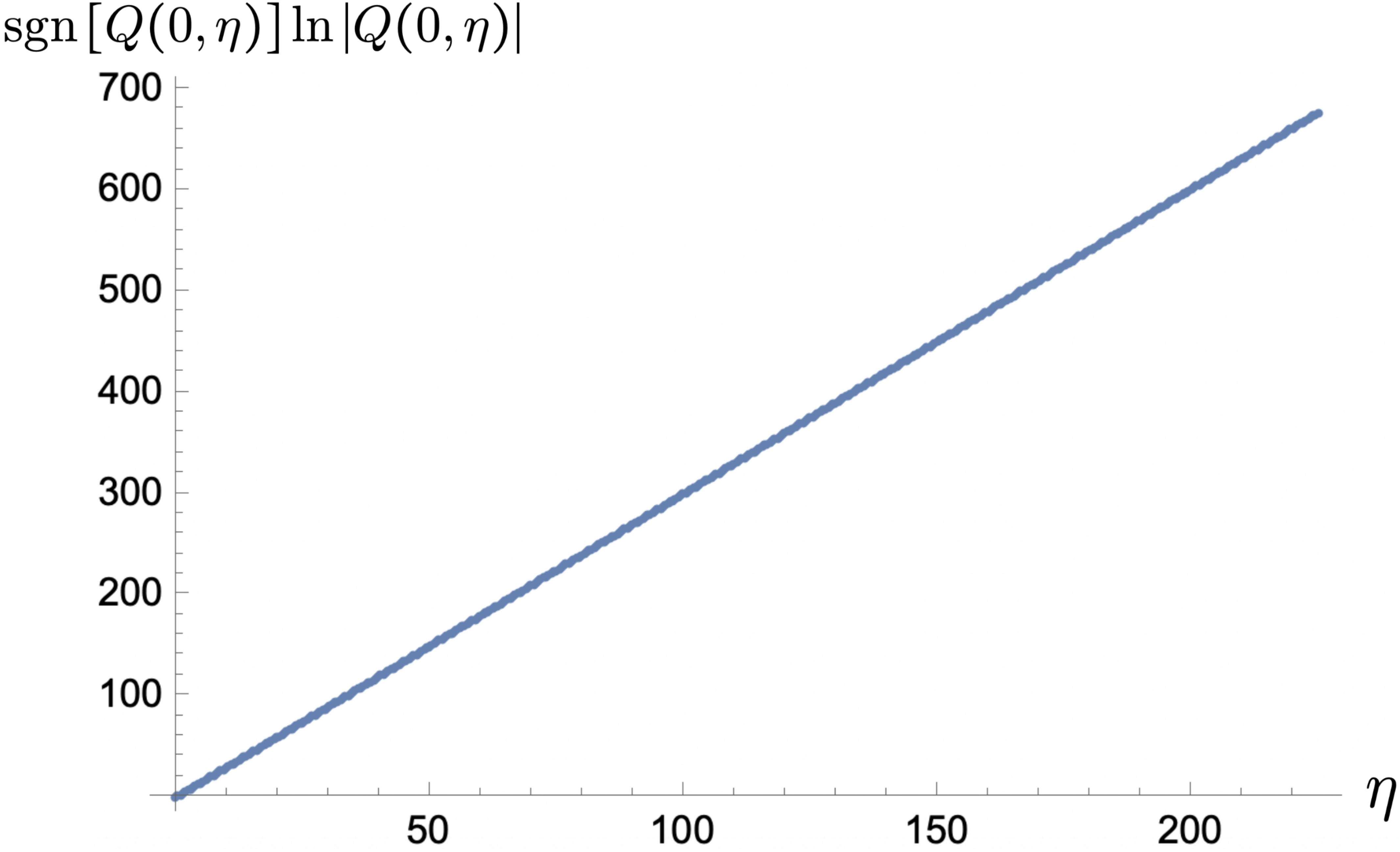}
         \caption{$\text{sgn}\left[Q(0,\eta)\right]\ln\left|Q(0,\eta)\right|$}
         \label{fig:ln2dLargeEta_Q}
     \end{subfigure}  \;
     \begin{subfigure}[b]{0.32\textwidth}
         \centering
         \includegraphics[width=\textwidth]{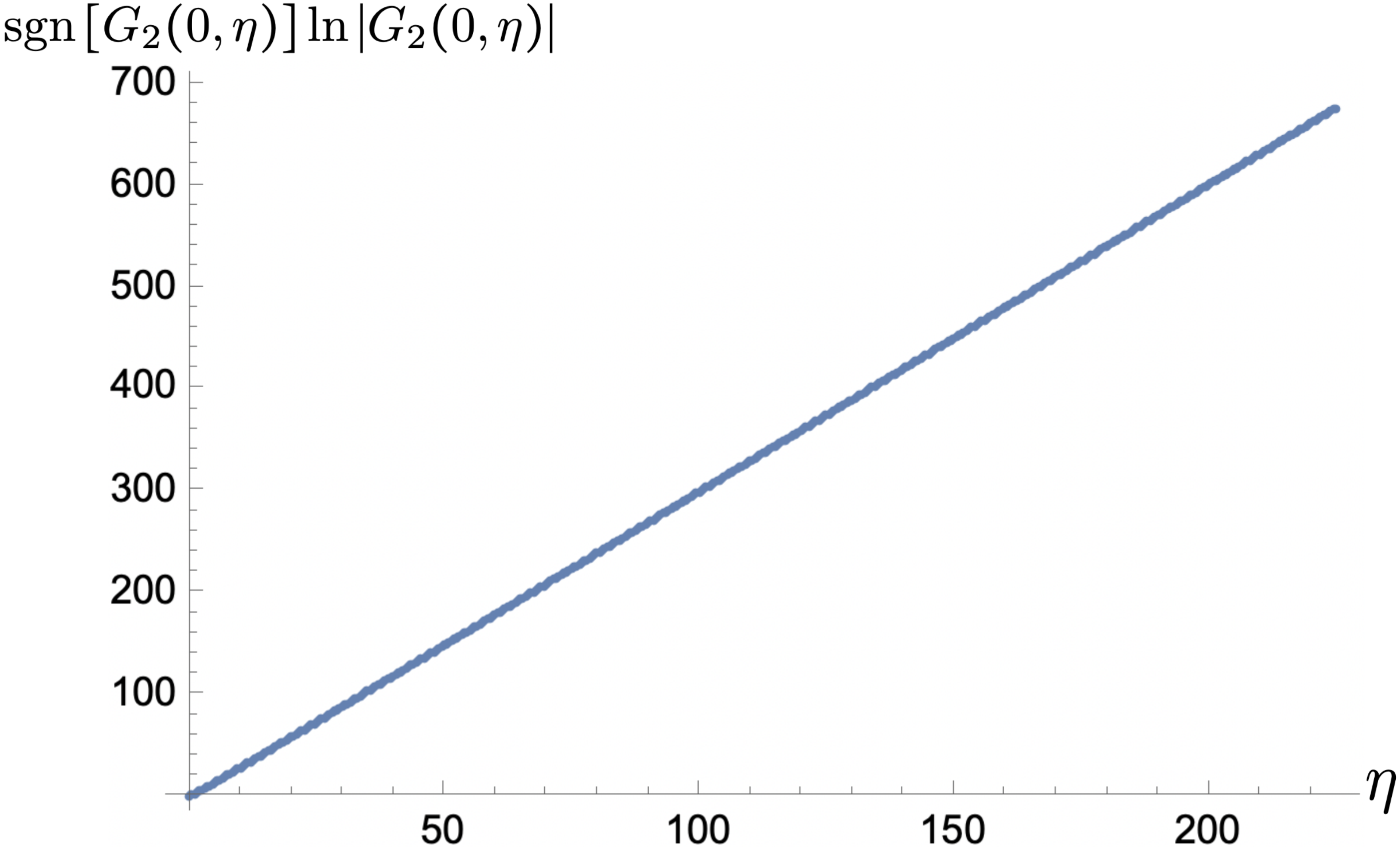}
         \caption{$\text{sgn}\left[G_2(0,\eta)\right]\ln\left|G_2(0,\eta)\right|$}
         \label{fig:ln2dLargeEta_G2}
     \end{subfigure} \;
     \begin{subfigure}[b]{0.32\textwidth}
         \centering
         \includegraphics[width=\textwidth]{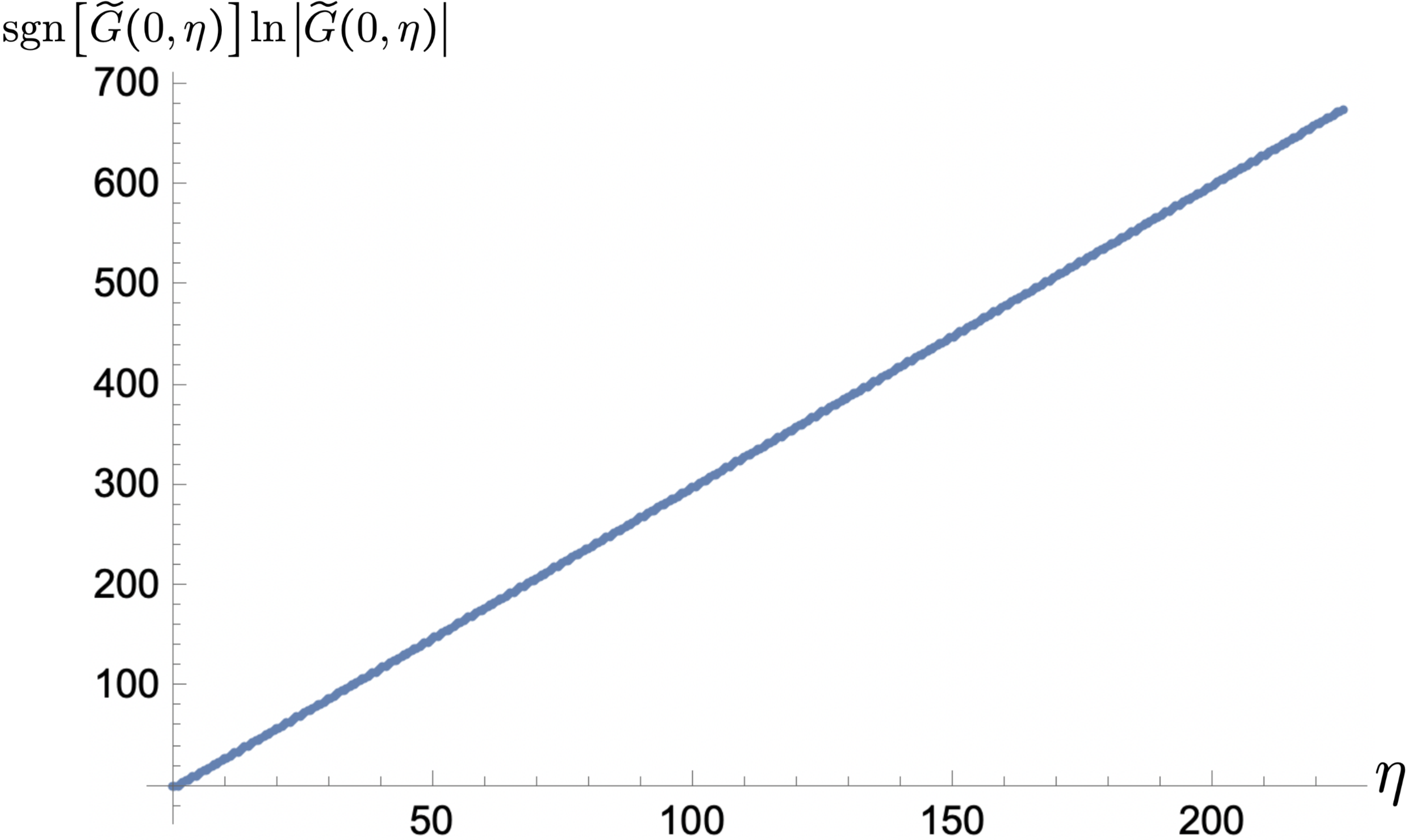}
         \caption{$\text{sgn}\,[{\widetilde G}(0,\eta)]\ln |{\widetilde G}(0,\eta)|$}
         \label{fig:ln2dLargeEta_G}
     \end{subfigure}
	\caption{The plots of logarithms of the absolute values of polarized dipole amplitudes $Q$, $G_2$ and ${\widetilde G}$, multiplied by their signs, along $s_{10}=0$ line, versus the rapidity, $\eta$. The amplitudes are computed numerically in the range $0\leq\eta\leq\eta_{\max} =225$ using step size $\delta = 0.5$ at $N_f=4, N_c =3$.}
\label{fig:ln2dLargeEta}
\end{figure}

With the qualitatively different results between $N_f\leq 5$ and $N_f=6$, we examine the possibility that the amplitudes also oscillate at $N_f\leq 5$ but with much longer periods than $\eta_{\max}=70$, which is the largest rapidity value we used in our calculations to this point for $N_f=2,3,4$. In particular, we repeat the computation at $N_f=4$ with $\delta=0.5$ and $\eta_{\max}=225$. As shown in \fig{fig:ln2dLargeEta} up to the rapidity of $\eta=225$, the logarithms of the absolute values of the dipole amplitudes at $N_f=4$ still grow linearly with $\eta$, displaying no sign of oscillation or other non-exponential behavior. The mathematical reason behind a radical change in the asymptotic behavior of the amplitudes as $N_f$ reaches the value of $2 N_c = 6$ remains unclear. It is likely that an analytic solution is necessary to offer a clear explanation of the transition. This is beyond the scope of this paper.

Now that the large-$\eta$ asymptotics have been established for the polarized dipole amplitudes, we are able to deduce the small-$x$ asymptotics for the gluon helicity PDF, $\Delta G(x,Q^2)$, by reading off the parameter estimates for $G_2(0,\eta)$ from Table~\ref{tab:Nf6resultsCont}. In particular, we have
\begin{align}
    &\Delta G(x,Q^2)\Big|_{Q^2=\Lambda^2} \sim \left(\frac{1}{x}\right)^{\left(2.83 \pm 0.04\right)\sqrt{\frac{\alpha_sN_c}{2\pi}}} \cos\left[\left(0.15041 \pm 0.00008\right)\sqrt{\frac{\alpha_sN_c}{2\pi}}\;\ln\frac{1}{x} - \left(0.90 \pm 0.10\right)\right]  . \label{glPDFNf6}
\end{align}

\begin{figure}
\begin{center}
\includegraphics[width=0.5\textwidth]{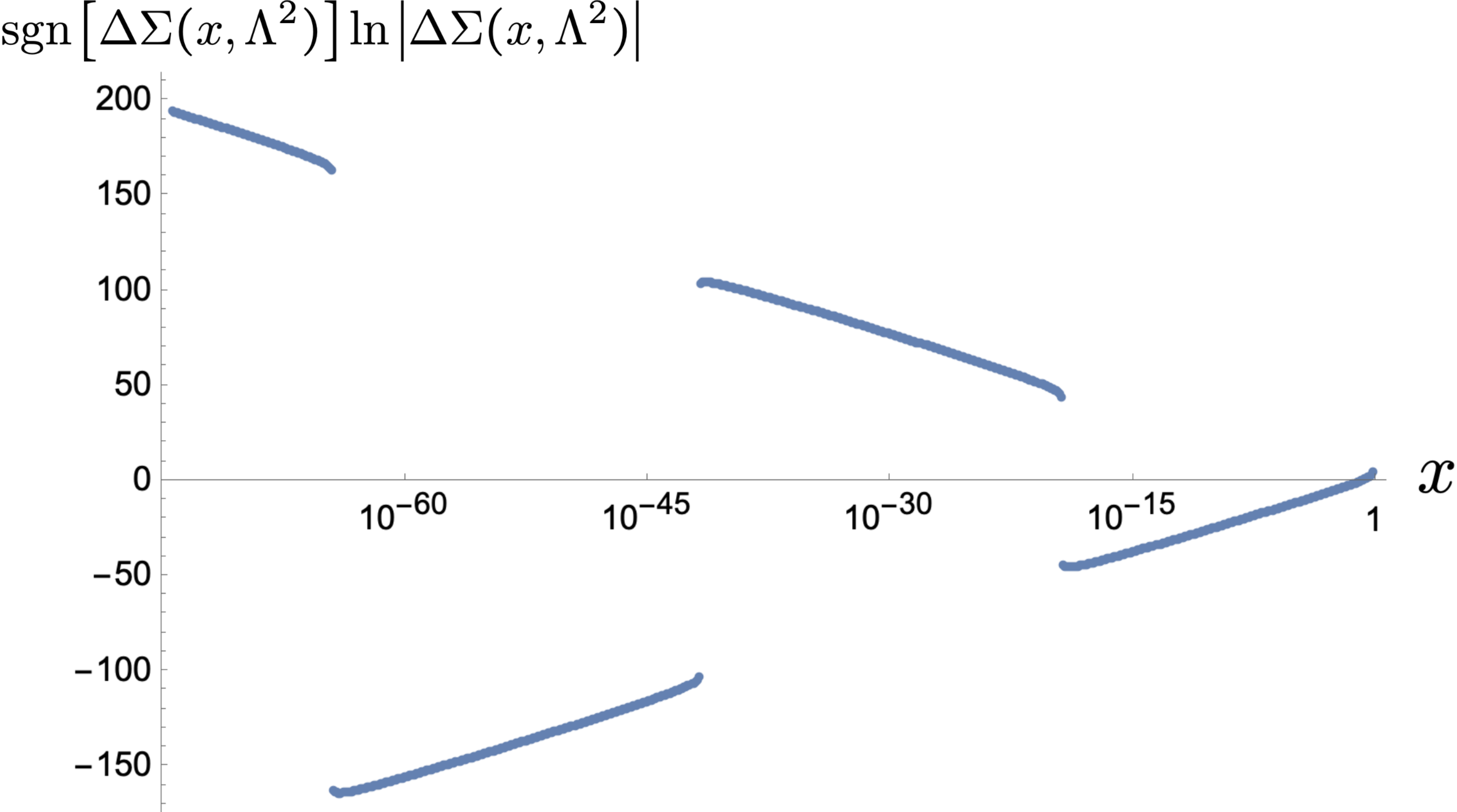}
\caption{The plot of sgn$\left[\Delta\Sigma(x,Q^2)\right]\ln\left|\Delta\Sigma(x,Q^2)\right|$, numerically computed at $Q^2=\Lambda^2$ using equation \eqref{asym15}, as a function of Bjorken $x$. In the calculation, we used the step size of $\delta=0.1$ and maximum rapidity of $\eta_{\max}=70$. Other constants are set such that $N_c=3$, $N_f=6$ and $\alpha_s=0.35$.}
\label{fig:qkhPDFNf6}
\end{center}
\end{figure}

As for the quark helicity PDF, which we have argued in Section \ref{ssec:lowNf} to have the same small-$x$ asymptotics as the $g_1$ structure function, we repeat the numerical calculation for $\Delta\Sigma(x,Q^2)\Big|_{Q^2=\Lambda^2}$ at $N_f=6$ using the recursive discretized formula \eqref{asym15} with $\Delta\Sigma_0=0$ (see Eq.~\eqref{asym14} for the definition of $\Delta\Sigma_j$). Here, we begin with step size $\delta=0.1$ and $\eta_{\max}=70$. This results in the quark helicity PDF plotted in \fig{fig:qkhPDFNf6}, which displays similar oscillations together with the exponential growth in $\ln(1/x)$ as we saw previously for the polarized dipole amplitudes as a function of $\eta$. We apply the process described earlier in this Section to the quark helicity PDF obtaining 
\begin{align}\label{asym16}
&\Delta\Sigma(x,Q^2)\Big|_{Q^2=\Lambda^2} \sim g_1(x,Q^2)\Big|_{Q^2=\Lambda^2} \\
&\sim \left(\frac{1}{x}\right)^{\left(2.801 \pm 0.007\right)\sqrt{\frac{\alpha_sN_c}{2\pi}}}  \cos\left[\left(0.14689 \pm 0.00002\right)\sqrt{\frac{\alpha_sN_c}{2\pi}}\;\ln\frac{1}{x} + \left(2.080 \pm 0.008\right)\right] .  \notag
\end{align}
Comparing these parameters to the results in Table~\ref{tab:Nf6results} for $\delta=0.1$ and $\eta_{\max}=70$, we see that the intercept and oscillation frequency for the quark helicity PDF are similar to those for the polarized fundamental dipole amplitudes, $Q$ and $G_2$. This allows us to use the continuum-limit estimates for the intercept and the frequency, 
\begin{subequations}\label{asym16a}
\begin{align}
\alpha_h^{(6)} &= 2.83 \, \sqrt{\frac{\alpha_s N_c}{2\pi}} \,,  \label{asym16aa} \\
\omega_h^{(6)} &= 0.150 \, \sqrt{\frac{\alpha_s N_c}{2\pi}} \, ,  \label{asym16ab} 
\end{align}
\end{subequations}
respectively, to obtain the small-$x$ asymptotics of the quark helicity PDF. As for the initial phase, besides the expected offset by $\pi$ due to the overall sign flip, which in turn follows from the leading negative sign in Eq.~\eqref{asym12}, the initial phase estimate for the quark helicity PDF is slightly off from those for both $Q$ and $G_2$. This makes it inaccurate to directly deduce the initial phase for quark helicity PDF asymptotics without performing the actual calculation. However, as discussed in \cite{Kovchegov:2020hgb}, the initial phase has a strong dependence on the value of $x$ where our small-$x$ evolution begins to dominate, which makes it less important for studies of low-$x$ asymptotics.

Another observation we can make using \fig{fig:qkhPDFNf6} is that the oscillation period is large in term of $x$. Depending on the initial condition and/or the value of $x$ where the evolution begins to dominate, one should be able to observe at most one sign flip in the quark helicity PDF or the $g_1$ structure function in the kinematics of the future Electron-Ion Collider (EIC) \cite{Accardi:2012qut,Boer:2011fh,Proceedings:2020eah,AbdulKhalek:2021gbh}. In fact, Eq.~\eqref{glPDFNf6} implies the same conclusion for the gluon helicity PDF at $N_f=6$ as well. With the range of measurements at the EIC, which will not be lower than $x\sim 10^{-4}$ \cite{Accardi:2012qut,AbdulKhalek:2021gbh}, we will not be able to observe a full oscillation period in a foreseeable future. (And we have not even mentioned the fact that to reach $N_f =6$ in helicity measurements one would need to perform double spin asymmetry measurements at unprecedentedly high values of the photon virtuality $Q^2$.) Furthermore, as $x$ decreases, single-logarithmic effects start to significantly mix in, coming both from the helicity evolution\footnote{See \cite{Kovchegov:2021lvz} for a partial derivation of the small-$x$ helicity evolution at singlet-logarithmic order, without the type-2 polarized dipole amplitude.} and from the unpolarized BK/JIMWLK evolution. (The situation is further complicated by the impact of running coupling corrections, which also come in at the single-logarithmic order.)  Saturation corrections are highly likely to significantly modify all of the small-$x$ helicity asymptotics we have derived above in the linearized approximation, most likely suppressing the contributions to the proton spin coming from very low $x$. The interplay of all these phenomena needs to be better understood in order to determine if or how the oscillatory pattern we observed in this Section will exhibit itself in actual experimental measurements at small $x$.


\section{Cross-checks}
\label{sec:discuss}

\subsection{Effect of Different Initial Conditions}
\label{ssec:icresults}

In this Section we consider two different approximations to the initial conditions discussed in Sec.~\ref{ssec:icintro}. The first approximation comes from the Born-level amplitude given in Eqs.~\eqref{dip0}. With $i$ and $j$ defined in Eq.~\eqref{dipij}, the Born-level initial condition for $G_2$, Eq.~\eqref{G20}, discretizes to 
\begin{align}\label{Nf61}
G^{(0)}_{2,ij} &= - \frac{\alpha_s^2 C_F}{2N_c} \pi \sqrt{\frac{2\pi}{\alpha_sN_c}} \, i  \delta \,.
\end{align}
(For simplicity, we have put the $\theta$-function in \eq{G20} to one.)
As for the type-1 dipole amplitudes, the initial condition involves the true infrared cutoff, $\Lambda_{\text{IR}}$, such that $1/\Lambda_{\text{IR}}$ must be greater than any transverse separation encountered in the calculation. This warrants the definition of $s_{\min}$, such that 
\begin{align}\label{Nf62}
s_{\min} &= \sqrt{\frac{\alpha_s N_c}{2\pi}}\,\ln\frac{\Lambda^2}{\Lambda^2_{\text{IR}}} \, .
\end{align}
In term of $s_{\min}$, the infrared cutoff condition, $x_{10}\ll \frac{1}{\Lambda_{\text{IR}}}$, becomes $s_{10} > - s_{\min}$. Then, the discretized Born-level initial condition for the type-1 dipole amplitudes is
\begin{align}\label{Nf63}
&Q^{(0)}_{ij} = {\widetilde G}^{(0)}_{ij}= \frac{\alpha_s^2 C_F}{2N_c} \pi\sqrt{\frac{2\pi}{\alpha_sN_c}} \, \delta \left[ C_F \left(j+i_{\min}\right) - 2 \,\min\left\{j-i,\,j\right\} \right] ,
\end{align}
where we defined $i_{\min}$ such that $s_{\min} =  i_{\min}\delta$. 

The other constant (or the ``all-one") approximation to the initial conditions is 
\begin{align}\label{Nf64}
&Q^{(0)}_{ij} = {\widetilde G}^{(0)}_{ij} = G^{(0)}_{2,ij} = 1\,.
\end{align} 
Relying on the results from the previous Section that the dipole amplitudes grow exponentially in magnitude with $\eta$, one would expect that the difference between the dipole amplitudes sourced by this initial condition and by its Born-level counterpart, which grows at most linearly with $\eta$, should be negligible, reduced perhaps to the overall normalization factor at large $\eta$. This was shown to be the case at large $N_c$ in \cite{Cougoulic:2022gbk, Kovchegov:2016weo, Kovchegov:2017jxc}. 

However, at large $N_c\& N_f$, it was argued in \cite{Kovchegov:2020hgb} for the previous version of our small-$x$ helicity evolution, which did not include the type-2 dipole amplitude, that different initial conditions can lead to a significant difference in detailed behavior of the solution.\footnote{In \cite{Kovchegov:2020hgb}, the solution also takes the similar form of an exponential in $\eta$ multiplied by a sinusoidal function of $\eta$. There, different initial conditions result in the same intercept and oscillation frequency, but they lead to different initial phases for the oscillation.} It is worth noting that \cite{Kovchegov:2020hgb} compared the initial condition in Eq.~\eqref{Nf64} against the Born-level initial condition with $\Lambda$ still taken to be an IR cutoff. 

In this Section we show that the difference is unlikely to persist for the revised helicity evolution \cite{Cougoulic:2022gbk} once we select the initial conditions that correctly treat $\Lambda$ as a scale corresponding to the target's transverse size and use a different scale for the infrared cutoff. Specifically, we show numerically that the initial condition given by Eq.~\eqref{Nf64} and that given by Eqs.~\eqref{Nf61} and \eqref{Nf63} only result in small differences in the parameters of the asymptotic solutions \eqref{Nf101} and \eqref{asym2}. Furthermore, the shapes of the amplitudes are qualitatively the same.

The correct treatment of $\Lambda$ in the Born-level initial condition is of physical importance. When the target size, $1/\Lambda$, also acts as an infrared cutoff for the projectile dipole's size, $x_{10}$, the target-projectile symmetry of the (linear) evolution is explicitly broken. The target-projectile symmetry is the symmetry under the transformation $x_{10}\leftrightarrow\frac{1}{\Lambda}$ while keeping the center-of-mass energy squared $s$ fixed. Ultimately, if the asymptotic solutions had no significant dependence on the choices of initial conditions, as long as the latter respect the target-projectile symmetry and do not grow faster than a polynomial of $\eta$ and $s_{10}$, the conclusion of the target-projectile symmetry of our helicity evolution would generalize to the exact initial condition derived from the experimental results at moderate $x$. Consequently, the asymptotic results from our numerical calculation obtained with the simpler initial condition \eqref{Nf64} could be applied in the small-$x$ region for rigorous large-$N_c\& N_f$ phenomenological studies.

Another consequence of the fact that initial conditions negligibly affect the solution is that they can be linearly combined without any significant change to the results. A useful consequence of this is our freedom in choosing the fixed value of $s_{\min}$ from Eq.~\eqref{Nf62}, as any change in $s_{\min}$ has the same result as adding a multiple of initial condition,
\begin{align}\label{Nf65}
&Q^{(0)}_{ij} = {\widetilde G}^{(0)}_{ij} = 1\;\;\;\;\;\text{and}\;\;\;\;\; G^{(0)}_{2,ij} = 0 \, ,
\end{align} 
which respects the target-projectile symmetry, to the Born-level initial condition from Eqs.~\eqref{Nf61} and \eqref{Nf63}.  

To compare the two choices of initial conditions, we perform the numerical computation as described in Sec.~\ref{sec:evoleqn} at $N_f=4,6$ and $N_c=3$, using the step size $\delta = 0.1$ and maximum rapidity $\eta_{\max}=50$. First we employ the initial condition \eqref{Nf64}. The method is the same as the one described in Sec.~\ref{sec:numecalc}, resulting in the parameter estimates given in Table~\ref{tab:iconesNf4} for $N_f=4$ and Table~\ref{tab:icones} for $N_f=6$.

\begin{table}[h]
\begin{center}
\begin{tabular}{|c|c|}
\hline
\;Dipole Amplitudes\; 
& Intercept ($\alpha$)
\\ \hline 
$Q(0,\eta)$
& \;$3.28966 \pm 0.00008$\;
\\ \hline 
$G_2(0,\eta)$
& $3.28963 \pm 0.00008$
\\ \hline 
${\widetilde G}(0,\eta)$
& $3.28975 \pm 0.00008$
\\ \hline 
\end{tabular}
\caption{Summary of the parameter estimates and uncertainties for all types of polarized dipole amplitudes along the $s_{10}=0$ line. Here, the number of quark flavors and colors are taken to be $N_f=4$ and $N_c=3$, respectively. The computation is performed with step size $\delta=0.1$, maximum rapidity $\eta_{\max}=50$, and the all-one initial condition \eqref{Nf64}.}
\label{tab:iconesNf4}
\end{center}
\end{table}

\begin{table}[h]
\begin{center}
\begin{tabular}{|c|c|c|c|}
\hline
\;Dipole Amplitudes\; 
& Intercept ($\alpha$)
& Frequency ($\omega$)
& \;Initial phase ($\varphi$)\;
\\ \hline 
$Q(0,\eta)$
& \;$2.79 \pm 0.01$\;
& \;$0.146549\pm 0.000004$\;
& $-0.947 \pm 0.007$
\\ \hline 
$G_2(0,\eta)$
& $2.79 \pm 0.01$
& $0.146604\pm 0.000004$
& $-0.978 \pm 0.007$
\\ \hline 
${\widetilde G}(0,\eta)$
& $2.80 \pm 0.01$
& $0.145510\pm 0.000004$
& $0.783 \pm 0.007$
\\ \hline 
\end{tabular}
\caption{Summary of the parameter estimates and uncertainties for all types of polarized dipole amplitudes along the $s_{10}=0$ line. Here, the number of quark flavors and colors are taken to be $N_f=6$ and $N_c=3$, respectively. The computation is performed with the step size $\delta=0.1$, maximum rapidity $\eta_{\max}=50$, and the all-one initial condition \eqref{Nf64}.}
\label{tab:icones}
\end{center}
\end{table}

For the second part, we repeat the calculation with the same numbers of flavors, step size ($\delta=0.1$) and maximum rapidity ($\eta_{\max}=50$). However, this time, we employ the Born-level initial conditions given in Eqs.~\eqref{Nf61} and \eqref{Nf63}, with $s_{\min} = 50$. Recall from above that the arbitrary choice of $s_{\min}$ merely amounts to adding multiples of initial condition \eqref{Nf65} to the Born-level initial condition. Performing the parameter estimation process described in Sec.~\ref{sec:evoleqn}, we obtain the results shown in Table~\ref{tab:icbornNf4} for $N_f=4$ and Table~\ref{tab:icborn} for $N_f=6$. 

\begin{table}[h]
\begin{center}
\begin{tabular}{|c|c|}
\hline
\;Dipole Amplitudes\; 
& Intercept ($\alpha$)
\\ \hline 
$Q(0,\eta)$
& \;$3.28968 \pm 0.00008$\;
\\ \hline 
$G_2(0,\eta)$
& $3.28958 \pm 0.00008$
\\ \hline 
${\widetilde G}(0,\eta)$
& $3.28984 \pm 0.00007$
\\ \hline 
\end{tabular}
\caption{Summary of the parameter estimates and uncertainties for all types of polarized dipole amplitudes along the $s_{10}=0$ line. Here, the number of quark flavors and colors are taken to be $N_f=4$ and $N_c=3$, respectively. The computation is performed with the step size $\delta=0.1$, maximum rapidity $\eta_{\max}=50$, and the Born-level initial condition from Eqs.~\eqref{Nf61} and \eqref{Nf63}.}
\label{tab:icbornNf4}
\end{center}
\end{table}

\begin{table}[h]
\begin{center}
\begin{tabular}{|c|c|c|c|}
\hline
\;Dipole Amplitudes\; 
& Intercept ($\alpha$)
& Frequency ($\omega$)
& \;Initial phase ($\varphi$)\;
\\ \hline 
$Q(0,\eta)$
& \;$2.79 \pm 0.01$\;
& \;$0.146895\pm 0.000004$\;
& $-1.003 \pm 0.007$
\\ \hline 
$G_2(0,\eta)$
& $2.79 \pm 0.01$
& $0.146841\pm 0.000004$
& $-1.014 \pm 0.007$
\\ \hline 
${\widetilde G}(0,\eta)$
& $2.80 \pm 0.01$
& $0.145141\pm 0.000004$
& $0.753 \pm 0.007$
\\ \hline 
\end{tabular}
\caption{Summary of the parameter estimates and uncertainties for all types of polarized dipole amplitudes along the $s_{10}=0$ line. Here, the number of quark flavors and colors are taken to be $N_f=6$ and $N_c=3$, respectively. The computation is performed with the step size $\delta=0.1$, maximum rapidity, $\eta_{\max}=50$, and the Born-level initial condition from Eqs.~\eqref{Nf61} and \eqref{Nf63}.}
\label{tab:icborn}
\end{center}
\end{table}

\begin{figure} 
	\centering
     \begin{subfigure}[b]{0.32\textwidth}
         \centering
         \includegraphics[width=\textwidth]{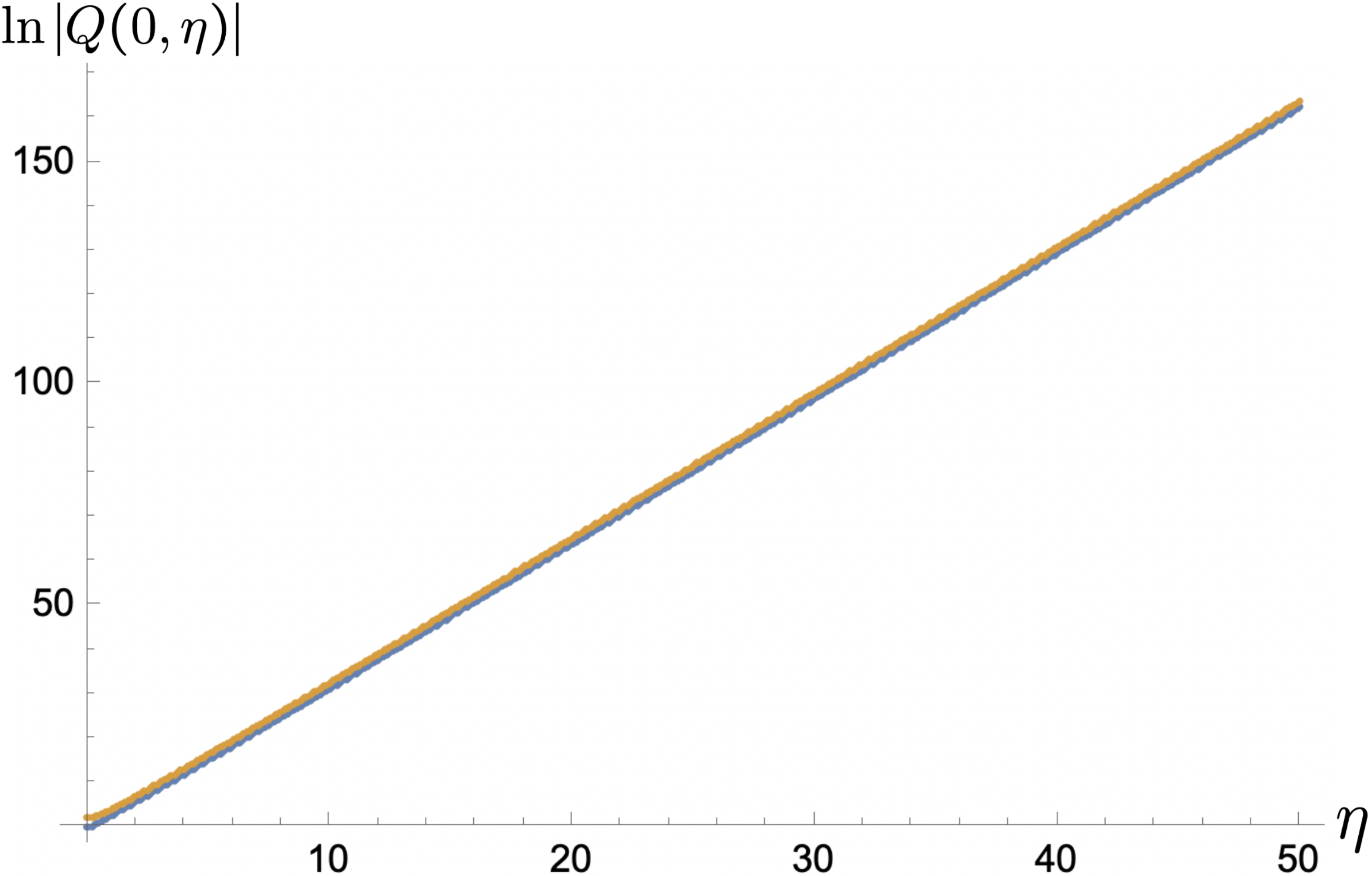}
         \caption{$\ln\left|Q(0,\eta)\right|$}
         \label{fig:ICcomparisonNf4_Q}
     \end{subfigure}  \;
     \begin{subfigure}[b]{0.32\textwidth}
         \centering
         \includegraphics[width=\textwidth]{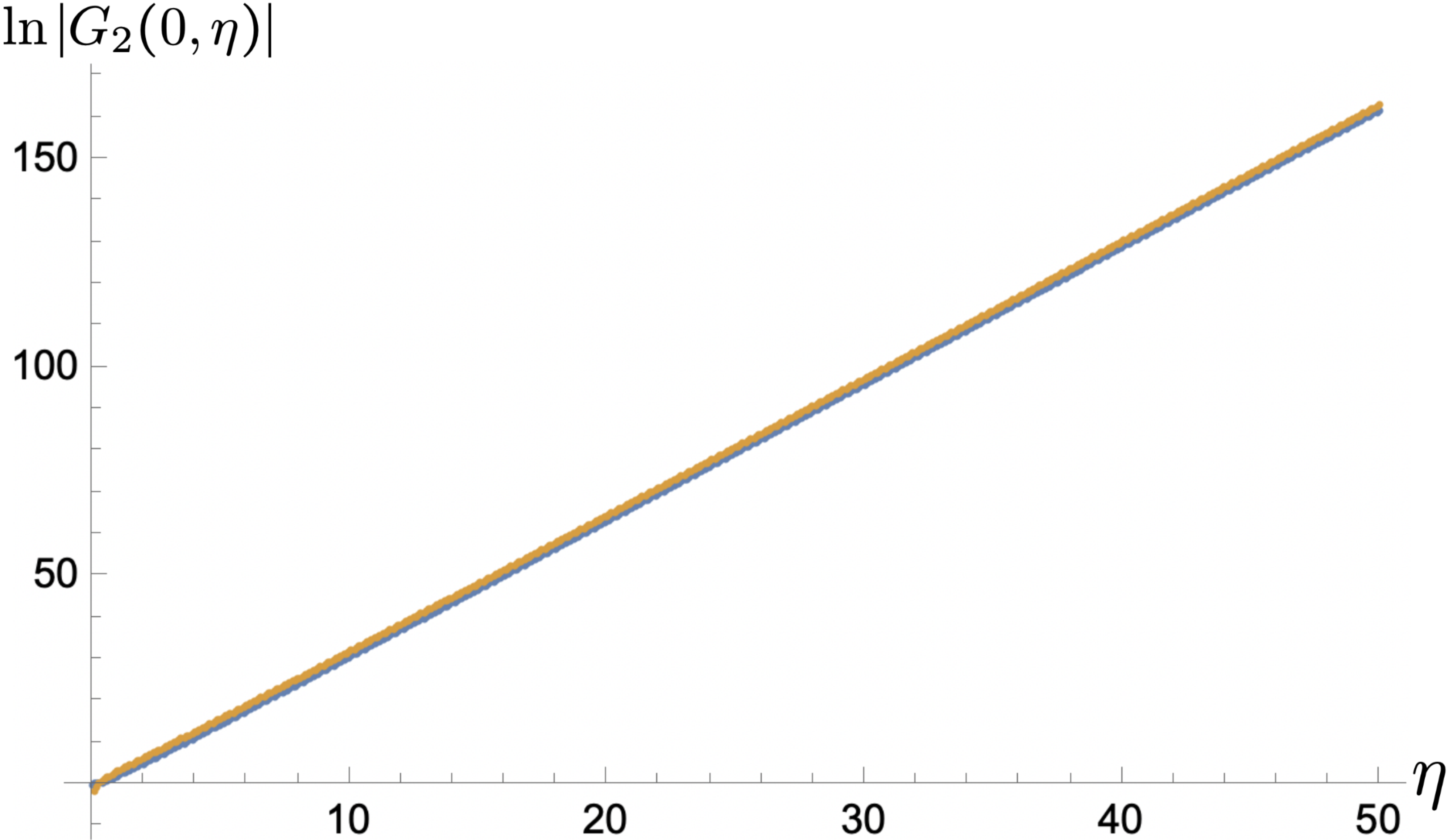}
         \caption{$\ln\left|G_2(0,\eta)\right|$}
         \label{fig:ICcomparisonNf4_G2}
     \end{subfigure}  \;
     \begin{subfigure}[b]{0.32\textwidth}
         \centering
         \includegraphics[width=\textwidth]{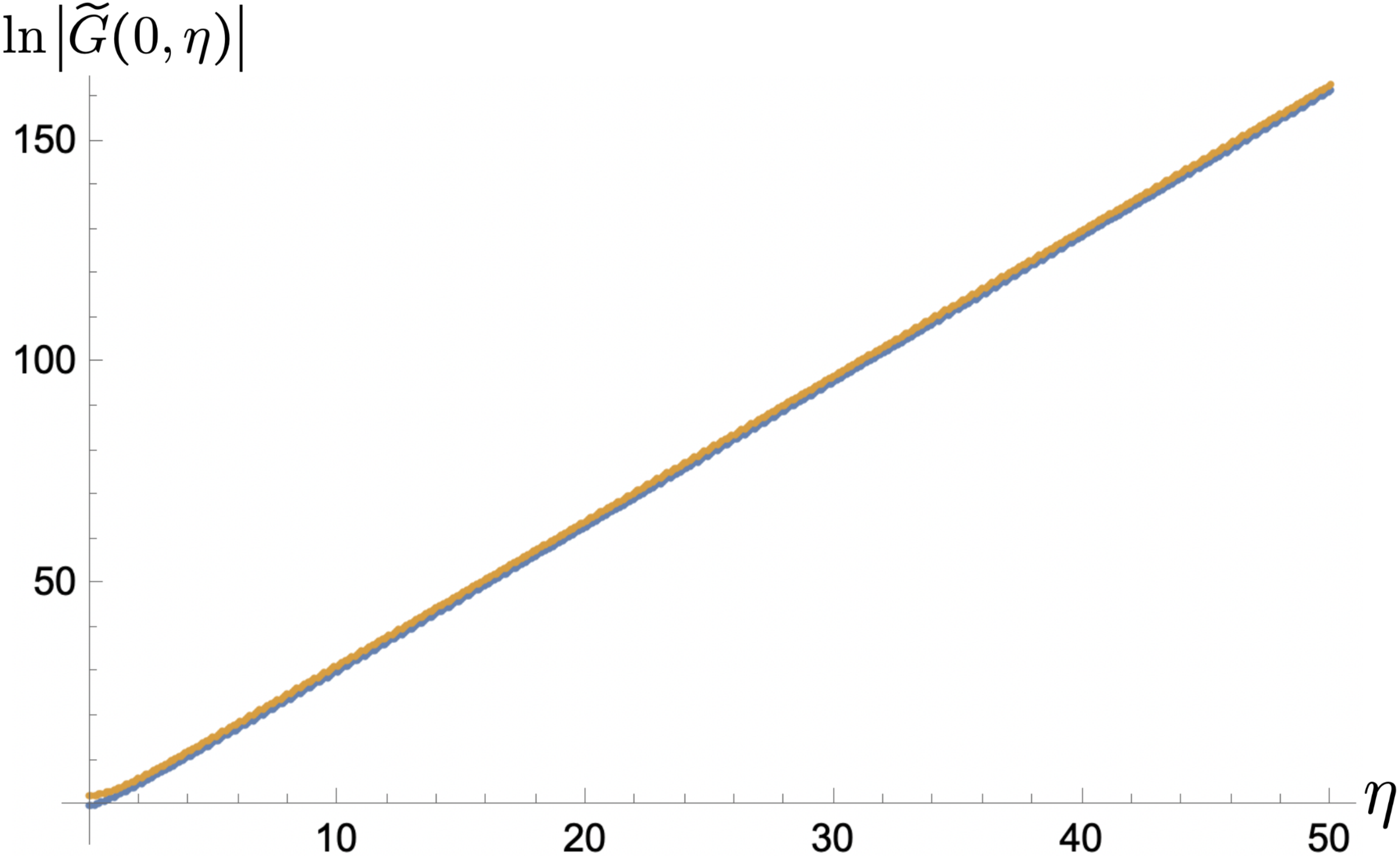}
         \caption{$\ln\left|{\widetilde G}(0,\eta)\right|$}
         \label{fig:ICcomparisonNf4_G}
     \end{subfigure}
	\caption{Plots of $\ln\left|Q(0,\eta)\right|$, $\ln\left|G_2(0,\eta)\right|$ and $\ln\left|{\widetilde G}(0,\eta)\right|$ versus $\eta$ at $N_f=4$ and $N_c =3$. All the graphs are numerically computed with step size $\delta = 0.1$ and $\eta_{\max} = 50$. In each plot, the blue dots are computed using the all-one initial condition \eqref{asym1}, while the orange dots are computed using the Born-level initial conditions \eqref{Nf61} and \eqref{Nf63}.}
\label{fig:ICcomparisonNf4}
\end{figure}

For $N_f=4$, we compare Table~\ref{tab:iconesNf4} to Table~\ref{tab:icbornNf4}. The resulting intercepts for all amplitudes are the same up to their uncertainties, implying no significant difference in parameter estimates regardless of the choice of initial conditions. This numerically justifies our choice of using the all-one initial condition \eqref{Nf64}, which we decided to use for simplicity in Sec.~\ref{sec:evoleqn}, instead of the Born-level approximation given in Eqs.~\eqref{Nf61} and \eqref{Nf63}.

\begin{figure}[h!]
	\centering
     \begin{subfigure}[b]{0.32\textwidth}
         \centering
         \includegraphics[width=\textwidth]{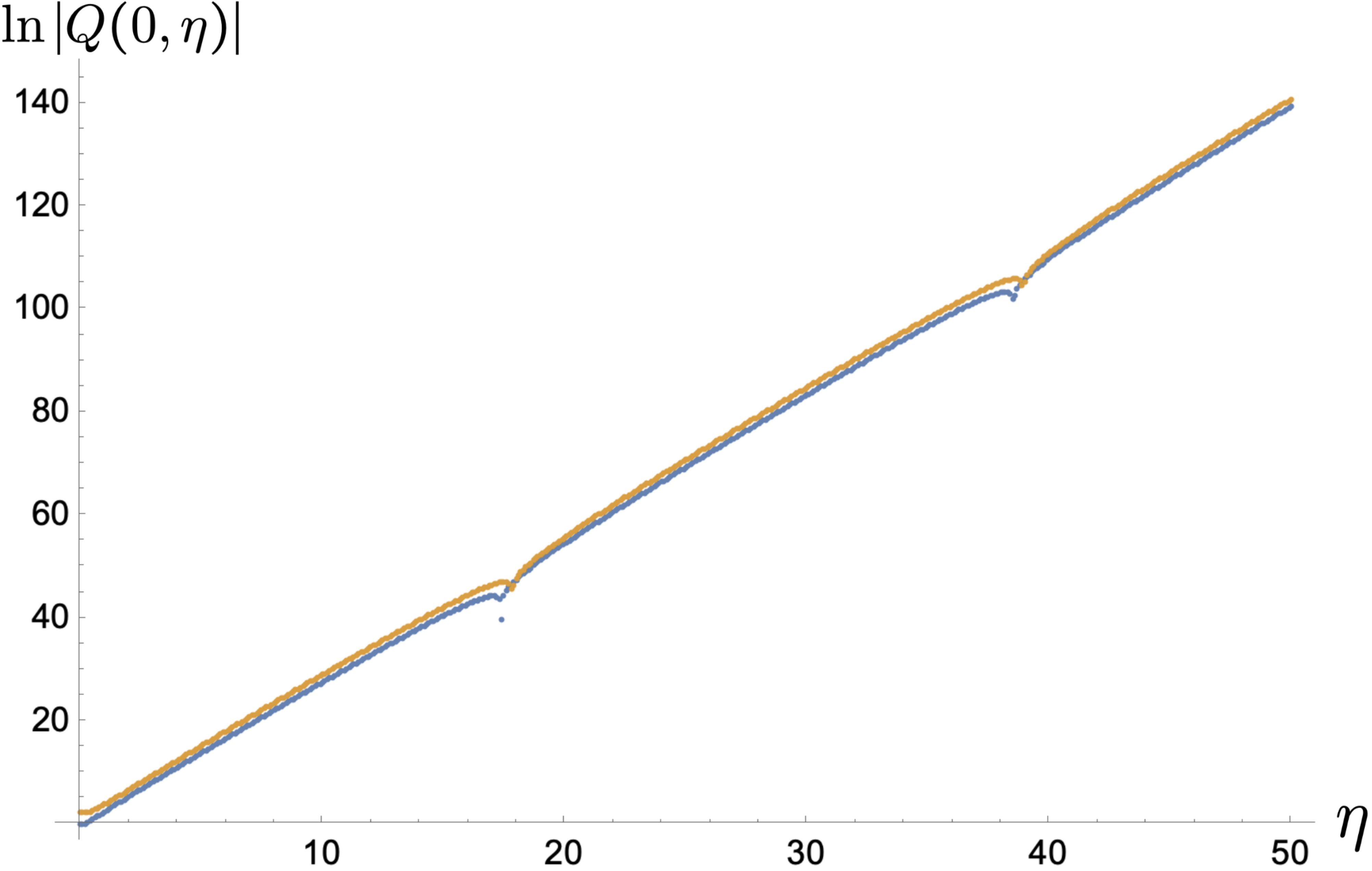}
         \caption{$\ln\left|Q(0,\eta)\right|$}
         \label{fig:ICcomparison_Q}
     \end{subfigure}  \;
     \begin{subfigure}[b]{0.32\textwidth}
         \centering
         \includegraphics[width=\textwidth]{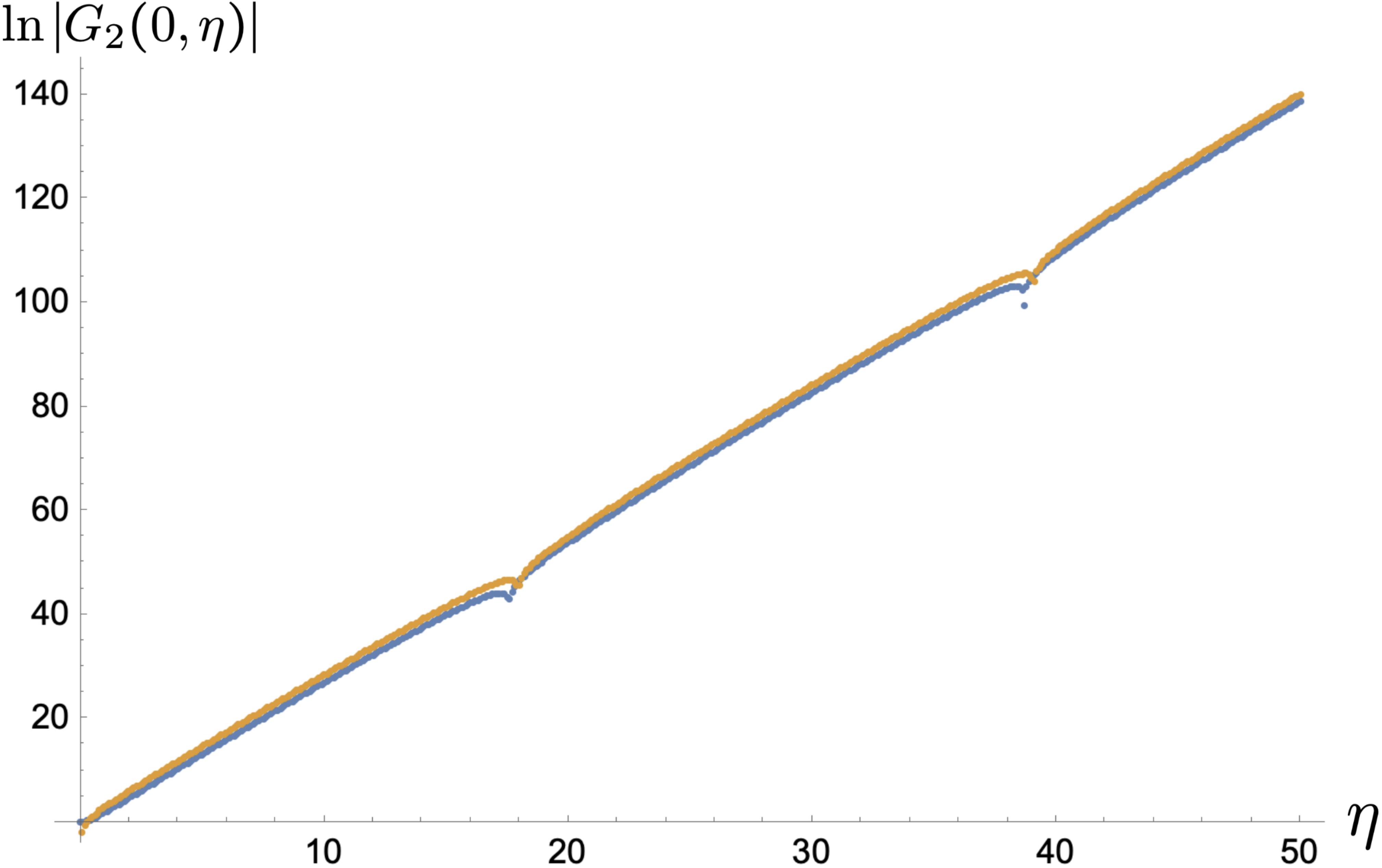}
         \caption{$\ln\left|G_2(0,\eta)\right|$}
         \label{fig:ICcomparison_G2}
     \end{subfigure}  \;
     \begin{subfigure}[b]{0.32\textwidth}
         \centering
         \includegraphics[width=\textwidth]{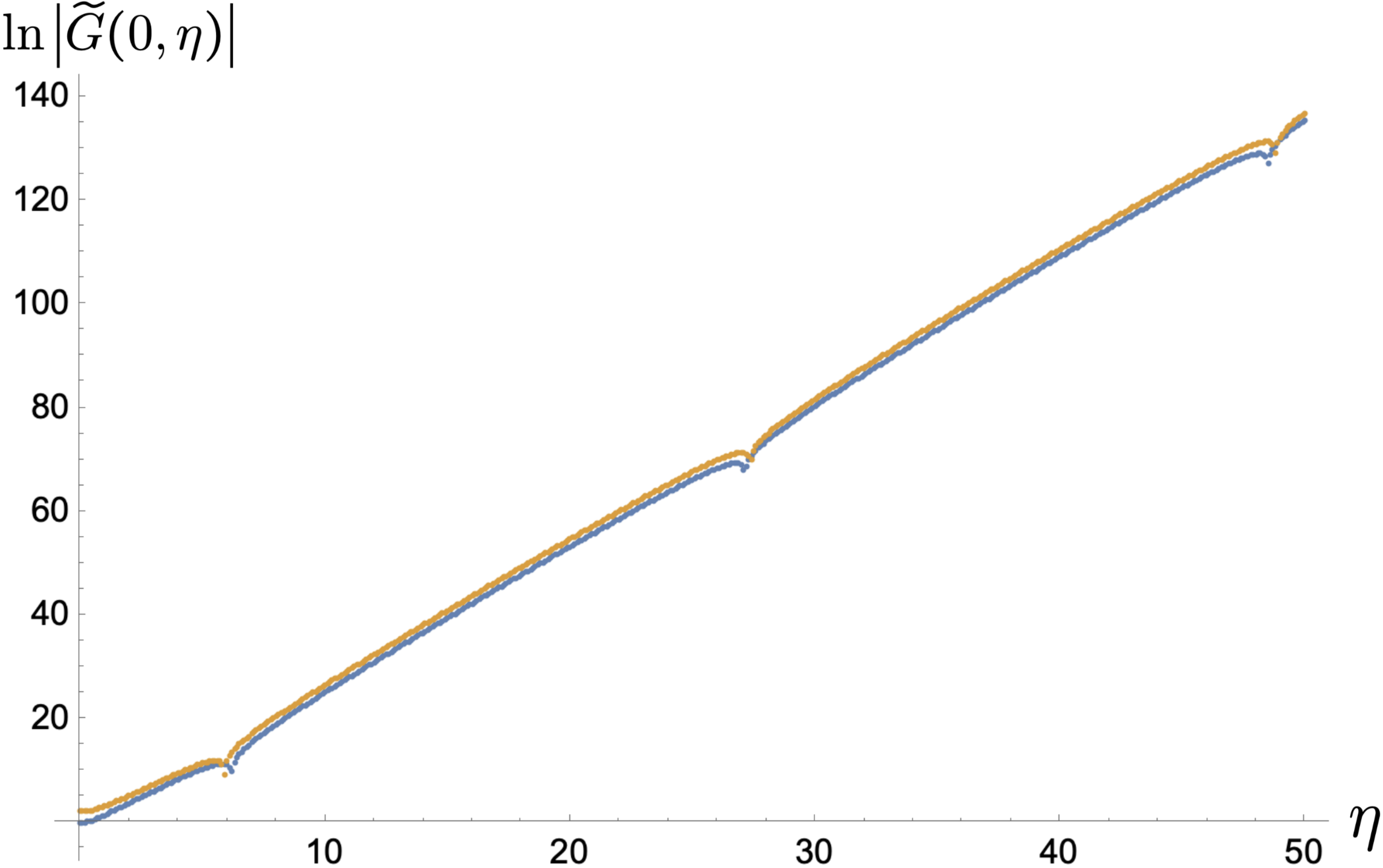}
         \caption{$\ln\left|{\widetilde G}(0,\eta)\right|$}
         \label{fig:ICcomparison_G}
     \end{subfigure}
	\caption{Plots of $\ln\left|Q(0,\eta)\right|$, $\ln\left|G_2(0,\eta)\right|$ and $\ln\left|{\widetilde G}(0,\eta)\right|$ versus $\eta$ at $N_f=6$ and $N_c =3$. All the graphs are numerically computed with step size $\delta = 0.1$ and $\eta_{\max} = 50$. In each plot, the blue dots are computed using the all-one initial condition \eqref{asym1}, while the orange dots are computed using the Born-level initial conditions \eqref{Nf61} and \eqref{Nf63}.}
\label{fig:ICcomparison}
\end{figure}

As for $N_f=6$, comparing Table~\ref{tab:icones} to Table~\ref{tab:icborn}, we see that the frequencies and the phases differ from their respective counterparts by greater amounts than the associated uncertainties. However, once we compare the differences to the continuum-limit uncertainties, c.f. Table~\ref{tab:Nf6resultsCont}, the discrepancies become insignificant for the initial phase. As for the frequency, there is still a statistically significant but very small difference.

Finally, we show in Figs.~\ref{fig:ICcomparisonNf4} (for $N_f=4$) and \ref{fig:ICcomparison} (for $N_f=6$) the plots of the logarithms of the absolute values of the dipole amplitudes along the $s_{10}=0$ line. In each plot, corresponding to the specified polarized dipole amplitude, the blue dots are made of the values at discrete steps computed using the all-one initial condition \eqref{Nf64}, while the orange dots resulted from the Born-level initial conditions \eqref{Nf61} and \eqref{Nf63}. All six plots show minimal differences in the values of the dipole amplitudes at $N_f=4$ and $N_f=6$. The only significant difference visible from the plots is on the initial phases of the oscillation at $N_f=6$ in \fig{fig:ICcomparison}, which also seem to be minor themselves. Most importantly, all qualitative features in the amplitudes are the same regardless of the choices of initial conditions. From this observation, we conclude that the all-one initial condition \eqref{asym1} is a viable simplification for all the computation in the large-$N_c\& N_f$ limit aimed at determining the small-$x$ asymptotics, assuming any possible error caused by the choice of initial condition to be negligible.\footnote{This simplifying assumption was also employed in \cite{Kovchegov:2020hgb} where the large-$N_c\& N_f$ equations without the type-2 dipole amplitude were solved numerically.}


\subsection{Target-Projectile Symmetry} 
\label{ssec:tp}

In this Section, we examine the target-projectile symmetry, which we briefly discussed in Sec.~\ref{ssec:icresults}, in the context of the Born-level initial condition. Target--projectile symmetry is indeed only possible if one treats the target and projectile on equal footing. Perhaps the cleanest process is to consider the double-spin asymmetry in the scattering of two transversely polarized virtual photons, $\gamma^* + \gamma^*$, each of which splits into a $q \bar q$ dipole: the dipoles then interact with each other in a polarization-dependent way. In \cite{Cougoulic:2022gbk} it was shown that a single virtual photon generates a dipole which interacts with the polarized target via the $Q+2G_2$ linear combination of the dipole amplitudes. This combination gives the $g_1$ structure function, which in turn relates to the cross section of the helicity-dependent DIS process. Since our goal here is to verify the target-projectile symmetry of our helicity evolution \eqref{evoleq}, we will not consider the full $\gamma^* + \gamma^*$ scattering with the corresponding Born-level initial conditions, and will instead employ our evolution with the all-one initial conditions \eqref{Nf64}, concentrating on studying the properties of the $Q+2G_2$ linear combination of dipole amplitudes under the target-projectile interchange.

Under the exchange between target and projectile, we switch $x_{10} \leftrightarrow \frac{1}{\Lambda}$, while keeping the center-of-mass energy squared, $s$, fixed. In terms of $\eta$ and $s_{10}$, c.f. Eq.~\eqref{eta_s}, this corresponds to the transformation
\begin{align}\label{TP1}
Q(s_{10},\eta) \to Q'(s_{10},\eta)\equiv Q(-s_{10},\eta-s_{10})\,, \ \ \ G_2 (s_{10},\eta) \to G'_2 (s_{10},\eta)\equiv G_2 (-s_{10},\eta-s_{10}).
\end{align}
Thus, to check for target-projectile symmetry in the asymptotic solution, we need to check whether $Q+2G_2=Q'+2G'_2$. 

\begin{figure}[ht]
	\centering
         \includegraphics[width=0.5\textwidth]{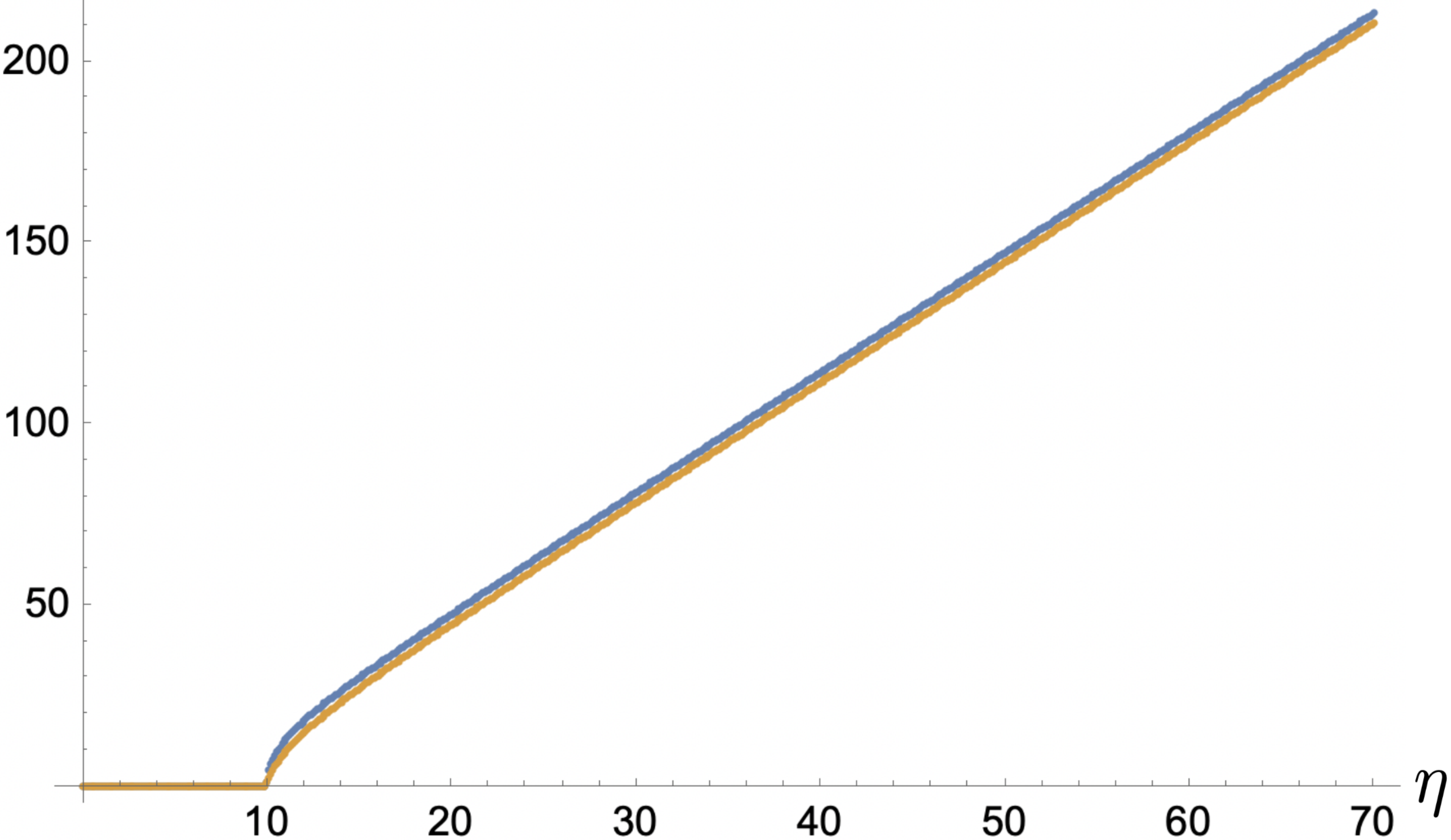}
	\caption{Plot of $\ln\left|Q(10,\eta) + 2G_2(10,\eta)\right|$ (blue) and $\ln\left|Q'(10,\eta) + 2G'_2(10,\eta)\right|$ (orange) versus $\eta$ at $N_f=4$, $N_c =3$ and $s_{10}=10$. Both curves are numerically computed using the all-one initial condition with step size $\delta = 0.1$ and $\eta_{\max} = 70$. }
\label{fig:tpNf4}
\end{figure}

We start with the qualitative check through plots. First, notice that the dipole amplitudes and their primed counterparts are trivially equal along the $s_{10}=0$ line, since $s_{10}=0$ implies that $x_{10} = 1/\Lambda$. As a result, we need to examine the amplitudes at $s_{10} \neq 0$ in order to check for the target-projectile symmetry. In particular, we plot the logarithms of the absolute values of the dipole amplitudes along the $s_{10}=10$ line. For the $N_f\leq 5$ case, we consider $N_f=4$, which is qualitatively the same as the cases where $N_f=2$, 3 or 5. The results are plotted in \fig{fig:tpNf4}, where the blue dots describe the original amplitudes, while the orange dots describe the primed amplitudes with the target and the projectile interchanged.

In \fig{fig:tpNf4}, we see that the curves are mostly parallel, implying that intercepts appear unchanged under the target-projectile exchange. However, the lines seem to shift slightly downward after the exchange, implying that $\frac{Q+2G_2}{Q'+2G'_2}$ approaches 1 for sufficiently large $\eta$, roughly for $\eta\gtrsim 20$. Hence, from the plot, the linear combination of primed and unprimed amplitudes seem to have the same leading asymptotic behavior.

To make the comparison more quantitative, we employ the same method as in Sec.~\ref{ssec:lowNf} to estimate the intercept at large $\eta$ along the $s_{10}=10$ line. Repeating the process for the values of the step size $\delta$ and maximum rapidity $\eta_{\max}$ listed in Table~\ref{tab:M_delta_Nf234}, we obtain the intercepts for each amplitude and each $N_f$ in the continuum limit ($\delta\to 0$ and $\eta_{\max}\to\infty$) listed in Table~\ref{tab:tpNf4}. Similar to Sec.~\ref{ssec:lowNf}, the quadratic model fits the best with the intercept results. In Table~\ref{tab:tpNf4} the uncertainty accounts for the residue from the quadratic model. It is slightly higher than its counterpart in Sec.~\ref{ssec:lowNf} because we have fewer data points at $s_{10}=10$ than at $s_{10}=0$, as our helicity evolution only takes place at $\eta\geq s_{10}$. For each $N_f$, we see from Table~\ref{tab:tpNf4} that the intercepts for the primed and unprimed amplitudes are the same within the uncertainty. This implies that $Q+2G_2$ respects the target-projectile symmetry in their large-$\eta$ asymptotics for the $N_f\leq 5$ cases where there is no oscillation.

\begin{table}[h]
\begin{center}
\begin{tabular}{|c|c|c|}
\hline
\;$N_f$\;
& \;$Q(10,\eta)+2G_2(10,\eta)$\;
& \;$Q'(10,\eta)+2G'_2(10,\eta)$\;
\\ \hline 
$2$
& $3.52\pm 0.02$
& $3.52\pm 0.03$
\\ \hline 
$3$
& $3.42\pm 0.02$
& $3.43\pm 0.02$
\\ \hline 
$4$
& $3.32\pm 0.01$
& $3.33\pm 0.01$
\\ \hline 
\end{tabular}
\caption{Summary of the estimates and uncertainties of the intercepts, $\alpha$, for $Q+2G_2$ and $Q'+2G'_2$ along the $s_{10}=10$ line. Here, the number of quark colors is taken to be $N_c=3$. The computation is performed with the all-one initial condition \eqref{Nf64}.}
\label{tab:tpNf4}
\end{center}
\end{table}

\begin{figure} 
         \centering
         \includegraphics[width=0.5\textwidth]{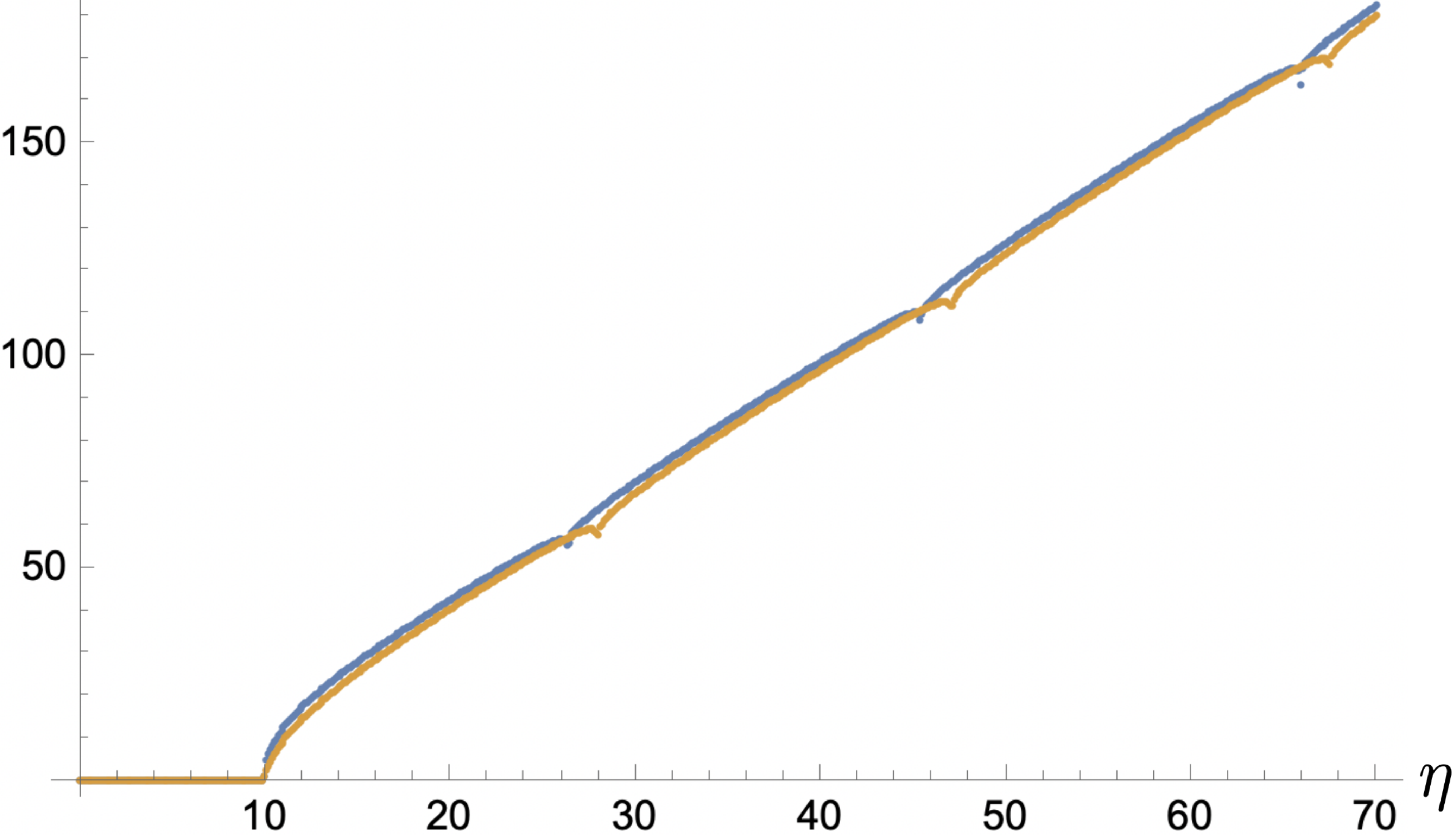}
	\caption{Plot of $\ln\left|Q(10,\eta) + 2G_2(10,\eta)\right|$ (blue) and $\ln\left|Q'(10,\eta) + 2G'_2(10,\eta)\right|$ (orange) versus $\eta$ at $N_f=6$, $N_c =3$ and $s_{10}=10$. Both curves are numerically computed using the all-one initial condition with step size $\delta = 0.1$ and $\eta_{\max} = 70$. }
\label{fig:tp}
\end{figure}

Now, we move on to consider the case where $N_f=6$, for which the plot for $\ln\left|Q(10,\eta) + 2G_2(10,\eta)\right|$ and its primed counterparts is shown in \fig{fig:tp}. Qualitatively, the only parameter that may significantly violate target-projectile symmetry is the initial phase. 

To see the potential violation more clearly, we compute the parameter estimates for the large-$\eta$ asymptotics of the amplitudes and their primed counterparts along the $s_{10}=10$ line. Repeating the computation and the parameter evaluation steps outlined in Sec.~\ref{ssec:highNf} for several values of step size $\delta$ and maximum rapidity $\eta_{\max}$, we deduce the continuum-limit estimates, $\delta = 1/\eta_{\max}=0$, through the weighted polynomial regression method described in Sec.~\ref{ssec:highNf}. At the end, we obtain the continuum-limit parameter estimates listed in Table~\ref{tab:tp}. Surprisingly, the initial phase discrepancies are within the uncertainties, but the frequencies do have significant discrepancies. However, the difference itself is only within $0.15\%$. Altogether, we conclude that $Q+2G_2$, which is the object that yields the $g_1$ structure function, respects the target-projectile symmetry for any general $N_f$. 

\begin{table}[h]
\begin{center}
\begin{tabular}{|c|c|c|c|}
\hline
Dipole Amplitudes 
& Intercept ($\alpha$)
& Frequency ($\omega$)
& \;Initial phase ($\varphi$)\;
\\ \hline 
$Q(10,\eta)+2G_2(10,\eta)$
& $2.81 \pm 0.04$
& $0.16146\pm 0.00008$
& $-1.62 \pm 0.07$
\\   \hline
\;$Q'(10,\eta)+2G'_2(10,\eta)$\;
& \;$2.82 \pm 0.04$\;
& \;$0.16169\pm 0.00008$\;
& $-1.55 \pm 0.09$
\\ \hline 
\end{tabular}
\caption{Summary of the parameter estimates and uncertainties at the continuum limit ($\delta\to 0$ and $\eta_{\max}\to\infty$) for $Q(10,\eta) + 2G_2(10,\eta)$ and $Q'(10,\eta) + 2G'_2(10,\eta)$ along the $s_{10}=10$ line. Here, the number of quark flavors and colors are taken to be $N_f=6$ and $N_c=3$, respectively. The computation is performed with the all-one initial condition \eqref{Nf64}.}
\label{tab:tp}
\end{center}
\end{table}


\section{Comparison with the Polarized DGLAP Evolution}
\label{sec:DGLAP}

Similar to what was done in \cite{Cougoulic:2022gbk} for the large-$N_c$ version of our helicity evolution, let us now attempt an iterative cross-check of the large-$N_c \& N_f$ version of helicity evolution. While it would be better to solve Eqs.~\eqref{evoleq} analytically (cf. \cite{Borden:2023ugd} for the large-$N_c$ limit), in absence of such a solution at present we will solve Eqs.~\eqref{evoleq} iteratively and compare the results to the finite-order calculations based on the Dokshitzer-Gribov-Lipatov-Altarelli-Parisi (DGLAP) evolution equations \cite{Gribov:1972ri,Altarelli:1977zs,Dokshitzer:1977sg}.

The polarized DGLAP evolution equation \cite{Gribov:1972ri,Altarelli:1977zs,Dokshitzer:1977sg} is the renormalization group equation for the quark and gluon helicity PDFs. Physically, each step of the evolution corresponds to an emission of a daughter parton in such a way that the transverse momenta of the partons are strongly ordered. As a result, the small-$x$ limit of DGLAP equations should be contained in the KPS-CTT evolution, particularly in the terms driven by the type-2 polarized dipole amplitude, $G_2(x^2_{10},zs)$. Hence, it is useful to further cross-check the large-$N_c \& N_f$ KPS-CTT evolution by investigating whether or not its iteration reproduces the small-$x$ limit of the polarized DGLAP kernels. There are several caveats in such a comparison, which we will outline below. 

Throughout this Section, we re-introduce the role of $\Lambda$ as the infrared cutoff for the transverse dipole size in the KPS-CTT evolution, in order to be in line with how the polarized DGLAP evolution is set up. Treating $\Lambda$ as the IR cutoff was important in the comparison of the iterative solution of the large-$N_c$ helicity evolution to polarized DGLAP equation in \cite{Cougoulic:2022gbk}. This results in the following large-$N_c\& N_f$ evolution equations with the IR cutoff $\Lambda$:
\begin{subequations}\label{eq_LargeNcNf}
\begin{align}
& Q(x^2_{10},zs) = Q^{(0)}(x^2_{10},zs) + \frac{\alpha_sN_c}{2\pi} \int_{1/s x^2_{10}}^{z} \frac{dz'}{z'}   \int_{1/z's}^{x^2_{10}}  \frac{dx^2_{21}}{x_{21}^2}    \left[ 2 \, {\widetilde G}(x^2_{21},z's) + 2 \, {\widetilde \Gamma}(x^2_{10},x^2_{21},z's) \right. \label{eq_LargeNcNfa} \\
&\hspace*{5cm}\left.+ \; Q(x^2_{21},z's) -  \overline{\Gamma}(x^2_{10},x^2_{21},z's) + 2 \, \Gamma_2(x^2_{10},x^2_{21},z's) + 2 \, G_2(x^2_{21},z's)   \right] \notag \\
&\hspace*{3cm}+ \frac{\alpha_sN_c}{4\pi} \int_{\Lambda^2/s}^{z} \frac{dz'}{z'}   \int_{1/z's}^{\min \{ x^2_{10}z/z', 1/\Lambda^2 \}}  \frac{dx^2_{21}}{x_{21}^2} \left[Q(x^2_{21},z's) + 2 \, G_2(x^2_{21},z's) \right] ,  \notag  \\
&\overline{\Gamma}(x^2_{10},x^2_{21},z's) = Q^{(0)}(x^2_{10},z's) + \frac{\alpha_sN_c}{2\pi} \int_{1/s x^2_{10}}^{z'} \frac{dz''}{z''}   \int_{1/z''s}^{\min\{x^2_{10}, x^2_{21}z'/z''\}}  \frac{dx^2_{32}}{x_{32}^2}    \left[ 2\, {\widetilde G} (x^2_{32},z''s)  \right. \label{eq_LargeNcNfb} \\
&\hspace*{2.5cm}\left.+ \; 2\, {\widetilde \Gamma} (x^2_{10},x^2_{32},z''s) +  Q(x^2_{32},z''s) -  \overline{\Gamma}(x^2_{10},x^2_{32},z''s) + 2 \, \Gamma_2(x^2_{10},x^2_{32},z''s) + 2 \, G_2(x^2_{32},z''s) \right] \notag \\
&\hspace*{3cm}+ \frac{\alpha_sN_c}{4\pi} \int_{\Lambda^2/s}^{z'} \frac{dz''}{z''}   \int_{1/z''s}^{\min \{ x^2_{21}z'/z'', 1/\Lambda^2 \}}  \frac{dx^2_{32}}{x_{32}^2} \left[Q(x^2_{32},z''s) + 2 \, G_2(x^2_{32},z''s) \right] , \notag \\
& {\widetilde G}(x^2_{10},zs) = {\widetilde G}^{(0)}(x^2_{10},zs) + \frac{\alpha_s N_c}{2\pi}\int_{1/s x^2_{10}}^z\frac{dz'}{z'}\int_{1/z's}^{x^2_{10}} \frac{dx^2_{21}}{x^2_{21}} \left[3 \, {\widetilde G}(x^2_{21},z's) + {\widetilde \Gamma}(x^2_{10},x^2_{21},z's) \right. \label{eq_LargeNcNfc} \\
&\hspace*{4cm}\left.  + \; 2\,G_2(x^2_{21},z's)  +  \left(2 - \frac{N_f}{2N_c}\right) \Gamma_2(x^2_{10},x^2_{21},z's) - \frac{N_f}{4N_c}\,\overline{\Gamma}(x^2_{10},x^2_{21},z's)\right] \notag \\
&\hspace*{3cm}- \frac{\alpha_sN_f}{8\pi}  \int_{\Lambda^2/s}^z \frac{dz'}{z'}\int_{\max\{x^2_{10},\,1/z's\}}^{\min \{ x^2_{10}z/z', 1/\Lambda^2 \}} \frac{dx^2_{21}}{x^2_{21}}  \left[   Q(x^2_{21},z's) +     2 \, G_2(x^2_{21},z's)  \right] , \notag \\
& {\widetilde \Gamma} (x^2_{10},x^2_{21},z's) = {\widetilde G}^{(0)}(x^2_{10},z's) + \frac{\alpha_s N_c}{2\pi}\int_{1/s x^2_{10}}^{z'}\frac{dz''}{z''}\int_{1/z''s}^{\min\{x^2_{10},x^2_{21}z'/z''\}} \frac{dx^2_{32}}{x^2_{32}} \left[3 \, {\widetilde G} (x^2_{32},z''s) \right. \label{eq_LargeNcNfd} \\
&\hspace*{2.5cm}\left. + \; {\widetilde \Gamma}(x^2_{10},x^2_{32},z''s) + 2 \, G_2(x^2_{32},z''s)  +  \left(2 - \frac{N_f}{2N_c}\right) \Gamma_2(x^2_{10},x^2_{32},z''s) - \frac{N_f}{4N_c} \,\overline{\Gamma}(x^2_{10},x^2_{32},z''s)  \right] \notag \\
&\hspace*{3cm}- \frac{\alpha_sN_f}{8\pi}  \int_{\Lambda^2/s}^{z'x^2_{21}/x^2_{10}} \frac{dz''}{z''}\int_{\max\{x^2_{10},\,1/z''s\}}^{\min \{ x^2_{21}z'/z'', 1/\Lambda^2 \} } \frac{dx^2_{32}}{x^2_{32}}  \left[   Q(x^2_{32},z''s) +  2  \,  G_2(x^2_{32},z''s)  \right] , \notag \\
& G_2(x_{10}^2, z s)  =  G_2^{(0)} (x_{10}^2, z s) + \frac{\as N_c}{\pi} \, \int\limits_{\frac{\Lambda^2}{s}}^z \frac{d z'}{z'} \, \int\limits_{\max \left[ x_{10}^2 , \frac{1}{z' s} \right]}^{\min \{\frac{z}{z'} x_{10}^2, 1/\Lambda^2 \}} \frac{d x^2_{21}}{x_{21}^2} \left[ {\widetilde G} (x^2_{21} , z' s) + 2 \, G_2 (x_{21}^2, z' s)  \right] , \label{eq_LargeNcNfe} \\
& \Gamma_2 (x_{10}^2, x_{21}^2, z' s)  =  G_2^{(0)} (x_{10}^2, z' s) + \frac{\as N_c}{\pi}  \int\limits_{\frac{\Lambda^2}{s}}^{z' \frac{x_{21}^2}{x_{10}^2}} \frac{d z''}{z''}  \int\limits_{\max \left[ x_{10}^2 , \frac{1}{z'' s} \right]}^{\min \{ \frac{z'}{z''} x_{21}^2, 1/\Lambda^2 \}} \frac{d x^2_{32}}{x_{32}^2} \left[ {\widetilde G} (x^2_{32} , z'' s) + 2 \, G_2(x_{32}^2, z'' s)  \right] .  \label{eq_LargeNcNff}
\end{align}
\end{subequations}


\subsection{The setup}\label{sec:DGLAP_setup}

The polarized DGLAP equation written in the integral form reads
\begin{align}\label{hDGLAP}
    \begin{pmatrix} 
\Delta\Sigma\left(x, Q^2\right) \\
\Delta G\left(x, Q^2\right)
\end{pmatrix} 
&= \begin{pmatrix} 
\Delta\Sigma\left(x, \Lambda^2\right) \\
\Delta G\left(x, \Lambda^2\right)
\end{pmatrix}  + \int\limits_{\Lambda^2}^{Q^2} \frac{d\mu^2}{\mu^2} \int\limits_x^1 \frac{dz}{z}
\begin{pmatrix} 
\Delta P_{qq}(z) & \Delta P_{qG}(z) \\
\Delta P_{Gq}(z) & \Delta P_{GG}(z)
\end{pmatrix} 
\begin{pmatrix} 
\Delta\Sigma\left(\frac{x}{z}, \mu^2\right) \\
\Delta G\left(\frac{x}{z}, \mu^2\right)
\end{pmatrix} 
,
\end{align}
where the polarized splitting functions, $\Delta P_{ij}(z)$, depend on the longitudinal momentum fraction $z$. Generally, the splitting functions depend on the renormalization scheme. In the $\overline{\text{MS}}$ scheme, their small-$x$ limits at large-$N_c \& N_f$ are \cite{Altarelli:1977zs,Dokshitzer:1977sg,Mertig:1995ny,Moch:2014sna},
\begin{subequations}\label{Pfuncs}
\begin{align}
&\Delta \tilde{P}_{qq}(x) = \left(\frac{\alpha_s}{4\pi}\right)N_c + \left(\frac{\alpha_s}{4\pi}\right)^2\frac{N_c}{2}(N_c-4N_f)\ln^2\frac{1}{x} + \left(\frac{\alpha_s}{4\pi}\right)^3\frac{N_c^2}{12}(N_c-20N_f)\ln^4\frac{1}{x} + {\cal O} (\alpha_s^4) \,,     \label{Pqq} \\
&\Delta \tilde{P}_{qG}(x) =  - \left(\frac{\alpha_s}{4\pi}\right)2N_f - \left(\frac{\alpha_s}{4\pi}\right)^2 5N_cN_f\ln^2\frac{1}{x} - \left(\frac{\alpha_s}{4\pi}\right)^3\frac{N_cN_f}{3}(17N_c-3N_f)\ln^4\frac{1}{x} + {\cal O} (\alpha_s^4) \,,     \label{PqG} \\
&\Delta \tilde{P}_{Gq}(x) =   \left(\frac{\alpha_s}{4\pi}\right)2N_c + \left(\frac{\alpha_s}{4\pi}\right)^2 5N_c^2\ln^2\frac{1}{x} + \left(\frac{\alpha_s}{4\pi}\right)^3\frac{2}{3}N_c^2(9N_c-N_f)\ln^4\frac{1}{x} + {\cal O} (\alpha_s^4)  \,,   \label{PGq} \\
&\Delta \tilde{P}_{GG}(x) =    \left(\frac{\alpha_s}{4\pi}\right)8N_c + \left(\frac{\alpha_s}{4\pi}\right)^2 2N_c(8N_c-N_f)\ln^2\frac{1}{x} + \left(\frac{\alpha_s}{4\pi}\right)^3\frac{N_c^2}{3}(56N_c-11N_f)\ln^4\frac{1}{x} + {\cal O} (\alpha_s^4) \,.   \label{PGG} 
\end{align}
\end{subequations}
From now on, we use tildes to denote the polarized splitting functions and the helicity PDFs ($\Delta {\tilde \Sigma}$, $\Delta {\tilde G}$) in the $\overline{\text{MS}}$ scheme.

In general, the $g_1$ structure function can be written as a linear combination of the quark and gluon helicity PDFs convoluted (over $x$) with the coefficient functions. While both the hPDFs and the coefficient functions are renormalization scheme-dependent, the $g_1$ structure function is indeed a physical observable and is independent of a scheme. In our small-$x$ formalism, defining $\Delta q_f^+ = \Delta q_f + \Delta {\bar q}_f$ \cite{Ethier:2017zbq,Adamiak:2021ppq} with $\Delta q_f$ and $\Delta {\bar q}_f$ the quark and anti-quark helicity distributions for each quark flavor $f$, we can write \cite{Cougoulic:2022gbk} (also cf. Eqs.~\eqref{g1} and \eqref{qkhPDF} above)
\begin{subequations}
\begin{align}
& g_1 (x, Q^2)  = \frac{1}{2} \sum_f \, Z^2_f \, \Delta q_f^+ (x, Q^2), \label{g1_scheme} \\
& \Delta\Sigma(x,Q^2) = \sum_{f} \Delta q_f^+ (x, Q^2), 
\end{align}
\end{subequations}
with
\begin{align}\label{q+}
\Delta q_f^+ (x, Q^2) = - \frac{N_c}{2 \pi^3} \:  \int\limits_{\Lambda^2/s}^1 \frac{d z}{z} \,  \int\limits_{\frac{1}{zs}}^{\min \left\{ \frac{1}{z Q^2} , \frac{1}{\Lambda^2} \right\}} \frac{d x^{2}_{10}}{x_{10}^2}  \, \left[  Q (x^2_{10} , zs) + 2 \, G_2 (x^2_{10} , zs) \right].
\end{align}
These relations are valid to all orders in $\as \, \ln^2 (1/x)$ and $\as \, \ln (1/x) \ln (Q^2 / \Lambda^2)$. Note the particularly simple relation \eqref{g1_scheme} between the $g_1$ structure function and quark hPDFs. This is similar to the DIS scheme \cite{Altarelli:1979ub}, in which the $F_2(x,Q^2)$ structure function is also related to the unpolarized quark PDFs, $q_f^+ = q_f +{\bar q}_f$, to all orders in $\as$, via the simple leading-order (LO) relation, $F_2 (x, Q^2 ) = \sum_f Z_f^2 \, x \, q_f^+ (x,Q^2)$. Therefore, our calculation, preserving the LO relation \eqref{g1_scheme} to all orders in $\as$, can be thought of as being in the ``polarized DIS scheme" or ``pDIS scheme". Below, the helicity PDFs and polarized splitting functions in our pDIS scheme will be written without the tilde, in contrast to their counterparts in the $\overline{\text{MS}}$ scheme. 

If we choose flavor-independent initial conditions for the evolution equations \eqref{evoleq}, we obtain a simple relation between the $g_1$ structure function and $\Delta \Sigma$,
\begin{align}\label{g1Sigma}
    g_1 (x, Q^2)  = \frac{1}{2 \, N_f} \left( \sum_f \, Z^2_f \right) \, \Delta\Sigma(x,Q^2). 
\end{align}
We conclude that, when comparing to the fixed-order calculations, our $\Delta \Sigma$ should be compared to the results for $g_1$ structure function, while taking the proportionality factor between the two from \eq{g1Sigma} into account. 

The factorized structure function $g_1$ can be written as the convolution of the coefficient functions and hPDFs (see the discussion in \cite{Davies:2022ofz}).
We, therefore, write the quark hPDF in the pDIS scheme at next-to-leading order (NLO) as
\begin{align}\label{Sigma_tildeSigmaG}
\Delta \Sigma (x, Q^2) = \Delta {\tilde \Sigma} (x, Q^2) + \int\limits_x^1 \frac{dz}{z} \, \left[ \Delta c_q (z) \,  \Delta {\tilde \Sigma} \left( \frac{x}{z} , Q^2\right) +  \Delta c_G (z) \,  \Delta {\tilde G} \left( \frac{x}{z} , Q^2\right) \right],
\end{align}
in terms of the quark and gluon hPDFs $\Delta {\tilde \Sigma}$ and $\Delta {\tilde G}$ in the $\overline{\text{MS}}$ scheme. Here the small-$x$ large-$N_c \& N_f$ coefficient functions are \cite{Zijlstra:1993sh}
\begin{subequations}\label{coef_functions}
\begin{align}
& \Delta c_q (z) = \frac{\as N_c}{4 \pi} \, \ln \frac{1}{z} + \frac{5}{12} \, \left( \frac{\as N_c}{4 \pi} \right)^2 \left[ 1 - 4 \, \frac{N_f}{N_c} \right] \, \ln^3 \frac{1}{z} + {\cal O} (\as^3), \label{Cq_full}  \\
& \Delta c_G (z) = - \frac{\as \, N_f}{2 \pi} \, \ln \frac{1}{z} - \frac{11}{2} \, \left( \frac{\as }{4 \pi} \right)^2 \, N_c \, N_f \, \ln^3 \frac{1}{z} + {\cal O} (\as^3). \label{dCg}
\end{align}
\end{subequations}

However, the order-by-order in $\as$ solution of the small-$x$ helicity evolution equation we are about to perform should be compared to the unfactorized partonic structure function ${\hat g}_1$. That means, instead of \eq{Sigma_tildeSigmaG} we should write \cite{Zijlstra:1993sh,Davies:2022ofz}
\begin{align}
    \Delta {\hat \Sigma}_j (x, \epsilon ) = \sum_{i=q,G} \left[ \Delta C_i (\epsilon) \otimes Z_{ij} (\epsilon) \right] (x)
\end{align}
for the quark hPDF sourced by the parton $j= q,G$. Here the transition matrix is $Z_{ij} = 1 + \tfrac{1}{\epsilon} \, \Delta {\tilde P}^{(1)}_{ij} (z) +\ldots$ with $\Delta {\tilde P}^{(1)}_{ij} (z)$ the order-$\as$ contribution to the splitting functions from Eqs.~\eqref{Pfuncs}. The dimensional regularization parameter is $\epsilon = d-4$ with $d$ the number of space-time dimensions. We defined the $\otimes$ operation such that
\begin{align}\label{CrossNotation}
&\left[ f \otimes g \right](x,Q^2) = \int_x^1\frac{dz}{z} \left[f(z)\right]\left[g\left(\frac{x}{z}\right)\right] ,
\end{align}
for any pair of functions $f$ and $g$. 

The coefficient functions now are (see, e.g., \cite{Vermaseren:2005qc})
\begin{align}\label{coef_functions_e}
    \Delta C_i (z, \epsilon) = \delta_{iq} \, \delta (1-z) + \sum_{l=1}^\infty 
    \left[ \Delta c_i^{(l)} (z) + \epsilon \, \Delta a_i^{(l)} (z) + \epsilon^2 \, \Delta b_i^{(l)} (z) + \ldots \right] ,
\end{align}
with the coefficient functions in \eq{coef_functions} given by the $\epsilon \to 0$ limit of \eq{coef_functions_e},
\begin{align}
    \Delta c_i (z) = \delta_{iq} \, \delta (1-z) + \sum_{l=1}^\infty 
    \Delta c_i^{(l)} (z) .
\end{align}
Here the superscript $l$ denotes the order of $\as$ in each term. 

The coefficients $\Delta a_i^{(l)} (z)$ and $\Delta b_i^{(l)} (z)$ contribute at finite orders in $\as$. Below we will find it useful that for the ${\hat g}_1$ structure function and the corresponding quark hPDF sourced by a gluon, the finite (in the $\epsilon \to 0$ limit) term at the order-$\as^2$ is \cite{Zijlstra:1993sh}
\begin{align}\label{caa}
    \Delta {\hat \Sigma}_G (x) \supset \Delta c_g^{(2)} + \Delta P_{GG}^{(1)} \otimes \Delta a_G^{(1)} + \Delta P_{qG}^{(1)} \otimes \Delta a_q^{(1)}
\end{align}
with
\cite{Zijlstra:1993sh}
\begin{align}
\Delta a_G^{(1)}  (z) = - \frac{\as \, N_f}{8 \pi} \, \ln^2 z , \ \ \ \Delta a_q^{(1)} = \frac{\as \, N_c}{16 \pi} \, \ln^2 z 
\end{align}
in the $z \ll 1$ and the large-$N_c \& N_f$ limits.

As discussed earlier, the DGLAP-type parton emission is driven in the KPS-CTT evolution by the terms involving the type-2 polarized dipole amplitude. In order to perform the crosscheck, it is then reasonable to work, for simplicity, with the initial condition such that only $G_2(x^2_{10},zs)$ is non-zero (cf. \cite{Cougoulic:2022gbk}). In particular, we take
\begin{align}\label{IChPDF}
G_2^{(0)} &= \frac{\alpha_s\pi^2}{2N_c}\;\;\;\;\;\text{and}\;\;\;\;\;Q^{(0)} = {\widetilde G}^{(0)} = 0
\end{align}
as the initial condition of our small-$x$ evolution. With the help of Eqs.~\eqref{hPDFs}, this translates to the following helicity PDFs in our pDIS scheme:
\begin{align}\label{ICdG}
\Delta G^{(0)}(x,Q^2) &= 1\;\;\;\;\;\text{and}\;\;\;\;\;\Delta\Sigma^{(0)}(x,Q^2) = 0\,.
\end{align}

Below we will calculate $\Delta G$ and $\Delta \Sigma$ order-by-order in $\as$, such that 
\begin{subequations}\label{pert_exp}
    \begin{align}
        & \Delta \Sigma (x, Q^2) = \Delta\Sigma^{(0)}(x,Q^2) + \Delta\Sigma^{(1)}(x,Q^2) + \Delta\Sigma^{(2)}(x,Q^2) + \ldots , \\
        & \Delta G (x, Q^2) = \Delta G^{(0)}(x,Q^2) + \Delta G^{(1)}(x,Q^2) + \Delta G^{(2)}(x,Q^2) + \ldots , 
    \end{align}
\end{subequations}
where the index in the superscript corresponds to the power of $\as$ correction to $\Delta\Sigma^{(0)}$ and $\Delta G^{(0)}$ from \eq{ICdG}.


\subsection{Order-$\as$ Corrections}

Before iterating the small-$x$ evolution, we substitute the initial condition \eqref{IChPDF} into Eq.~\eqref{qkhPDF} for the quark helicity PDF. This gives
\begin{align}\label{qk1}
\Delta\ord{\Sigma}{1}(x,Q^2) &=  - \frac{\alpha_sN_f}{2\pi}  \left[ \frac{1}{2} \ln^2\frac{1}{x}  + \ln\frac{1}{x} \ln\frac{Q^2}{\Lambda^2}  \right]  ,
\end{align}
which is of order-$\alpha_s$. Hence, the result \eqref{qk1} is more properly associated with the first-order quark helicity PDF. This shift in the order is consistent with the fact that the operator definition of $\Delta\Sigma$ already involves one parton loop at small $x$ \cite{Cougoulic:2022gbk,Kovchegov:2018znm}. Finally, notice that the additional power of $\alpha_s$ brings two additional logarithmic factors, each of which is either transverse ($\ln\frac{Q^2}{\Lambda^2}$) or longitudinal ($\ln\frac{1}{x}$). This is consistent with the DLA nature of the small-$x$ helicity evolution.

Now, we are ready to iterate the evolution. First, we substitute Eqs.~\eqref{IChPDF} into the KPS-CTT evolution equations \eqref{eq_LargeNcNf}. From \eq{eq_LargeNcNfe}, we obtain the first-order polarized dipole amplitude,
\begin{align}\label{G2ord1}
\ord{G}{1}_2(x^2_{10},zs) &= \frac{2 \, \alpha_sN_c}{\pi} \, \ln \left(zsx^2_{10}\right)\ln \left( \frac{1}{x^2_{10}\Lambda^2} \right) \, G_2^{(0)} \, ,
\end{align}
which, via \eq{glhPDF}, yields the first-order gluon helicity PDF of
\begin{align}\label{gl1}
&\Delta\ord{G}{1}(x,Q^2) =\frac{2 \, \alpha_sN_c}{\pi} \, \ln\frac{1}{x} \ln\frac{Q^2}{\Lambda^2} \, .
\end{align}

Let us compare the results in Eqs.~\eqref{qk1} and \eqref{gl1} to the predictions of the finite-order DGLAP-based calculations. First we note that the initial conditions \eqref{IChPDF} for small-$x$ evolution in pDIS scheme may not map precisely onto the initial conditions for the DGLAP evolution \eqref{hDGLAP} in the $\overline{\text{MS}}$ scheme. We thus write, in full generality,
\begin{subequations}\label{ICDGLAP1}
\begin{align}
&\Delta {\tilde G} (x,\Lambda^2) = 1 + \sum_{n=1}^{\infty}a_n\left(\alpha_s\ln^{2}\frac{1}{x}\right)^n ,     \label{ICDGLAP1gl} \\
&\Delta {\tilde \Sigma} (x,\Lambda^2) = \sum_{n=1}^{\infty}b_n\left(\alpha_s\ln^{2}\frac{1}{x}\right)^n , \label{ICDGLAP1qk} 
\end{align}
\end{subequations}
with $a_n$'s and $b_n$'s some unknown parameters.

Employing Eqs.~\eqref{ICDGLAP1} along with Eqs.~\eqref{Pfuncs} in \eq{hDGLAP} yields the order-$\as$ hPDFs in the $\overline{\text{MS}}$ scheme:
\begin{subequations}\label{DGLAPord1}
\begin{align}
&\Delta {\tilde G}^{(1)}(x,Q^2) = \alpha_s\left[ a_1\ln^2\frac{1}{x} + \frac{2N_c}{\pi}\ln\frac{1}{x}\ln\frac{Q^2}{\Lambda^2} \right] ,     \label{DGLAPord1gl} \\
&\Delta {\tilde \Sigma}^{(1)}(x,Q^2) = \alpha_s\left[ b_1\ln^2\frac{1}{x} - \frac{N_f}{2\pi}\ln\frac{1}{x}\ln\frac{Q^2}{\Lambda^2} \right] .  \label{DGLAPord1qk} 
\end{align}
\end{subequations}
While a direct comparison of Eqs.~\eqref{DGLAPord1} to the hPDFs in Eqs.~\eqref{qk1} and \eqref{gl1} is impossible, since the two sets of hPDFs are potentially in different schemes, we can employ \eq{Sigma_tildeSigmaG}, which readily yields 
\begin{align}\label{Sigma_tildeSigmaG1}
\Delta \Sigma^{(1)} (x, Q^2) = \Delta {\tilde \Sigma}^{(1)} (x, Q^2) + \int\limits_x^1 \frac{dz}{z} \, \Delta c_G^{(1)} (z) \,  \Delta {\tilde G}^{(0)} \left( \frac{x}{z} , Q^2\right)
\end{align}
for $b_1 =0$. Here $\Delta c_g^{(1)} (z)$ is the order-$\as$ term in \eq{dCg}. It appears that for the scheme-independent quantity, $\Delta \Sigma \sim g_1$, we have an agreement between the two approaches at this leading non-trivial order in $\as$. The agreement requires that $b_1 =0$.

We also note that \eq{G2ord1} is in agreement with \eq{DGLAPord1gl} for $a_1 =0$ in the latter, though at this point we cannot equate the two gluon distributions.


\subsection{Order-$\as^2$ Corrections}

Similar to the above, substituting Eqs.~\eqref{IChPDF} into \eq{eq_LargeNcNfa}, we obtain
\begin{align}\label{Qord1}
\ord{Q}{1}(x^2_{10},zs) &= \frac{\alpha_s \,N_c}{\pi} \left[\frac{5}{4}\ln^2(zsx^2_{10}) + \frac{1}{2}\ln \left( \frac{1}{x^2_{10}\Lambda^2} \right) \ln(zsx^2_{10}) \right] \, G_2^{(0)} .    
\end{align}
Further, substituting Eqs.~\eqref{G2ord1} and \eqref{Qord1} into \eq{qkhPDF} yields the second-order quark helicity PDF,
\begin{align}\label{qk2}
&\Delta\ord{\Sigma}{2}(x,Q^2) =  - \left(\frac{\alpha_sN_c}{\pi}\right)\left(\frac{\alpha_sN_f}{4\pi}\right) \left[ \frac{7}{24}\ln^4\frac{1}{x} + \frac{7}{6}\ln^3\frac{1}{x}\ln\frac{Q^2}{\Lambda^2} + \frac{9}{8}\ln^2\frac{1}{x}\ln^2\frac{Q^2}{\Lambda^2} \right]   .    
\end{align}

Once again, using Eqs.~\eqref{IChPDF} in \eq{eq_LargeNcNfc} gives us the first-order type-1 adjoint dipole amplitude, 
\begin{align}
    {\widetilde G}^{(1)} = \left[ \frac{\as N_c}{2 \pi} \left( 2 - \frac{N_f}{4 \, N_c} \right) \, \ln^2 (z s x_{10}^2) - \frac{\as N_f}{4 \pi} \, \ln (z s x_{10}^2) \, \ln \frac{1}{x_{10}^2 \, \Lambda^2} \right] \, G_2^{(0)} \, .
\end{align}
We then substitute this result, together with \eq{G2ord1}, into \eq{eq_LargeNcNfe} in order to obtain the order-$\as^2$ type-2 dipole amplitude, $\ord{G}{2}_2$. With the help of \eq{glhPDF}, this result gives us the second-order gluon helicity PDF,
\begin{align}\label{gl2}
\Delta\ord{G}{2}(x,Q^2) &=   \left( \frac{\alpha_s \, N_c}{\pi} \right)^2   \left[  \frac{1}{3}\left(1 - \frac{N_f}{8N_c}\right) \ln^3\frac{1}{x} \ln\frac{Q^2}{\Lambda^2}   + \left(1 - \frac{N_f}{16N_c}\right) \ln^2\frac{1}{x}\ln^2\frac{Q^2}{\Lambda^2} \right] . 
\end{align}

Employing the initial conditions \eqref{ICDGLAP1} with $b_1 =0$ in \eq{hDGLAP} we obtain the following $\overline{\text{MS}}$ hPDFs at the order-$\as^2$:
\begin{subequations}\label{DGLAPord2}
\begin{align}
&\Delta {\tilde G}^{(2)}(x,Q^2) = \alpha_s^2 \left\{ a_2\ln^4\frac{1}{x} + \frac{2 N_c}{3 \pi}\left[\frac{N_c}{16\pi}\left(8-\frac{N_f}{N_c}\right) + a_1 \right]\ln^3\frac{1}{x}\ln\frac{Q^2}{\Lambda^2} + \frac{N_c^2}{\pi^2}\left(1 - \frac{N_f}{16N_c}\right)\ln^2\frac{1}{x}\ln^2\frac{Q^2}{\Lambda^2} \right\}   ,  \label{DGLAPord2gl} \\
&\Delta {\tilde \Sigma}^{(2)}(x,Q^2) =  \alpha_s^2\left[ b_2\ln^4\frac{1}{x}  - \frac{N_f}{6 \pi} \, \left( a_1 + \frac{5 N_c}{8 \pi}  \right) \, \ln^3\frac{1}{x}\ln\frac{Q^2}{\Lambda^2} - \frac{9N_cN_f}{32\pi^2}\ln^2\frac{1}{x}\ln^2\frac{Q^2}{\Lambda^2} \right] .   \label{DGLAPord2qk} 
\end{align}
\end{subequations}

The agreement between \eq{qk2} and \eq{DGLAPord2qk} is obtained by using \eq{Sigma_tildeSigmaG} augmented by \eq{caa}, which at this order in $\as$ gives
\begin{align}\label{Sigma_tildeSigmaG2}
\Delta \Sigma^{(2)} (x, Q^2) = \Delta {\tilde \Sigma}^{(2)} (x, Q^2) + \int\limits_x^1 \frac{dz}{z} \, \left\{ \Delta c_q^{(1)} (z) \,  \Delta {\tilde \Sigma}^{(1)} \left( \frac{x}{z} , Q^2\right) + \Delta c_G^{(1)} (z) \,  \Delta {\tilde G}^{(1)} \left( \frac{x}{z} , Q^2\right) \right. \notag \\ \left. +  \left[ \Delta c_G^{(2)} + \Delta P_{GG}^{(1)} \otimes \Delta a_G^{(1)} + \Delta P_{qG}^{(1)} \otimes \Delta a_q^{(1)} \right] (z) \  \Delta {\tilde G}^{(0)} \left( \frac{x}{z} , Q^2\right) \right\}.
\end{align}
The agreement requires that
\begin{align}
    a_1 =0 , \ \ \ b_2 = \frac{23}{48} \, \frac{N_c \, N_f}{8 \pi^2} . 
\end{align}
Further, putting $a_2 =0$ would make the $\overline{\text{MS}}$ gluon hPDF contribution in \eq{DGLAPord2gl} agree with \eq{gl2}, though we do not expect these two hPDFs to be necessarily equal, as they may be in different renormalization schemes.


\subsection{Order-$\as^3$ Corrections}

Further iterations of small-$x$ evolution equations Eqs.~\eqref{eq_LargeNcNf}, involving the calculation of the first-order neighbor dipole amplitudes and the second-order ordinary dipole amplitudes of all types lead to the third-order quark and gluon helicity PDFs,
\begin{subequations}\label{qkgl3}
\begin{align}
&\Delta\ord{\Sigma}{3}(x,Q^2) = - \frac{\alpha_s^3N_c^2N_f}{16\pi^3} \left[\frac{1}{1440}\left(221 - 20\frac{N_f}{N_c}\right)\ln^6\frac{1}{x} + \frac{1}{240}\left(221 - 20\frac{N_f}{N_c}\right)\ln^5\frac{1}{x}\ln\frac{Q^2}{\Lambda^2} \right. \label{qk3} \\
&\;\;\;\;\;+ \left. \frac{1}{96}\left(151 - 12\frac{N_f}{N_c}\right)\ln^4\frac{1}{x}\ln^2\frac{Q^2}{\Lambda^2} + \frac{1}{72}\left(73 - 4\frac{N_f}{N_c}\right)\ln^3\frac{1}{x}\ln^3\frac{Q^2}{\Lambda^2}\right]    , \notag \\
&\Delta\ord{G}{3}(x,Q^2) = \left( \frac{\alpha_s N_c}{\pi} \right)^3 \left[ \frac{1}{960}\left( 56 - 13\frac{N_f}{N_c} \right)\ln^5\frac{1}{x}\ln\frac{Q^2}{\Lambda^2}   \right. \label{gl3} \\
&\;\;\;\;\;+ \left. \frac{1}{384}\left( 64 - 15\frac{N_f}{N_c} \right)\ln^4\frac{1}{x}\ln^2\frac{Q^2}{\Lambda^2} + \frac{1}{576}\left( 128 - 17\frac{N_f}{N_c} \right)\ln^3\frac{1}{x}\ln^3\frac{Q^2}{\Lambda^2}  \right] . \notag
\end{align}
\end{subequations}
Thus, starting with the initial condition \eqref{IChPDF}, we have determined the helicity PDFs at small $x$ up to the third order in $\alpha_s$ through an iterative calculation of the evolution kernel in the KPS-CTT equations.

The corresponding $\overline{\text{MS}}$ hPDFs at the order-$\as^3$ are 
\begin{subequations}
\begin{align}
\Delta {\tilde \Sigma}^{(3)} (x,Q^2) = & \, - \frac{\as \, N_f}{4 \pi} \, \left( \frac{\as \, N_c}{2 \pi} \right)^2 \, \left[ \frac{1}{60} \,  \left( 17 - 3 \, \frac{N_f}{N_c} \right)  \, \ln  \frac{Q^2}{\Lambda^2} \, \ln^5 \frac{1}{x} + \frac{1}{48} \, \left( 39 - 4 \, \frac{N_f}{N_c} \right)  \, \ln^2  \frac{Q^2}{\Lambda^2} \, \ln^4 \frac{1}{x} \right. \\ 
& \left. + \frac{1}{72} \, \left( 73 - 4 \, \frac{N_f}{N_c} \right)  \, \ln^3  \frac{Q^2}{\Lambda^2} \, \ln^3 \frac{1}{x} \right] + \as^3 \, b_3 \, \ln^6\frac{1}{x}  + \frac{\as^3 }{20 \pi} \, ( N_c \, b_2 - 2 N_f \, a_2) \, \ln  \frac{Q^2}{\Lambda^2} \, \ln^5 \frac{1}{x} , 
\notag \\
\Delta {\tilde G}^{(3)} (x,Q^2) = & \left( \frac{\alpha_s N_c}{\pi} \right)^3 \left[ \frac{1}{3840}\left( 224 - 39 \frac{N_f}{N_c} \right)\ln^5\frac{1}{x}\ln\frac{Q^2}{\Lambda^2}  + \frac{1}{384}\left( 64 - 13\frac{N_f}{N_c} \right)\ln^4\frac{1}{x}\ln^2\frac{Q^2}{\Lambda^2} \right. \label{gl3} \\
&\;\;\;\;\;+ \left.
 \frac{1}{576}\left( 128 - 17\frac{N_f}{N_c} \right)\ln^3\frac{1}{x}\ln^3\frac{Q^2}{\Lambda^2}  \right] + \as^3 \, a_3 \, \ln^6\frac{1}{x} + \frac{\as^3 N_c}{10 \pi} \, ( b_2 + 4 \, a_2) \, \ln  \frac{Q^2}{\Lambda^2} \, \ln^5 \frac{1}{x}. \notag
\end{align}
\end{subequations}
Concentrating on the cubic and quadratic terms in $\ln (Q^2/\Lambda^2)$ contributing to $\Delta \Sigma^{(3)}$, we observe that
\begin{align}\label{DSigma3_subtr}
& \Delta \Sigma^{(3)} - \Delta {\tilde \Sigma}^{(3)} - \delta c_q^{(1)} \otimes \Delta {\tilde \Sigma}^{(2)}  - \delta c_G^{(1)} \otimes \Delta G^{(2)} = - \frac{\as \, N_f}{4 \pi}  \, \left( \frac{\as \, N_c}{2 \pi} \right)^2 \\ & \times \, \left[ \frac{1}{1440} \, \left( 221 - 20 \, \frac{N_f}{N_c} \right)  \, \ln^6 \frac{1}{x} + \frac{1}{60} \,  \left( 29 -  \frac{N_f}{N_c} \right)  \, \ln  \frac{Q^2}{\Lambda^2} \, \ln^5 \frac{1}{x}  \right] . \notag
\end{align}
The comparison of the linear terms in $\ln (Q^2/\Lambda^2)$ and the terms independent of $Q^2$ would require knowledge of $\Delta a_G^{(2)}$, $\Delta a_q^{(2)}$, $\Delta b_G^{(1)}$ and $\Delta b_q^{(1)}$, which can be found in \cite{Blumlein:2022gpp}. However, it is clear that any values of those coefficients can be accommodated by an appropriate choice of $b_3$ and $a_2$. This may lead to disagreement between $\Delta G^{(2)} + \Delta G^{(3)}$ and $\Delta {\tilde G}^{(2)} + \Delta {\tilde G}^{(3)}$, but it is possible that such disagreement could be ascribed to the scheme dependence of the gluon hPDF.


\subsection{Exploring the Scheme Dependence}

At this point it appears that there is a difference between the hPDFs resulting from the iterative solution of our large-$N_c \& N_f$ evolution equations \eqref{eq_LargeNcNf} and the $\overline{\text{MS}}$ hPDFs. Here we explore the possibility that the difference is solely due to the scheme dependence.  

Suppose the hPDFs we obtained using Eqs.~\eqref{eq_LargeNcNf} satisfy DGLAP evolution equations
\begin{align}\label{DGLAP_diff}
\frac{\pd}{\pd \ln Q^2}
\begin{pmatrix}
 \Delta {\Sigma} (x, Q^2) \\
 \Delta {G} (x, Q^2)
\end{pmatrix}
= 
\int\limits_x^1 \frac{dz}{z} \, 
\begin{pmatrix}
\Delta P_{qq} (z) & \Delta P_{qG} (z) \\
\Delta P_{Gq} (z)  & \Delta P_{GG} (z) 
\end{pmatrix}
\,
\begin{pmatrix}
 \Delta {\Sigma} \left( \frac{x}{z} , Q^2 \right) \\
 \Delta {G} \left( \frac{x}{z} , Q^2 \right)
\end{pmatrix}
\equiv 
\left[\Delta\mathbf{P} \otimes
\begin{pmatrix} 
\Delta {\Sigma}   \\
\Delta {G} 
\end{pmatrix} \right] (x, Q^2 )
\end{align}
with some unknown helicity splitting functions $\Delta P_{ij} (z)$. If \eq{DGLAP_diff} holds for our hPDFs, then perturbative expansions of the splitting functions $\Delta P_{ij} (z)$ can be obtained using the results of the above iterative solution. 

Substituting Eqs.~\eqref{ICdG}, \eqref{qk1}, \eqref{gl1}, \eqref{qk2}, \eqref{gl2}, and \eqref{qkgl3} into \eq{pert_exp}, and using it in \eq{DGLAP_diff} order-by-order in $\as$,
after some algebra we end up with the following splitting functions (at small $x$ and large $N_c \& N_f$) in the pDIS scheme:
\begin{subequations}\label{P'}
\begin{align}
& \Delta P_{qq} (z) = \frac{\as N_c}{4 \pi} + \frac{1}{8} \, \left(1 - 8 \, \frac{N_f}{N_c} \right) \, \left( \frac{\as N_c}{2 \pi} \right)^2 \, \ln^2 \frac{1}{z} + {\cal O} (\as^3) , \\
& \Delta P_{qG} (z) = - \frac{\as N_f}{2 \pi} - \frac{13}{16} \, \frac{\as^2 N_c N_f}{\pi^2} \, \ln^2 \frac{1}{z} + {\cal O} (\as^3)  , \label{P'_qG} \\
& \Delta P_{Gq} (z) = \frac{\as N_c}{2 \pi} + \frac{1}{2} \, \left( \frac{\as N_c}{\pi} \right)^2 \, \ln^2 \frac{1}{z} + {\cal O} (\as^3)  , \\
& \Delta P_{GG} (z) = 2 \, \frac{\as N_c}{\pi} + \left( \frac{\as N_c}{\pi} \right)^2 \, \ln^2 \frac{1}{z} + {\cal O} (\as^3) .
\end{align}
\end{subequations}
Note that the coefficients $\Delta a_i^{(l)}$ and $\Delta b_i^{(l)}$ do not contribute to the splitting functions \eqref{P'} at the shown orders in $\as$: at the orders $\as$ and $\as^2$ these coefficients do not contribute to \eq{DGLAP_diff}, while at the order $\as^3$ they only affect the $Q^2$-independent terms, which contribute to the polarized splitting functions only at NNLO. Therefore, for the LO+NLO splitting functions \eqref{P'} we can ignore the difference between the factorized and unfactorized expressions for the $g_1$ structure function and $\Delta \Sigma$.

If our hPDFs were different from the ones in the $\overline{\text{MS}}$ scheme only by the scheme dependence, then the two would be related by (see e.g. \cite{vanNeerven:2000uj,Moch:2014sna})
\begin{align}\label{hPDFscheme}
\begin{pmatrix} 
\Delta \Sigma \left(x, Q^2\right)  \\
\Delta G \left(x, Q^2\right)
\end{pmatrix}
&= \left[\Delta\mathbf{C} \otimes
\begin{pmatrix} 
\Delta\tilde{\Sigma}   \\
\Delta\tilde{G} 
\end{pmatrix} \right] (x, Q^2 ) \, 
,
\end{align}
with some unknown matrix of coefficient functions, 
\begin{align}\label{coef_matrix}
\Delta {\mathbf C} (z) = 
\begin{pmatrix}
 z \delta (1-z) + \as \, \Delta C_{qq}^{(1)} \, \ln \frac{1}{z} + \as^2 \, \Delta C_{qq}^{(2)} \, \ln^3 \frac{1}{z} + \ldots  & \as \, \Delta C_{qG}^{(1)} \, \ln \frac{1}{z} + \as^2 \, \Delta C_{qG}^{(2)} \, \ln^3 \frac{1}{z} + \ldots   \\
\as \, \Delta C_{Gq}^{(1)} \, \ln \frac{1}{z} + \as^2 \, \Delta C_{Gq}^{(2)} \, \ln^3 \frac{1}{z} + \ldots  & z \delta (1-z) + \as \, \Delta C_{GG}^{(1)} \, \ln \frac{1}{z} + \as^2 \, \Delta C_{GG}^{(2)} \, \ln^3 \frac{1}{z} + \ldots  
\end{pmatrix}
.
\end{align}

Equation \eqref{hPDFscheme} implies that 
\begin{align}\label{P'P}
\Delta {\mathbf P} = \Delta {\mathbf C} \otimes \Delta {\mathbf {\tilde P}} \otimes \Delta {\mathbf C}^{-1} ,
\end{align}
with ${\mathbf {\tilde P}}$ the splitting function matrix in the $\overline{\text{MS}}$ scheme. Rewriting \eq{P'P} as
\begin{align}\label{P'CCP}
\Delta {\mathbf P} \otimes \Delta {\mathbf C} = \Delta {\mathbf C} \otimes \Delta {\mathbf {\tilde P}}
\end{align}
allows one to construct the coefficients \eqref{coef_matrix} order-by-order in $\as$.

At the order-$\as$ the splitting functions are the same in pDIS and $\overline{\text{MS}}$ schemes, cf. \eq{P'} and \eq{Pfuncs}. Hence \eq{P'CCP} is trivially satisfied at this order. 

At the order-$\as^2$, \eq{P'CCP} gives the following conditions,
\begin{subequations}\label{CCC1}
\begin{align}
& N_f \, \Delta C_{Gq}^{(1)}  + N_c \, \Delta C_{qG}^{(1)} = - \frac{N_c \, N_f}{2 \pi}, \\
& \frac{N_f}{2} \, \Delta C_{qq}^{(1)} - \frac{7}{4} \, N_c \, \Delta C_{qG}^{(1)} - \frac{N_f}{2} \, \Delta C_{GG}^{(1)} =   \frac{N_c \, N_f}{\pi}, \\
&\Delta C_{qq}^{(1)} + \frac{7}{2} \, \Delta C_{Gq}^{(1)} -  \Delta C_{GG}^{(1)} = - \frac{3 \, N_c}{4 \pi}.
\end{align}
\end{subequations}
Equations~\eqref{CCC1} do not have a solution for $N_f \neq 0$. Therefore, the difference between our hPDFs and the ones in the $\overline{\text{MS}}$ scheme cannot be entirely attributed to the scheme dependence, putting our interpretation of the earlier cross-checks in question. 


\subsection{Further Discussion}\label{sec:DGLAPdiscussion}

The origin of the discrepancies found above is not clear to us at the moment. Indeed, since all the discrepancies come with a factor of $N_f$, one may suspect that some quark contributions are missing in the evolution of \cite{Cougoulic:2022gbk}. A possible missing piece could be due to the terms where an $s$-channel quark becomes a gluon (and {\sl vice versa}) after interacting with the shock wave. Such contributions were considered in \cite{Kovchegov:2015pbl}, see Sec.~IV there, where they were argued not to contribute to the flavor-singlet helicity evolution at DLA. These contributions appear to contribute to the flavor non-singlet helicity evolution, as was argued in \cite{Kovchegov:2018znm}, but at the subleading-$N_c$ order. The contribution of quark-to-gluon and gluon-to-quark transition terms has been studied in detail in \cite{Chirilli:2021lif}. If such terms do contribute to the flavor-singlet helicity evolution, perhaps due to some loophole in the arguments of \cite{Kovchegov:2015pbl,Kovchegov:2018znm}, it appears difficult to include them into any closed-form evolution equations at large-$N_c \& N_f$.     

However, let us explore another possibility here, by considering the exact solution of small-$x$ helicity evolution equations in the large-$N_c$ case from \cite{Borden:2023ugd}. For our initial condition \eqref{IChPDF}, we substitute Eqs.~(63) from \cite{Borden:2023ugd} into Eq.~(58) in the same reference, multiply the result by $\as \pi^2/(2 N_c)$,  obtaining
\begin{align}\label{DS_exact}
\Delta \Sigma (x, Q^2) = - \frac{N_f}{4 N_c} \, \int \frac{d \omega}{2 \pi i} \, e^{\omega \, \ln \tfrac{1}{x}} \, \left[ \frac{1}{\omega - \Delta \gamma_{GG}(\omega)} \, e^{\Delta \gamma_{GG}(\omega) \, \ln \left( \tfrac{Q^2}{\Lambda^2} \right) }  - \frac{1}{\omega} \right].
\end{align}
Similarly, Eq.~(64) in \cite{Borden:2023ugd}, multiplied by $\as \pi^2/(2 N_c)$, gives
\begin{align}\label{DG_exact}
\Delta G (x, Q^2) = \int \frac{d \omega}{2 \pi i} \, e^{\omega \, \ln \tfrac{1}{x} + \Delta \gamma_{GG}(\omega) \, \ln \left( \tfrac{Q^2}{\Lambda^2} \right) } \, \frac{1}{\omega} .
\end{align}
The anomalous dimension $\Delta \gamma_{GG}(\omega)$ is given in Eq.~(65) of the same reference, 
\begin{align}\label{anomalous_dim}
\Delta \gamma_{GG}(\omega) = \frac{\omega}{2}\left[1 - \sqrt{1 - \frac{16\,\bas}{\omega^2}\sqrt{1-\frac{4\,\bas}{\omega^2}} } \ \right]
\end{align}
with
\begin{align}\label{bas_def}
    \bas \equiv \frac{\as \, N_c}{2 \pi} .
\end{align}

Next, let us compare this to the solution of DGLAP equations in the same approximation. At fixed coupling (as is the case in DLA), the solution of DGLAP equation \eqref{DGLAP_diff} is
\begin{align}\label{DGLAP_sol_1}
\begin{pmatrix}
 \Delta {\Sigma} (x, Q^2) \\
 \Delta {G} (x, Q^2)
\end{pmatrix}
= \int \frac{d \omega}{2 \pi i} \, e^{\omega \, \ln \tfrac{1}{x}} \, 
\begin{pmatrix}
 \Delta {\Sigma}_\omega (Q^2) \\
 \Delta {G}_\omega (Q^2)
\end{pmatrix}
= \int \frac{d \omega}{2 \pi i} \, e^{\omega \, \ln \tfrac{1}{x}} \, 
\exp \left\{ 
\begin{pmatrix}
\Delta \gamma_{qq} (\omega) & \Delta \gamma_{qG} (\omega) \\
\Delta \gamma_{Gq} (\omega)  & \Delta \gamma_{GG} (\omega) 
\end{pmatrix}
\, \ln \frac{Q^2}{\Lambda^2}
\right\}
\begin{pmatrix}
 \Delta {\Sigma}_\omega (\Lambda^2) \\
 \Delta {G}_\omega (\Lambda^2)
\end{pmatrix} .
\end{align}
Defining the eigenvalues
\begin{subequations}
\begin{align}
\lambda_1 = \frac{1}{2} \left[ \Delta \gamma_{qq} + \Delta \gamma_{GG} + \sqrt{(\Delta \gamma_{qq} - \Delta \gamma_{GG})^2 + 4 \, \Delta \gamma_{qG} \, \Delta \gamma_{Gq}} \right] \, \ln \frac{Q^2}{\Lambda^2}, \\
\lambda_2 = \frac{1}{2} \left[ \Delta \gamma_{qq} + \Delta \gamma_{GG} - \sqrt{(\Delta \gamma_{qq} - \Delta \gamma_{GG})^2 + 4 \, \Delta \gamma_{qG} \, \Delta \gamma_{Gq}} \right] \, \ln \frac{Q^2}{\Lambda^2},
\end{align}
\end{subequations}
we rewrite \eq{DGLAP_sol_1} as
\begin{align}\label{DGLAP_sol_2}
& \begin{pmatrix}
 \Delta {\Sigma} (x, Q^2) \\
 \Delta {G} (x, Q^2)
\end{pmatrix}
= \int \frac{d \omega}{2 \pi i} \, e^{\omega \, \ln \tfrac{1}{x}} \\
& \times \, 
\begin{pmatrix}
\frac{e^{\lambda_1} + e^{\lambda_2}}{2} +  (\Delta \gamma_{qq} - \Delta \gamma_{GG}) \frac{e^{\lambda_1} - e^{\lambda_2}}{2 (\lambda_1 - \lambda_2)} \, \ln \frac{Q^2}{\Lambda^2} & \Delta \gamma_{qG} \, \frac{e^{\lambda_1} - e^{\lambda_2}}{\lambda_1 - \lambda_2} \, \ln \frac{Q^2}{\Lambda^2} \\
\Delta \gamma_{Gq} \, \frac{e^{\lambda_1} - e^{\lambda_2}}{\lambda_1 - \lambda_2} \, \ln \frac{Q^2}{\Lambda^2} & \frac{e^{\lambda_1} + e^{\lambda_2}}{2} -  (\Delta \gamma_{qq} - \Delta \gamma_{GG}) \frac{e^{\lambda_1} - e^{\lambda_2}}{2 (\lambda_1 - \lambda_2)} \, \ln \frac{Q^2}{\Lambda^2}
\end{pmatrix}
\begin{pmatrix}
 \Delta {\Sigma}_\omega (\Lambda^2) \\
 \Delta {G}_\omega (\Lambda^2)
\end{pmatrix} . \notag
\end{align}

If we take $\Delta {\Sigma} (x, \Lambda^2) =0$,  $\Delta {G} (x, \Lambda^2) =1$ initial conditions, as we did above, then $ \Delta {\Sigma}_\omega (\Lambda^2) =0$ and $\Delta {G}_\omega (\Lambda^2) = 1/\omega$. Further, we remove all quarks except for the ``last one" giving $\Delta \Sigma$, that is put $\Delta \gamma_{qq} = \Delta \gamma_{Gq} = 0$ everywhere and $\Delta \gamma_{qG} =0$ everywhere except for the explicit factor of $\Delta \gamma_{qG}$ in the upper right corner of the matrix in \eq{DGLAP_sol_2}. This leads to $\lambda_1 =  \Delta \gamma_{GG} \, \ln \frac{Q^2}{\Lambda^2}$ and $\lambda_2 =0$. Using all this in \eq{DGLAP_sol_2} yields
\begin{align}\label{DGLAP_sol_3}
\begin{pmatrix}
 \Delta {\Sigma} (x, Q^2) \\
 \Delta {G} (x, Q^2)
\end{pmatrix}
= \int \frac{d \omega}{2 \pi i} \, e^{\omega \, \ln \tfrac{1}{x}} \, 
\begin{pmatrix}
1 & \Delta \gamma_{qG} \, \frac{e^{\Delta \gamma_{GG} \, \ln \frac{Q^2}{\Lambda^2}} - 1}{\Delta \gamma_{GG} } \\
0 & e^{\Delta \gamma_{GG} \, \ln \frac{Q^2}{\Lambda^2}  }
\end{pmatrix}
\begin{pmatrix}
 0 \\
 \frac{1}{\omega}
\end{pmatrix} . 
\end{align}
Specifically, for $\Delta \Sigma$ we get
\begin{align}\label{DGLAP_sol_DS}
\Delta {\Sigma} (x, Q^2) = \int \frac{d \omega}{2 \pi i} \, \frac{\Delta \gamma_{qG} (\omega)}{\Delta \gamma_{GG} (\omega)} \, e^{\omega \, \ln \tfrac{1}{x}} \, \left[  e^{\Delta \gamma_{GG} (\omega) \, \ln \frac{Q^2}{\Lambda^2}} - 1 \right]  \,  \frac{1}{\omega} ,
\end{align}
while for $\Delta G$ we get exactly \eq{DG_exact}.

Equation \eqref{DGLAP_sol_DS} is not the same as \eq{DS_exact}. The anomalous dimensions in the exponents are the same, $\Delta \gamma_{GG} (\omega)$ and $0$. However, the prefactors of the exponents are different. If we expand the prefactors in Eqs.~\eqref{DS_exact} and \eqref{DGLAP_sol_DS} to the lowest order in $\as$ we would get
\begin{align}\label{DS_approx}
\Delta \Sigma (x, Q^2) \approx - \frac{N_f}{\as \, 2 \pi^2} \, \int \frac{d \omega}{2 \pi i} \, e^{\omega \, \ln \tfrac{1}{x}} \, \left[ e^{\gamma_\omega^- \, \ln \left( \tfrac{Q^2}{\Lambda^2} \right) }  - 1 \right] \, \frac{1}{\omega}
\end{align}
from both of them, since
\begin{align}
 \frac{\Delta \gamma_{qG}^{(1)} (\omega)}{\Delta \gamma_{GG}^{(1)} (\omega)} = - \frac{N_f}{4 N_c}.
\end{align}
 
However, the agreement between Eqs.~\eqref{DS_exact} and \eqref{DGLAP_sol_DS} ends beyond this leading order in $\as$ in the prefactor. Moreover, no scheme dependence, that is, no choice of $\Delta \gamma_{qG}$ and $\Delta \gamma_{GG}$ beyond LO would make Eqs.~\eqref{DS_exact} and \eqref{DGLAP_sol_DS} agree. Therefore, already at large $N_c$ we cannot attribute the difference between our result and that of DGLAP+coefficient functions calculation to a different scheme.

If we modify the initial conditions for DGLAP evolution to be (note the non-trivial quark hPDF)
\begin{align}\label{initial}
\begin{pmatrix}
 \Delta {\Sigma}_\omega (\Lambda^2) \\
 \Delta {G}_\omega (\Lambda^2)
\end{pmatrix} 
= 
\begin{pmatrix}
 - \frac{N_f}{4 N_c} \frac{1}{\omega} \frac{\Delta \gamma_{GG} (\omega)}{\omega - \Delta \gamma_{GG} (\omega)} \\
 \frac{1}{\omega}
\end{pmatrix} 
\end{align}
and rewrite \eq{DGLAP_sol_3} as
\begin{align}\label{DGLAP_sol_4}
\begin{pmatrix}
 \Delta {\Sigma} (x, Q^2) \\
 \Delta {G} (x, Q^2)
\end{pmatrix}
= \int \frac{d \omega}{2 \pi i} \, e^{\omega \, \ln \tfrac{1}{x}} \, 
\begin{pmatrix}
1 & \Delta \gamma_{qG} \, \frac{e^{\Delta \gamma_{GG} \, \ln \frac{Q^2}{\Lambda^2}} - 1}{\Delta \gamma_{GG} } \\
0 & e^{\Delta \gamma_{GG} \, \ln \frac{Q^2}{\Lambda^2}  }
\end{pmatrix}
\begin{pmatrix}
 - \frac{N_f}{4 N_c} \frac{1}{\omega} \frac{\Delta \gamma_{GG} (\omega)}{\omega - \Delta \gamma_{GG} (\omega)} \\
 \frac{1}{\omega}
\end{pmatrix} ,
\end{align}
then $\Delta G$ would remain the same, while $\Delta \Sigma$ would agree with  \eq{DS_exact} if the following relation holds for the large-$N_c$ small-$x$ anomalous dimensions:
\begin{align}\label{relation}
\Delta \gamma_{qG} (\omega) = - \frac{N_f}{4 N_c} \frac{\omega \, \Delta \gamma_{GG} (\omega)}{\omega - \Delta \gamma_{GG} (\omega)} .
\end{align}
However, this relation appears to be unlikely to hold to all orders in $\as$: it already breaks down beyond the leading order in $\as$ in the $\overline{\text{MS}}$ scheme.

Let us finally add that the gluon hPDF in \eq{DG_exact}, satisfying large-$N_c$ small-$x$ DGLAP evolution, is obtained in the exact solution from \cite{Borden:2023ugd} only for a very specific initial conditions, corresponding to $\Delta G^{(0)} (x, Q^2)=$~const and $\Delta \Sigma^{(0)} (x, Q^2)= 0$. Other initial conditions for the small-$x$ large-$N_c$ helicity evolution result in expressions for $\Delta G (x, Q^2)$ which, while still containing the anomalous dimension \eqref{anomalous_dim}, do not look quite as simple as \eq{DG_exact}. Therefore, our efforts to compare iterative solution of Eqs.~\eqref{eq_LargeNcNf} also appear to be strongly dependent on the choice of the initial conditions, making the choice in \eq{IChPDF} just one option, which does not necessarily make the resulting hPDFs agree with the DGLAP-based approaches. This conclusion is further corroborated by the fact that modifying the initial conditions \eqref{IChPDF} to 
\begin{align}
G_2^{(0)} = \frac{\as \pi^2}{2 N_c} \, \delta \left( z s x_{10}^2 -1 \right), \ \ \  Q^{(0)} = {\widetilde G}^{(0)} = 0 , 
\end{align}
results in disagreement between the polarized DGLAP and the iterative solution of Eqs.~\eqref{eq_LargeNcNf} at lower orders in $\as$ than what we observed above.


\section{Conclusions}
\label{sec:conclusion}

To conclude, let us re-iterate our main results. We have numerically solved the revised version of the large-$N_c \& N_f$ helicity evolution equations at small-$x$ \cite{Cougoulic:2022gbk}, given above in Eqs.~\eqref{evoleq}. The solution exhibited qualitatively different behavior depending on whether $N_f < 2 N_c$ or $N_f = 2 N_c$. For $N_f < 2 N_c$ all the polarized dipole amplitudes grow exponentially in rapidity (or, equivalently, with the logarithm of energy). The corresponding intercepts are summarized in Table~\ref{tab:lowNfinterceptsCont}. Comparing these intercepts with the earlier work by BER \cite{Bartels:1996wc} in Table~\ref{tab:BER_comparison}, we observe the discrepancy at the $2-3\%$ level, larger than our numerical precision but small enough not to be important for phenomenological applications. Similar, albeit smaller ($<0.1 \%$) discrepancy with BER has recently been observed in the analytic solution of the revised large-$N_c$ helicity evolution equations from \cite{Cougoulic:2022gbk} constructed in \cite{Borden:2023ugd}, where the origin of the discrepancy was traced down to the difference in the resummed polarized $GG$ anomalous dimensions.

For $N_f = 2 N_c$ our numerical solution of Eqs.~\eqref{evoleq} oscillates in $\ln (1/x)$ with the exponentially growing amplitude. This behavior of the solution for the revised evolution equations at $N_f = 2 N_c$ is reminiscent of the solution for the earlier version of the large-$N_c \& N_f$ helicity evolution equations constructed in \cite{Kovchegov:2020hgb}. The relevant parameters are summarized in the Table~\ref{tab:Nf6resultsCont} above. We noted that the oscillation period may be too high for the experimental detection of these oscillations. 

In addition, we have studied the dependence of the small-$x$ asymptotics on the initial conditions for the evolution. We found that the intercepts and the oscillation frequency (the latter for $N_f = 2 N_c$) are rather insensitive to the initial conditions (see Tables~\ref{tab:iconesNf4}, \ref{tab:icones}, \ref{tab:icbornNf4}, and \ref{tab:icborn}). The initial phase of the oscillations (also for $N_f = 2 N_c$) appears to depend rather strongly on the initial conditions (cf. \cite{Kovchegov:2020hgb}). We have also verified the target--projectile symmetry of the revised equations solution: this symmetry appeared to be missing in the earlier helicity evolution equations \cite{Kovchegov:2015pbl, Kovchegov:2016zex,  Kovchegov:2018znm}.

Finally, we have performed some limited cross-checks of the iterative solution of the revised large-$N_c \& N_f$ helicity evolution equations against the known finite-order calculations in the collinear factorization framework \cite{Altarelli:1977zs,Dokshitzer:1977sg,Zijlstra:1993sh,Mertig:1995ny,Moch:1999eb,vanNeerven:2000uj,Vermaseren:2005qc,Moch:2014sna,Blumlein:2021ryt,Blumlein:2021lmf,Davies:2022ofz,Blumlein:2022gpp}. The results were inconclusive: despite observing a good degree of agreement between the two approaches, we found some discrepancies as well, potentially related to our choice of the initial conditions for the helicity evolution. The investigation into the origin of these discrepancies is left for future work, which may involve an exact analytic solution of Eqs.~\eqref{eq_LargeNcNf} using the technique of \cite{Borden:2023ugd}. Using our disagreement with BER at large-$N_c \& N_f$ as an estimate of the potential error in our approach, and given the smallness of this disagreement, we are optimistic that our difference with polarized DGLAP evolution must also be comparably small. This would imply that our evolution should be sufficiently accurate to be used in further developing phenomenological predictions for the EIC along the lines of \cite{Adamiak:2021ppq}.


\section*{Acknowledgments}

\label{sec:acknowledgement}

The authors would like to thank Dr. Florian Cougoulic and Dr. Andrey Tarasov for the useful discussion about various aspects of this work. One of the authors (YK) is grateful to Professors Sven-Olaf Moch and Johannes Bluemlein for very helpful and informative discussions of perturbative QCD calculations of helicity parton distributions and the $g_1$ structure function. 

This material is based upon work supported by the U.S. Department of Energy, Office of Science, Office of Nuclear Physics under contract DE-AC05-06OR23177 and Award
Number DE-SC0004286. 

YT is supported by the Academy of Finland, the Centre of Excellence in Quark Matter and project 346567, under the European Union’s Horizon 2020 research and innovation programme by the European Research Council (ERC, grant agreement No. ERC-2018-ADG-835105 YoctoLHC) and by the STRONG-2020 project (grant agreement No. 824093). The content of this article does not reflect the official opinion of the European Union and responsibility for the information and views expressed therein lies entirely with the authors. 

The work is performed within the framework of the Saturated Glue (SURGE) Topical Theory Collaboration.




\providecommand{\href}[2]{#2}\begingroup\raggedright\endgroup

\end{document}